\documentclass[
 aps,
 prd,
 twocolumn,
 preprintnumbers,
 nofootinbib,
 floatfix,
 superscriptaddress,
 amsmath,amssymb,
longbibliography
]{revtex4-1}

\usepackage{amsmath}
\usepackage{amsfonts}
\usepackage{color}
\usepackage{tikz}
\tikzstyle{bag} = [align=center]

\usepackage{graphicx}
\usepackage[colorlinks=true,citecolor=blue,urlcolor=blue]{hyperref}

\usepackage[makeroom]{cancel}

\newcommand{\R}[1]{\textcolor{red}{#1}}

\DeclareMathOperator*{\argmin}{arg\,min}
\newcommand{\E}{\mathbb{E}\,}

\newcommand*{\defeq}{\mathrel{\vcenter{\baselineskip0.5ex \lineskiplimit0pt \hbox{\scriptsize.}\hbox{\scriptsize.}}}%
                     =}

\usepackage{bbold}

\begin{document}

\title{Tempered Multifidelity Importance Sampling for Gravitational Wave Parameter Estimation}

\author{Bassel Saleh}
 \email{bassel@utexas.edu}
 \affiliation{Oden Institute for Computational Engineering and Sciences, The University of Texas at Austin, Austin, TX 78712, USA}

\author{Aaron Zimmerman}
 \email{aaron.zimmerman@utexas.edu}
  \affiliation{Center for Gravitational Physics, The University of Texas at Austin, Austin, TX 78712, USA}

\author{Peng Chen}
 \email{pchen402@gatech.edu}
 \affiliation{College of Computing, Georgia Institute of Technology, Atlanta, GA 30332, USA}

\author{Omar Ghattas}%
 \email{omar@oden.utexas.edu}
 \affiliation{Oden Institute for Computational Engineering and Sciences, The University of Texas at Austin, Austin, TX 78712, USA}

\date{\today}

\begin{abstract}
Estimating the parameters of compact binaries which coalesce and produce gravitational waves is a challenging Bayesian inverse problem.
Gravitational-wave parameter estimation lies within the class of multifidelity problems, where a variety of models with differing assumptions, levels of fidelity, and computational cost are available for use in inference.
In an effort to accelerate the solution of a Bayesian inverse problem, cheaper surrogates for the best models may be used to reduce the cost of likelihood evaluations when sampling the posterior. 
Importance sampling can then be used to reweight these samples to represent the true target posterior, incurring a reduction in the effective sample size. 
In cases when the problem is high dimensional, or when the surrogate model produces a poor approximation of the true posterior, this reduction in effective samples can be dramatic and render multifidelity importance sampling ineffective. 
We propose a novel method of tempered multifidelity importance sampling in order to remedy this issue.
With this method the biasing distribution produced by the low-fidelity model is tempered, allowing for potentially better overlap with the target distribution. 
There is an optimal temperature which maximizes the efficiency in this setting, and we propose a low-cost strategy for approximating this optimal temperature using samples from the untempered distribution. 
In this paper, we motivate this method by applying it to Gaussian target and biasing distributions. 
Finally, we apply it to a series of problems in gravitational wave parameter estimation and demonstrate improved efficiencies when applying the method to real gravitational wave detections.
\end{abstract}

\maketitle

\section{Introduction}

The direct detection of gravitational waves (GWs)~\cite{LIGOScientific:2016aoc,LIGOScientific:2016sjg,LIGOScientific:2016dsl,LIGOScientific:2017bnn,LIGOScientific:2017ycc,LIGOScientific:2017vwq,LIGOScientific:2017vox,LIGOScientific:2018mvr,LIGOScientific:2020aai,LIGOScientific:2020stg,LIGOScientific:2020zkf,LIGOScientific:2020iuh,LIGOScientific:2020ibl,LIGOScientific:2021qlt,LIGOScientific:2020stg,LIGOScientific:2020zkf,LIGOScientific:2021usb,KAGRA:2021vkt,LIGOScientific:2024elc,Nitz:2018imz,Nitz:2020oeq,Nitz:2021uxj,Nitz:2021zwj,Zackay:2019tzo,Venumadhav:2019tad,Venumadhav:2019lyq,Zackay:2019btq,Olsen:2022pin,Mehta:2023zlk,Wadekar:2023gea} provides an unprecedented viewpoint on the most compact objects in the Universe.
Observations by the Advanced LIGO~\cite{LIGOScientific:2014pky}, Advanced Virgo~\cite{VIRGO:2014yos}, and KAGRA~\cite{KAGRA:2020tym} detectors reveal the properties of the black holes and neutron stars which emit GWs as they inspiral and coalesce.
Following detection of a GW event, the next step in applying GW data to a host of problems in fundamental physics and astrophysics is to measure the properties of the binary system that produced the signal from the noisy data.
The standard method for solving this inverse problem is through Bayesian inference, see e.g.~\cite{Veitch:2014wba,Thrane:2018qnx,Ashton:2018jfp,Romero-Shaw:2020owr}.

In GW data analysis, parameter estimation presents a number of challenges.
The parameters of the binary system such as the masses and spins of components and their location in the sky are highly correlated, and must be inferred with a high degree of accuracy for precision applications.
Further, the GW models needed for inference can be computationally expensive~\cite{Field:2013cfa}.
Full numerical simulations of the binary evolution provide the highest fidelity predictions, but are intractably expensive for use in most algorithms.
Large catalogs of simulations have been used together with likelihood interpolation and marginalization to infer the properties of GW sources, see e.g.~\cite{Healy:2020jjs}.
More commonly, numerical simulations are used together with analytical approximates to build a wide hierarchy of surrogate models, with different underlying approximations, fidelity, and speed.
They include phenomenological models~\cite{Ajith:2009bn,Hannam:2013oca,Pratten:2020fqn,Garcia-Quiros:2020qpx,Pratten:2020ceb,Estelles:2021gvs,Yu:2023lml,Thompson:2023ase}, Effective One Body models~\cite{Buonanno:1998gg,Buonanno:2000ef,Damour:2001tu,Taracchini:2013rva,Nagar:2021gss,Nagar:2023zxh,Pompili:2023tna,Ramos-Buades:2023ehm}, and surrogate models built directly on numerical simulations~\cite{Field:2013cfa,Blackman:2015pia,Blackman:2017dfb,Varma:2018mmi,Varma:2019csw,Pathak:2024zgo}
Cutting-edge models require fractions to tens of seconds to evaluate, see e.g.~\cite{Pratten:2020ceb,Ramos-Buades:2023ehm} despite attention paid to computational efficiency.
See also Ref.~\cite{LISAConsortiumWaveformWorkingGroup:2023arg} for a recent overview of GW models as well as methods for accelerating their evaluation.
The challenges of GW parameter estimation are compounded by the need of a large number of model evaluations due to the relatively low efficiency of sampling methods~\cite{Thrane:2018qnx}.
As the rate of GW detections rapidly increases with increasing detector sensitivity, it is important to explore methods of accelerating GW inference.

In this work we tackle the problem of accelerating GW inference from the standpoint of multifidelty methods.
The goal of using a multifidelity framework is to exploit the computational speed of low-fidelity models while retaining the accuracy of a high-fidelity model, in order to achieve fast, accurate solutions to many-query problems such as Bayesian inference~\cite{peherstorfer_survey_2018}.
GW models provide a variety of options for multifidelty approaches, since there are nested models in the sense that one model may be the limit of another in the case where some physical effect is neglected.
In this work we focus on the inclusion of higher modes (higher than quadrupolar radiative multipole moments) giving rise to our multifidelity hierarchy.
Accounting for these higher modes of emission can break degeneracies and improve the measurement of several parameters, especially the ratio of the masses and the inclination of the orbital plane to the line of sight~\cite{CalderonBustillo:2016rlt,Lange:2018pyp, Kumar:2018hml,LIGOScientific:2020stg,LIGOScientific:2020zkf,Huang:2020pba}.
Meanwhile, the evaluation of the higher mode contributions increases the computational cost of the forward model roughly in proportion to the number of modes used.
The expected differences between inferences with and without higher modes and their difference in computational expense make them a promising target for tempered importance sampling.

Specifically, we use the IMRPhenomXPHM~\cite{Pratten:2020ceb} family of waveform models for both the low and high fidelity signal model.
We treat the model as high fidelity when all of the available higher order modes are turned on, whereas for low fidelity runs, only the leading order mode is active.

One strategy to benefit from a multifidelity paradigm is through multifidelity importance sampling (MFIS)~\cite{peherstorfer2016multifidelity}, in which one samples from a low-fidelity posterior and reweights the samples using high-fidelity evaluations to obtain representative samples of the high-fidelity posterior, which can be used to compute Monte Carlo integrals. 
Importance sampling has been explored GW inference~\cite{Payne:2019wmy}, and has been applied successfully as one method to marginalize over detector calibration uncertainties~\cite{Payne:2020myg,LIGOScientific:2021usb,KAGRA:2021vkt}, to search for signatures of binary eccentricity~\cite{Romero-Shaw:2020owr}, and to improve the quality of samples from machine-learning based inference~\cite{Dax:2022pxd}.
However, these applications are often limited in practice by the efficiency of importance sampling.
If the two posteriors are too different from each other, the low efficiency of importance sampling means you need a huge number of low-fidelity samples to obtain desired accuracies~\cite{agapiou2017importance}.
More precisely, the effective sample size $N_\mathrm{eff}$ scales inversely with $\chi^2(p||q)$, the chi-squared divergence between the target and biasing probability distributions $p$ and $q$~\cite{alsup_context-aware_2021}.

In this work we propose a method, tempered multifidelity importance sampling, to improve the efficiency of importance sampling and broaden its domain of application.
A typical problem faced when deploying importance sampling is that the chi-squared divergence is very sensitive to the support of the two distributions in the tails. 
If the support of the biasing distribution does not cover the target well, $\chi^2(p||q)$ will be large and the effective sample size will thus be small. 
By tempering the biasing distribution, i.e.~raising the density function to some power smaller than one, the coverage expands and thus the overlap in the tails might improve as a result. 
It is not clear \textit{a priori} whether this improvement is to be expected in all cases, but we use this argument to motivate an exploration of the effect of tempering in a MFIS setting. We develop a principled procedure for selecting a good temperature, and we find that in several of our experiments this temperature resulted in appreciable improvements to the efficiency.

We emphasize that although we motivate the idea of tempered multifidelity importance sampling with the challenge of GW parameter estimation, our results have broad applicability.
MFIS can be useful in any problem for which many model evaluations are required and a hierarchy of varying fidelities exists. These are referred to as \textit{outer-loop applications} and include optimization, uncertainty propagation, data assimilation, control, and sensitivity analysis~\cite{peherstorfer_survey_2018}.

In the following sections, we present theoretical results demonstrating the impact of tempering on importance sampling efficiency and by extension on the Monte Carlo error in estimators using tempered samples. 
We argue that for arbitrary Gaussian probability densities $p$ and $q$, there exists a unique temperature which minimizes $\chi^2(p||q)$, and we provide a closed form expression for this optimal temperature. 
We derive a practical approximation for the optimal temperature for arbitrary distributions $p$ and $q$, under modest assumptions, and we validate this approximation in the Gaussian setting where it can be compared to the true solution. 
Finally, we apply our tempered multifidelity importance sampling algorithm to the Bayesian inference of both simulated GW observations and real GW observations.
We discuss the unique challenges of this problem and propose some directions for future work.

\section{Methodology}

\subsection{Bayesian Inference}
\label{sec:Bayesian}
In the context of an inverse problem, we start by considering the problem of finding some parameters of interest $\theta$, given some noisy measurements of an observable quantity related to those parameters through a forward model $h(\theta)$. Here $h$ is a vector of model predictions, i.e.~a time series of observed strain values for the GW case.
We assume that our observed data vector $d$ is related to $\theta$ through
\begin{align}
d = h(\theta) + n
\end{align}
where $n$ is a noise term which we assume to be additive in this manner.

In many settings, including standard GW analysis, it is reasonable to assume the noise is stationary and Gaussian, with zero mean and a known covariance $\Sigma_n$. 
In this case, $n \sim \mathcal{N}(0,\Sigma_n)$ and has a probability density $p_n$ given by
\begin{equation}
    p_n(n) = \frac{1}{\sqrt{|2 \pi \Sigma_n|}}\exp\left(-\frac12 n^T\Sigma_n^{-1}n\right) \,.
\end{equation}
Substituting $n = d - h(\theta)$ yields the \textit{likelihood function} $\mathcal{L}(d|\theta) = p_n(d -h(\theta))$, representing the probability of observing the data $y$ given the parameter $\theta$. 
We further assume a \textit{prior probability density} $\pi(\theta)$ on the parameters.

Bayes' rule allows us to relate the likelihood and prior probability densities to the \textit{posterior} density, which represents our knowledge of the parameter conditioned on a particular observation $y$. The posterior density is given by
\begin{equation} 
\label{eq:bayes-rule}
    p(\theta|d) = \frac{\mathcal{L}(d|\theta)\pi(\theta)}{Z}.
\end{equation}
The normalization constant $Z$ is called the \textit{evidence}, given by 
\begin{align}
    Z = \int \mathcal{L}(d|\theta)\pi(\theta)d\theta \,.
\end{align}

With the zero-mean Gaussian noise assumption and our expression for the likelihood, we have
\begin{align}
    \label{eq:post-exp}
    p(\theta|d) &\propto \mathcal{L}(d|\theta)\pi(\theta)
    \notag \\
    & \propto \exp\left[
        -\frac12(d-h(\theta))^T\Sigma_n^{-1}(d-h(\theta))
        \right]
        \pi(\theta) \,.
\end{align}
In GW data analysis the evaluation of the likelihood is typically carried out in the frequency domain, where the covariance matrix of stationary Gaussian noise is diagonal.
In terms of the one-sided power spectral density (PSD) $S_n(f)$, the Fourier transform of the model $\tilde h(\theta)$, and the frequency-domain data $\tilde d$, the standard GW likelihood for data taken by a single detector is (e.g.~\cite{Veitch:2014wba,Thrane:2018qnx})
\begin{align}
    \mathcal L(d|\theta) 
    & \propto
    \exp\left[
    -2 \, {\rm Re} \sum_k 
    \frac{|\tilde d_k- \tilde h(\theta; f_k)|^2}{S_n(f_k)} \Delta f 
    \right]\,,
\end{align}
where the sum is carried out at the discrete positive frequencies $f_k$ and $\Delta f$ is the frequency spacing.
In the case of multiple detectors observing the same GW event, the noise in each detector is assumed to be independent and so the individual detector likelihoods multiply, using the same model parameters $\theta$ for each detector while accounting for the light-travel time between the detectors.

It is important to recognize that the posterior density depends both on the observed data \textit{and} the model $h(\theta)$, precisely through the dependence of the likelihood on $h(\theta)$.
If one were to use a different model, the solution of the Bayesian inverse problem would be a different posterior probability density.

Thus, with data and model in hand, the task of Bayesian inference is to characterize the posterior given by Eq.~\eqref{eq:post-exp}. 
In many settings, the most useful thing to seek is a set of independent, identically distributed (i.i.d.) samples distributed according to the posterior. 
Once obtained, these samples can be used to compute Monte Carlo estimations of quantities that depend on the uncertain parameters, i.e.
\begin{align}
    \mathbb{E}_{p}[f(\theta)] 
    &= \int f(\theta)\, p(\theta|d)\,d\theta\; \approx\; \frac{1}{N}\sum_{i=1}^N f(\theta_i),\\
    & \{\theta_i\}_{i=1}^N \overset{\text{i.i.d.}}{\sim} p(\theta|d) \,.
\end{align}
The principal computational challenge of Bayesian inference is the need to evaluate the potentially expensive model $h(\theta)$ a large number of times in the process of generating samples.

\subsection{Multifidelity Importance Sampling}
\label{sec:Importance}
Sampling from a high-dimensional posterior distribution is, in general, a challenging task. 
The challenge of sampling from a posterior $p(\theta|d)$ is made worse when the model $h$ is computationally expensive to evaluate.
Importance sampling is one approach to exploit the speedup afforded by approximate models without sacrificing the accuracy of more sophisticated and expensive models. 
At its core, importance sampling involves estimating statistics of one distribution using samples drawn from another. 
To simplify notation, we let $p$ and $q$ refer to the high and low fidelity posterior distributions, respectively, as well as their corresponding probability density functions. 
Consider the mean of a function $f(\theta)$ with respect to the high fidelity posterior,
\begin{equation}
    \label{eq:Expectation}
    \mu \defeq \E_p[f(\theta)] = \int f(\theta)\,p(\theta)\,d\theta.
\end{equation}
We can introduce the low fidelity posterior by multiplying the integrand of Eq.~\eqref{eq:Expectation} by unity to obtain
\begin{align}
    & \mu = 
     \int f(\theta)\,\frac{p(\theta)}{q(\theta)}\,q(\theta)\,d\theta
    = \E_q\left[f(\theta)\,w(\theta)\right] \,, \\
    & w(\theta) \defeq \frac{p(\theta)}{q(\theta)} \,.
\end{align}
In general, we call $q$ the \textit{biasing distribution}. 
Note that we have rewritten the mean $\mu$ as an expectation with respect to the biasing probability density $q$. 
This allows us to define the importance sampling Monte Carlo estimator,
\begin{equation}
    \hat{\mu} \defeq \frac{1}{N}\sum_{i=1}^N f(\theta_i)\,w(\theta_i),\qquad \{\theta_i\}_{i=1}^N \overset{\text{i.i.d.}}{\sim} q,
\end{equation}
where now the $N$ i.i.d.~samples $\theta_i$ are drawn from the biasing distribution as opposed to the original posterior distribution.

In practice, the evidence $Z$ that appears in Eq.~\eqref{eq:bayes-rule} is often ignored, and we only have evaluations of the posterior densities up to a constant. 
Thus, we cannot compute $w(\theta)$ explicitly.
Instead we use \textit{self-normalized} importance sampling~\cite{mcbook}, where we define our estimate of $\mu$ as
     \begin{equation}
     \label{self-norm-estimate}
         \tilde{\mu} \defeq \frac{\sum_{i=1}^N f(\theta_i)\,w(\theta_i)}{\sum_{i=1}^N w(\theta_i)}.
     \end{equation}
Here, since the normalizing constants implicit in the weights cancel out, we can safely ignore them when computing $w(\theta)$.

Following the analysis in \cite{alsup_context-aware_2021, agapiou2017importance}, we see that the expected error of this self-normalized estimator is bounded by the chi-squared divergence between $p$ and $q$,  $\chi^2(p||q)$, given by
\begin{equation}
    \label{eq:ChiSquare}
    \chi^2(p||q) = 
    \int\frac{p(\theta)^2}{q(\theta)}\,d\theta 
    - 1 \, .
\end{equation}
Specifically, for a bounded measurable function $f$,
the mean-squared error of the estimate in Eq.~\eqref{self-norm-estimate} is bounded by
\begin{equation}
    \E[(\tilde{\mu} - \E_p[f(\theta)])^2] \leq 4||f||^2_{L^\infty}\frac{\chi^2(p||q) + 1}{N} \,,
\end{equation}
where $||f||_{L^\infty}$ is the $L^\infty$ norm of $f$.
This motivates us to search for biasing distributions that are as similar as possible to the target distribution, in the $\chi^2$ sense.

To assess the efficiency of importance sampling, it is not usually practical to compute or approximate the $\chi^2$ divergence directly. Instead we can compute the \textit{effective sample size} $N_{\rm eff}$,
\begin{equation}
    N_{\rm eff} = \frac{\left[\sum_{i=1}^N w(\theta_i)\right]^2}{\sum_{i=1}^N w(\theta_i)^2} \,.
\end{equation}
In the limit of a large number of samples $\theta_i$ drawn from the biasing distribution $q$, we can see that
\begin{align}
    \label{efficiency-limit}
    \frac{N_{\rm eff}}{N} \to \varepsilon \defeq \frac{1}{\chi^2(p||q) + 1}\,, 
\end{align}
so that minimizing $\chi^2$ has the effect of maximizing $N_{\rm eff}$.
Here we have defined the idealized efficiency $\varepsilon$ given $\chi^2$.

In practice we have access to only a limited number of choices for a reasonably accurate biasing distribution $q$.
For example, in GW data analysis $q$ may be defined by a signal model which is computationally cheaper to evaluate than the model implicit in the desired posterior $p$.
The development of accurate and computationally efficient signal models is itself a major challenge requiring significant effort.
To improve the efficiency of importance sampling without the flexibility to tune the biasing distribution through modeling, we turn to the idea of tempering $q$.

\subsection{Tempering}
\label{sec:Tempering}

In the setting of multifidelity importance sampling as outlined above, we propose introducing the \textit{tempered} biasing distribution, such that its density is
\begin{equation}
    q_T(\theta) \defeq \frac{q(\theta)^{1/T}}{Z_T} \,.
\end{equation}
Tempering is commonly used in Markov Chain Monte Carlo approaches to sample distributions through parallel tempering, e.g.~\cite{Swendsen:1986vqb,Geyer:1991,Earl:2005,Veitch:2014wba,Vousden:2015}.
Here our aim is to apply the concept of tempering to improve importance sampling.

By raising the density to the power $1/T$, $T$ acts as a ``temperature" controlling the width of the distribution. 
This additional free parameter allows us to improve the efficiency of importance sampling without changing the underlying density $q$.
A new constant $Z_T$ is needed to ensure the tempered density is normalized and is therefore
\begin{equation}
    Z_T = \int q(\theta)^{1/T}\,d\theta \,.
\end{equation}

To gain an intuition into how tempering improves the efficiency of importance sampling, we turn to a simple toy model.
Let $p$ and $q$ be one-dimensional Gaussians, and without loss of generality let $p$ be zero mean.
With
$p = \mathcal{N}(0,\sigma_p)$ and $q = \mathcal{N}(\mu,\sigma_q)$, the chi-squared divergence is
\begin{equation}
    \label{eq:chisq_gaussian}
    \chi^2(p||q_T) = \frac{T\sigma_q^2}{\sigma_p\sqrt{2T\sigma_q^2 - \sigma_p^2}}\exp\left(\frac{\mu^2}{2T\sigma_q^2 - \sigma_p^2}\right) - 1 \,.
\end{equation}
In this case, we can maximize the efficiency $\varepsilon$ by minimizing $\chi^2$ analytically.
The temperature that minimizes this expression, which we call $T_*$, is
\begin{align}
\label{eq:gauss_exact}
    T_* & = \frac{3\sigma_p^2 + 2\mu^2 + \sqrt{\sigma_p^4 + 12\sigma_p^2\mu^2 + 4\mu^4}}{4\sigma_q^2} \,.
\end{align}
We note that for this example $\chi^2$ diverges when $\sigma_q\sqrt{T}/\sigma_p < 1/\sqrt{2}$, and we expect in these cases that $q_T$ is too narrow to ever captures the tails of $p$; one benefit of tempering is that this can be evaded by increasing $T$.
Numerical experiments show that in practice when $q$ is too narrow, $N_{\rm eff}/N$ decreases towards zero with increasing $N$, although this appears to be a slow and highly stochastic process.

\begin{figure}[tb!]
    \includegraphics[width = 0.98 \columnwidth]{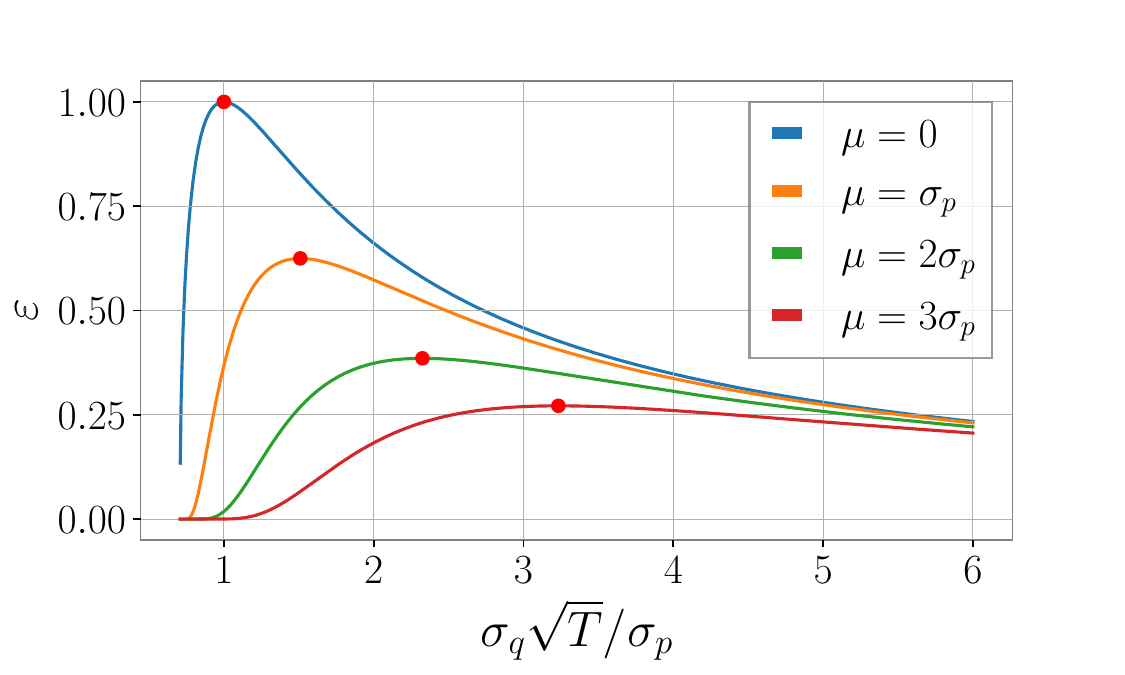}
    \includegraphics[width = 0.98 \columnwidth]{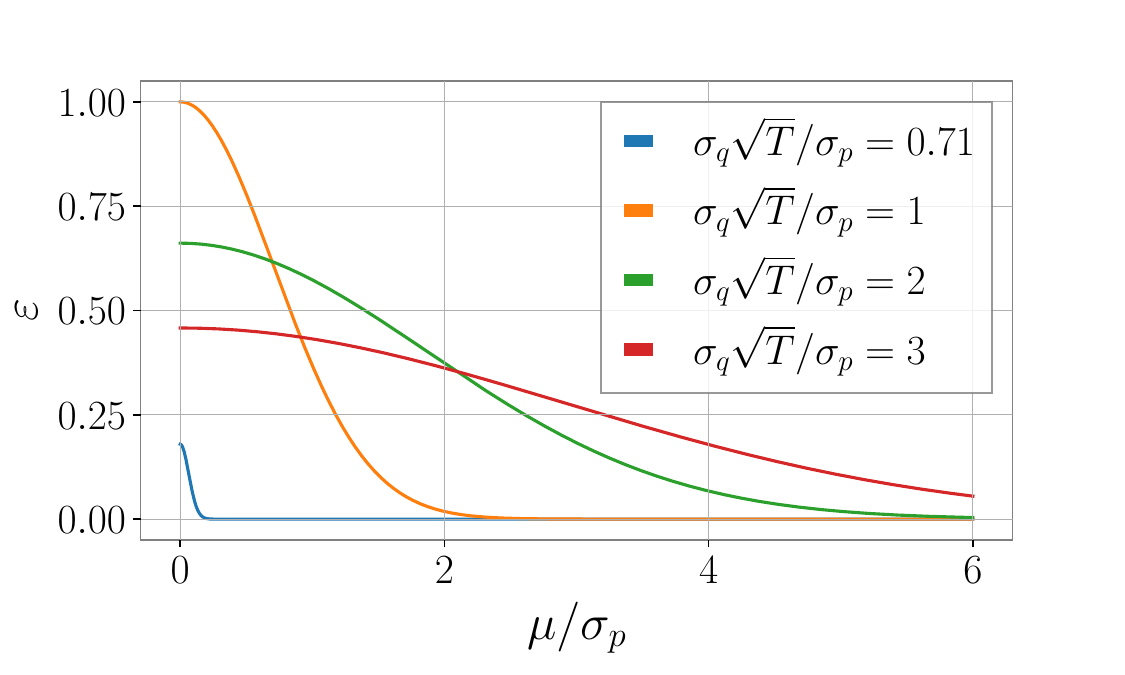}
    \caption{Example case of the idealized efficiency in terms of the $\chi^2$-divergence between target $p$ and tempered $q_T$ in the one-dimensional Gaussian case. 
    {\it Top:} The efficiency plotted against the ratio of widths rescaled by temperature, for various fixed relative biases. There is a unique maximum efficiency, given by using the optimal temperature of Eq.~\eqref{eq:gauss_exact}.
    {\it Bottom:} The efficiency plotted against the relative bias between the Gaussians, for various fixed relative widths and temperatures. 
    The results are proportional to a Gaussian in $\mu/\sigma_p$ in each case.
    }
    \label{fig:GaussianEff}
\end{figure}
We illustrate this simple example by plotting the idealized efficiency $\varepsilon$ in Fig.~\ref{fig:GaussianEff} for the one-dimensional Gaussians, where by \eqref{efficiency-limit} and \eqref{eq:chisq_gaussian},
\begin{align}
    \label{eq:GaussianEff}
    \varepsilon & = \frac{\sqrt{2T(\sigma_q^2/\sigma_p^2) - 1}}{T(\sigma_q^2/\sigma_p^2)}\exp\left(-\frac{\mu^2/\sigma_p^2}{2T(\sigma_q^2/\sigma_p^2) - 1}\right) \,.
\end{align}
The efficiency depends only on two parameters, the normalized bias $\mu/\sigma_p$ between the distributions and the ratio of the width of $p$ to the tempered width of $q_T$, specifically $\sigma_q \sqrt{T}/ \sigma_p$.
The dependence on the normalized bias is straightforward to understand: $\varepsilon$ is a Gaussian in $\mu/\sigma_p$ with variance $T(\sigma_q^2/ \sigma_p^2) - 1/2$.
Meanwhile, we see that for $T=1$ and $\mu = 1$, the efficiency becomes very poor for $\sigma_q/\sigma_p < 1$, and also becomes poor as $\sigma_q/\sigma_p$ becomes large. 
As the normalized bias increases, a wider biasing distribution gives better efficiency, as expected since otherwise the tails of $q$ cannot cover the bulk of the target distribution $p$.
We see that for each value of the bias, tempering allows us to tune the width of $q_T$ to attain an optimal efficiency.

To map this example onto inference for GW data analysis, consider the case where $p$ and $q$ result from two different GW signal models.
In the limit of high signal-to-noise (SNR), the widths of the posteriors scale inversely with SNR.
Thus while the bias $\mu$ and the ratio of the widths $\sigma_q/\sigma_p$ to be fixed by the differences in the signal models approximately independently of SNR, as the SNR increases the normaized bias $\mu/\sigma_p$ grows large. 
This situation is one where the efficiency is expected to be to the left of the peak of each curve in the top panel of Fig.~\ref{fig:GaussianEff}, and as SNR increases we traverse these curves towards larger $\mu/\sigma_p$ moving vertically down, with a severe loss of efficiency.
We illustrate this in the bottom panel of Fig.~\ref{fig:GaussianEff}, where we plot the efficiency for fixed $\sigma_q\sqrt{T}/\sigma_p$ versus $\mu/\sigma_p$.
If we imagine fixing $T=1$, then the efficiency decays monotonically with increasing $\mu/\sigma_p$.
Using tempering we can tune the value of $\sigma_q \sqrt{T}/\sigma_p$ to improve the efficiency, moving onto a different curve.

From this example we can take away another lesson, namely that some amount of tempering is expected to improve the efficiency in many situations, even if the optimal temperature is not known.
The danger is in moving to the right of the optimal efficiency, where in any case the decay in efficiency is less severe with increasing temperature.

This toy model can be readily extended to multi-dimensional Gaussians, $p = \mathcal N(0, \Sigma_p)$ and $q = \mathcal N(\mu, \Sigma_q)$.
The efficiency of importance sampling and effect of tempering depends on the details of the shapes of the covariance matrices and the direction of the bias $\mu$, but for isotropic Gaussians and a $\mu$ which is of order unity in all dimensions, the effect of increasing the dimension is that the efficiency is roughly that of the one-dimensional efficiency raised to the number of dimensions $n$, $\varepsilon \sim (\epsilon_\mathrm{1D})^n$. 
Thus we expect that the efficiency of importance sampling can be quite poor in a high number of dimensions.
Meanwhile, the analytic form of the optimal temperature is nearly the same as in the one-dimensional case.
Further details are given in Appendix~\ref{sec:Gaussian_appx}.

\subsection{Approximate optimal temperature}
\label{sec:OptimalAppx}

\begin{figure}[tb!]
    \includegraphics[width=0.98\columnwidth]{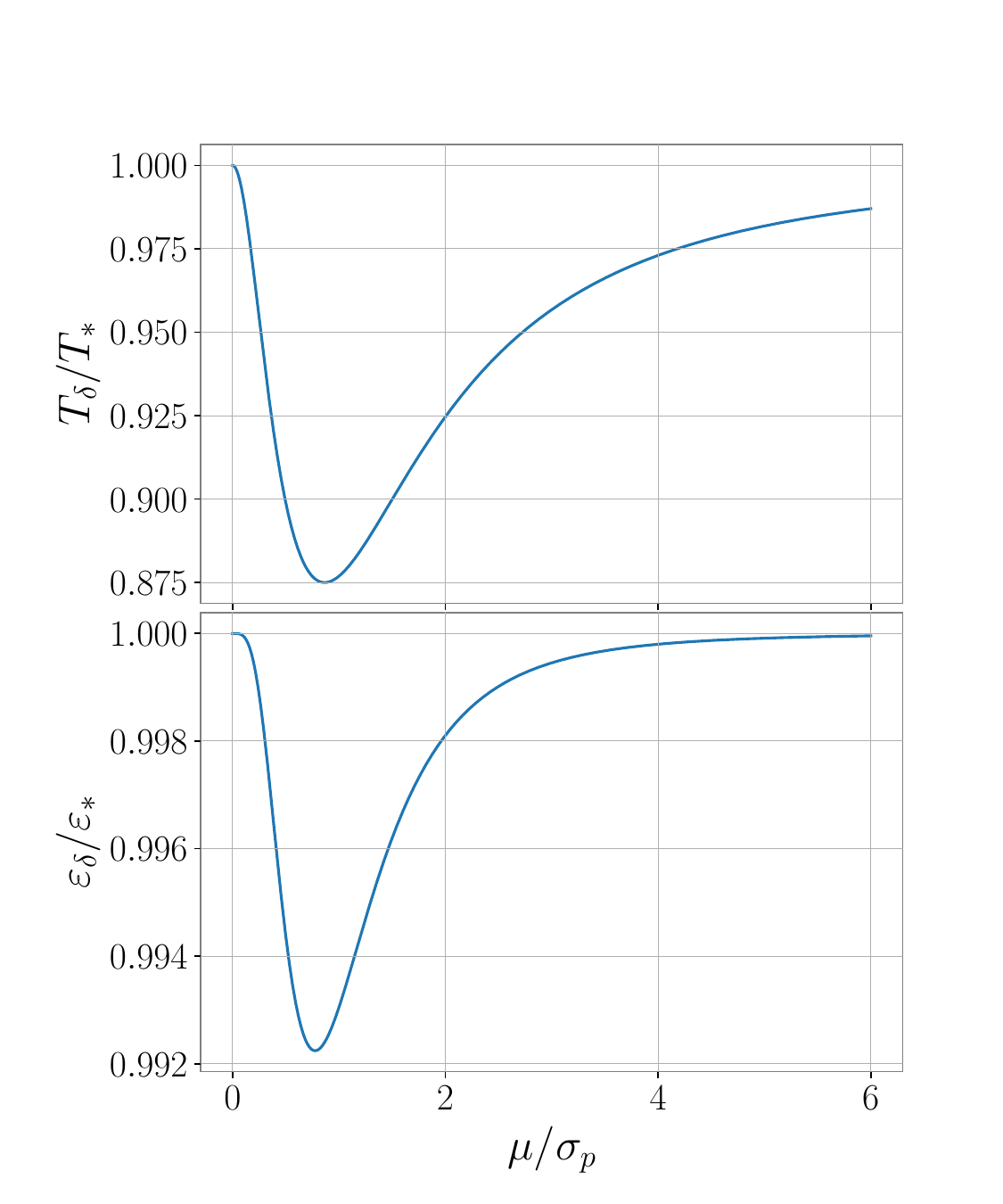}\\
     \caption{Comparisons between the optimal temperature $T_*$ from and the approximated temperature $T_\delta$ from for the one-dimensional Gaussian example, using Eqs.~\eqref{eq:gauss_exact} and~\eqref{eq:gauss_approx}.
     {\it Top:} Ratio of the temperatures $T_\delta/T_*$, which is a function of $\mu/\sigma_p$ only.
     {\it Bottom:} Ratio of the idealized efficiencies $\varepsilon_\delta/\varepsilon_*$ achieved for the optimal temperature $T_*$ and the approximated temperature $T_\delta$, which is a function of $\mu/\sigma_p$ only.}
     \label{fig:GaussAppxComps}
\end{figure}

In order to maximize the efficiency of importance sampling for a given $p$ and $q$, we would seek a temperature which minimizes $\chi^2(p||q_T)$.
In practice we cannot access the ideal optimal temperature $T_*$.
Performing a numerical search for $T_*$ is expected to be impractical, since each evaluation of a candidate temperature requires sampling from $q_T$.
Another challenge is that the efficiency must be estimated from samples, and so is inherently stochastic, as discussed in Sec.~\ref{sec:ErrorEstimation}, making numerical searches for $T_*$ unreliable.
However, we can work out an approximation to the optimal temperature under the condition that $q$ is sufficiently close to $p$ and assuming that $T$ is sufficiently close to 1. 
In this case we find an estimate for the optimal temperature $T_\delta \approx T_*$ which can be computed in practice using samples from $q$, provided that $q$ is normalized (and hence its evidence is known).
We find that
\begin{equation}
    T_\delta = 1 - \frac{\E_q[(p/q - 1)\log{q}]}{\text{Var}_q[\log{q}]}  \,.
    \label{eq:approx}
\end{equation}
The density $p$ does not to be normalized for this calculation, since the $p/q$ term can be estimated using self-normalized weights.
See Appendix~\ref{sec:approx_deriv} for a detailed derivation and more specific treatment of the assumptions.

Equation~\eqref{eq:approx} provides a practical path to estimating a good temperature for tempering.
The idea would be to first sample from $q$ to attain i.i.d.~ samples, use these to estimate $T_\delta$, and perform a second round of sampling from the tempered $q_{T}$.
As long as the computational expense of sampling twice using the model implicit in $q$ is less than that of sampling from $p$, tempering can lead to accurate inferences with less cost.

We can examine the accuracy of the approximation~\eqref{eq:approx} in the Gaussian case and compare it to the exact solution in Eq.~\eqref{eq:gauss_exact}. 
As before, when $p=\mathcal{N}(0,\sigma_p)$ and $q=\mathcal{N}(\mu,\sigma_q)$ we obtain
\begin{equation}
\label{eq:gauss_approx}
    T_\delta = \frac{\sigma_p^2 + \mu^2}{\sigma_q^2} \,.
\end{equation}
It is perhaps remarkable that the integrals involved give such a simple expression for $T_\delta$ in this case, and we give the full derivation in Appendix~\ref{sec:approx_gauss}.
We see that in the limit $\mu = 0$, our estimate $T_\delta$ agrees with the optimal temperature $T_*$. 
Their ratio depends only on the normalized bias $\mu/\sigma_p$, and is given in the top panel Fig.~\ref{fig:GaussAppxComps}.
Of greater interest is the effect of the approximation on the efficiency $\varepsilon$.
The lower panel of Fig.~\ref{fig:GaussAppxComps} gives the ratio of the efficiencies $\varepsilon_\delta/\varepsilon_*$ for the optimal and approximated temperatures. 
This ratio also only depends on the normalized bias, and we see that is is very close to unity for all values.

The fact that the idealized efficiencies depend only on the rescaled bias $\mu/\sigma_p$ for both temperature choices may initially be surprising, but this can be understood as follows.
The efficiency before tempering only depends on the normalized bias and the ratio of widths, and tempering only allows us to adjust the ratio of the widths without impacting the bias.
Our choice of temperature fixes the ratio of the widths, in a manner that depends on the value of the normalized bias.
The result is a tempered efficiency that depends only on the bias.

\section{Numerical Results}

With the notion of tempered importance sampling defined, we turn to applications of this method to multifidelity inference.
We carry out a sequence of computational experiments to test the effectiveness of tempered importance sampling in GW parameter estimation.
We use pairs of high- and low-fidelity GW models to recover the parameters of both simulated and real GW data.
Following the approach of~\cite{Payne:2019wmy}, our high-fidelity models incorporate higher radiative multipole moments, while our low fidelity models include only the dominant quadrupolar emission.
After reviewing our analysis setup, we provide examples of lower-dimensional GW inference that demonstrate the effects of tempering on importance sampling in controlled cases.
We then present results from simulated GW signals from aligned-spin systems (injections), as well as results from two real events from the third observing campaign of the LIGO, Virgo, KAGRA Collaborations.

\subsection{Analysis details}

We carry out two kinds of numerical experiments: injections of high-fidelity models into simulated data followed by Bayesian parameter estimation, and inference of real GW data.
In the injection-recovery experiments we used IMRPhenomXHM~\cite{Garcia-Quiros:2020qpx}, a model which assumes that the spin components are aligned with the orbital angular momentum of the binary, and hence neglects the effects of orbital precession. 
For these the waveforms were injected into zero noise using the high fidelity model and the posteriors sampled using the low fidelity model. 
We carry out a number of such experiments, in both restricted lower-dimensional cases as well as over the full 11 parameters.
The parameter choices for these injections are shown in Table~\ref{tab:injection_params}.
We also examine two events from the third gravitational wave transient catalog (GWTC-3)~\cite{KAGRA:2021vkt}, using open data from the Gravitational Wave Open Science Center~\cite{KAGRA:2023pio,GWOSC}. 
For these runs, we used IMRPhenomXPHM~\cite{Pratten:2020ceb}, which allows for generic spins and models orbital precession, resulting in a total of 15 parameters.
Tempering is carried out by scaling the power spectral density values by the appropriate temperature before sampling.

We use the following software tools for our computational experiments. 
The \texttt{bilby}~\cite{Ashton:2018jfp,Romero-Shaw:2020owr} Python package was employed to set up the inference problems in all the GW experiments we conducted.
We use the dynamic nested sampling algorithm~\cite{Higson2019} as implemented in the \texttt{dynesty} package~\cite{Speagle:2019ivv} as it is used in \texttt{bilby} to sample the posterior distributions. 
The use of nested sampling~\cite{Skilling2004,Skilling2006} is important for our chosen approach, since we need the evidence to normalize our biasing density $q$ in order to compute our temperature estimates $T_\delta$.
For processing the samples and visualizing the posteriors, we use \texttt{pesummary} \cite{Hoy:2020vys}.

For our injections, our priors are standard agnostic choices: uniform in detector-frame component masses, localization uniform in Euclidean volume (neglecting cosmological effects at the relatively low distances used in this study), inclination angle uniform in $\cos \iota$, polarization angle and coalescence phase uniform within their allowed ranges, and time of coalescence uniform in a window of $0.2$ s centered on the injection time.
We denote the dimensionless aligned-spin components of the spins as $S_{1z}$ and $S_{2z}$ in this work. 
Our priors in these components are the projection onto the orbital angular momentum of dimensionless spin vectors isotropic in orientation and with a magnitude uniform in $[0,0.99]$.
The result is a prior peaked around $S_iz = 0$ for each component, see e.g.~\cite{Ng:2018neg}.
For our GW likelihood we assume a two detector network composed of LIGO Hanford and LIGO Livingston.
We use the design noise curve \texttt{aLIGO\_ZERO\_DET\_high\_P\_psd}~\cite{ALIGODesignCurve} as our baseline PSD in both detectors.
We integrate the noise-weighted inner product from $f_{\rm low} = 20$ Hz to $f_{\rm high} = 1024 $ Hz except where noted.

For our analysis of real GW events, our priors and analysis settings mirror those used in GWTC-3~\cite{KAGRA:2021vkt,ligo_scientific_collaboration_and_virgo_2021_5546663}, with the following exceptions: 
we did not marginalize over calibration uncertainty~\cite{LIGOScientific:2016vlm,FarrCalMarg2014}, and we use the Euclidean distance prior $p(d_L) \propto d_L^2$ rather than accounting for cosmological expansion.
We also did not marginalize our likelihood over coalescence time or luminosity distance during sampling.

\begin{table}[t]
    \centering
    \begin{tabular}{c|c|c|c|c|c|c|c|c|c}
        \hline\hline
        $\mathcal{M}[M_\odot]$ & $q$ & $S_{1z}$ & $S_{2z}$ & $\alpha$ & $\delta$ & $\iota$ & $\psi$ & $\phi_c$ & $t_c[s]$ \\ 
        \hline
        30 & 0.5 & 0.4 & 0.3 & 1.3 & -1.21 & 1 & 2.6 & 2.3 & 1126259642.413 \\
        \hline\hline
    \end{tabular}
    \caption{Parameter values used in our injection studies. The component masses are $m_1$ and $m_2$, here expressed in terms of the chirp mass $\mathcal M = (m_1 m_2)^{3/5}/(m_1+m_2)^{1/5}$ and mass ratio $q = m_2/m_1$. Also listed are the components of the dimensionless spins aligned with the orbital angular momentum $S_{1z}$ and $S_{2z}$, the right ascension $\alpha$, the declination $\delta$, the inclination of the orbital angular momentum to the line of sight $\iota$, the polarization angle $\psi$, the phase of coalescence $\phi_c$, and the time of coalescence $t_c$. All angles are expressed in radians. We vary $d_L$ between $200$ Mpc and $1200$ Mpc in order to scan over SNR values between $108$ and $18$, respectively.}
    \label{tab:injection_params}
\end{table}

\subsection{Error estimation}
\label{sec:ErrorEstimation}

An important point to keep in mind when carrying out parameter estimation in practice is that we cannot access idealized quantities like the efficiency $\varepsilon$, the optimal $T$ to minimize $\chi^2$, or approximate temperatures defined using the distributions $p$ and $q$.
In all cases we instead must estimate these quantities through the samples we gather when carrying out parameter estimation.
This means that reported results, including our computed efficiency of importance sampling $N_{\rm eff}/N$ and our temperature estimate $T_\delta$ are Monte Carlo estimates and carry some uncertainty.

Usually parameter estimation routines generate a sufficient number of samples $N$ that these Monte Carlo uncertainties are small, and if this is not the case more samples can be gathered.
However we find that in practice importance sampling for high dimensional distributions can have poor efficiencies, resulting in a small number of effective samples.
The Monte Carlo error associated with these samples can be large, much larger than expected for a given $N$, and in some cases this prevents us from usefully estimating $T_\delta$, as discussed below in Sec.~\ref{sec:11DInjections}.

It is thus important to have a method for quantifying the uncertainties of out estimators.
Resampling methods provide simple and practical approaches for assessing the uncertainties and even biases associated with a set of samples.
In this study we use a bootstrap analysis~\cite{efron1982jackknife,Hogg:2010yz}, drawing a set of $N$ samples with replacement from the $N$ samples representing our distribution to get a new bootstrapped estimator. 
We repeat this 1000 times and compute the variance of our estimators.
This allows us to estimate the uncertainties in $N_{\rm eff}/N$, and display them as 1-$\sigma$ error bars on our plots.
Wherever practical we ensured that we had enough samples from $q$ and $q_T$ to reliably estimate the efficiency of our MFIS and tempered MFIS approaches.

\subsection{2D and 4D CBC parameter estimation}

\begin{figure}[tb!]
    \centering
    \includegraphics[scale=0.45]{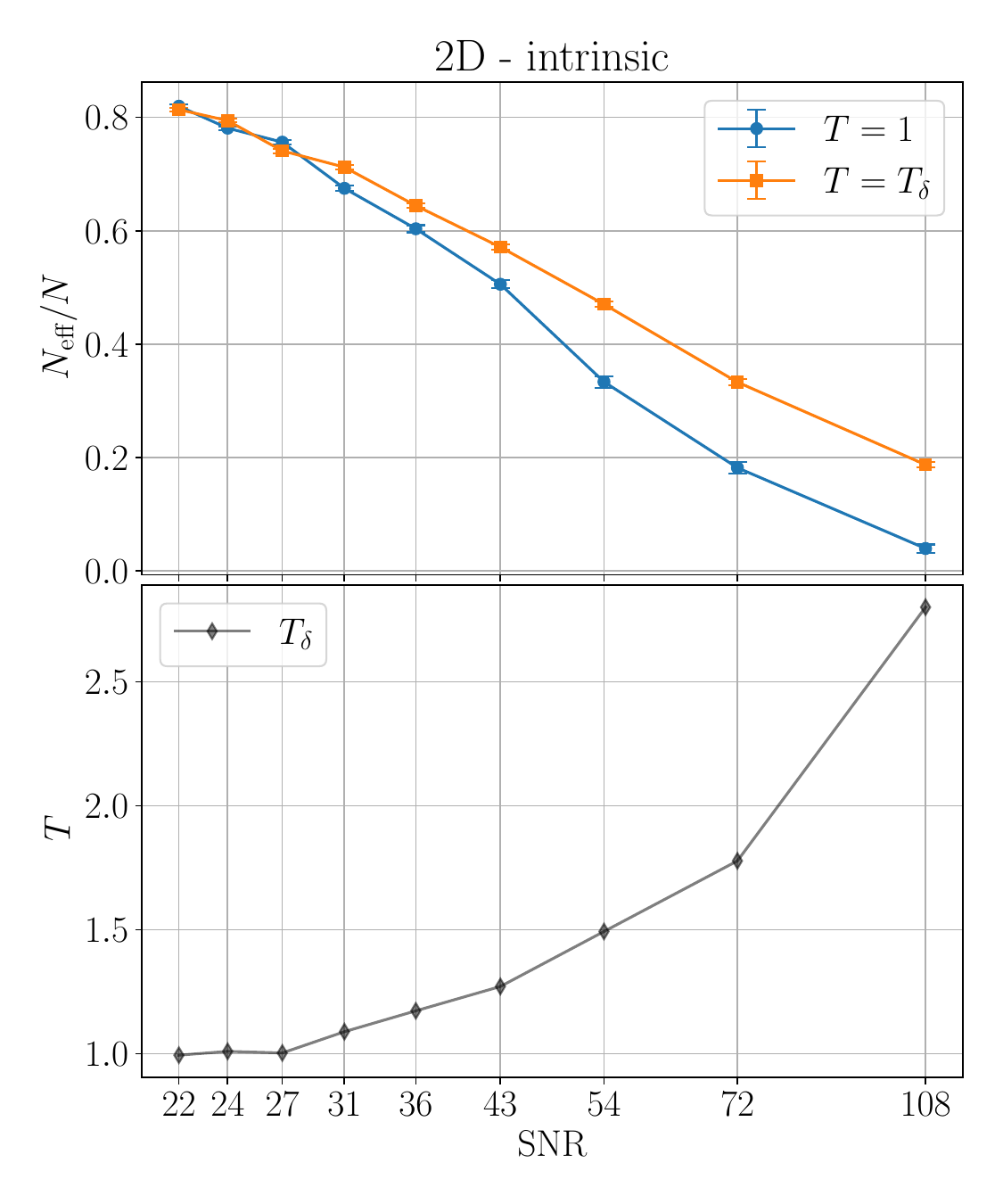}
    \caption{Result of tempering for the 2D intrinsic problem, where the inferred parameters are the mass ratio and chirp mass.}
    \label{fig:intrinsic-2D}
\end{figure}

\begin{figure}[tb!]
    \centering
    \includegraphics[scale=0.45]{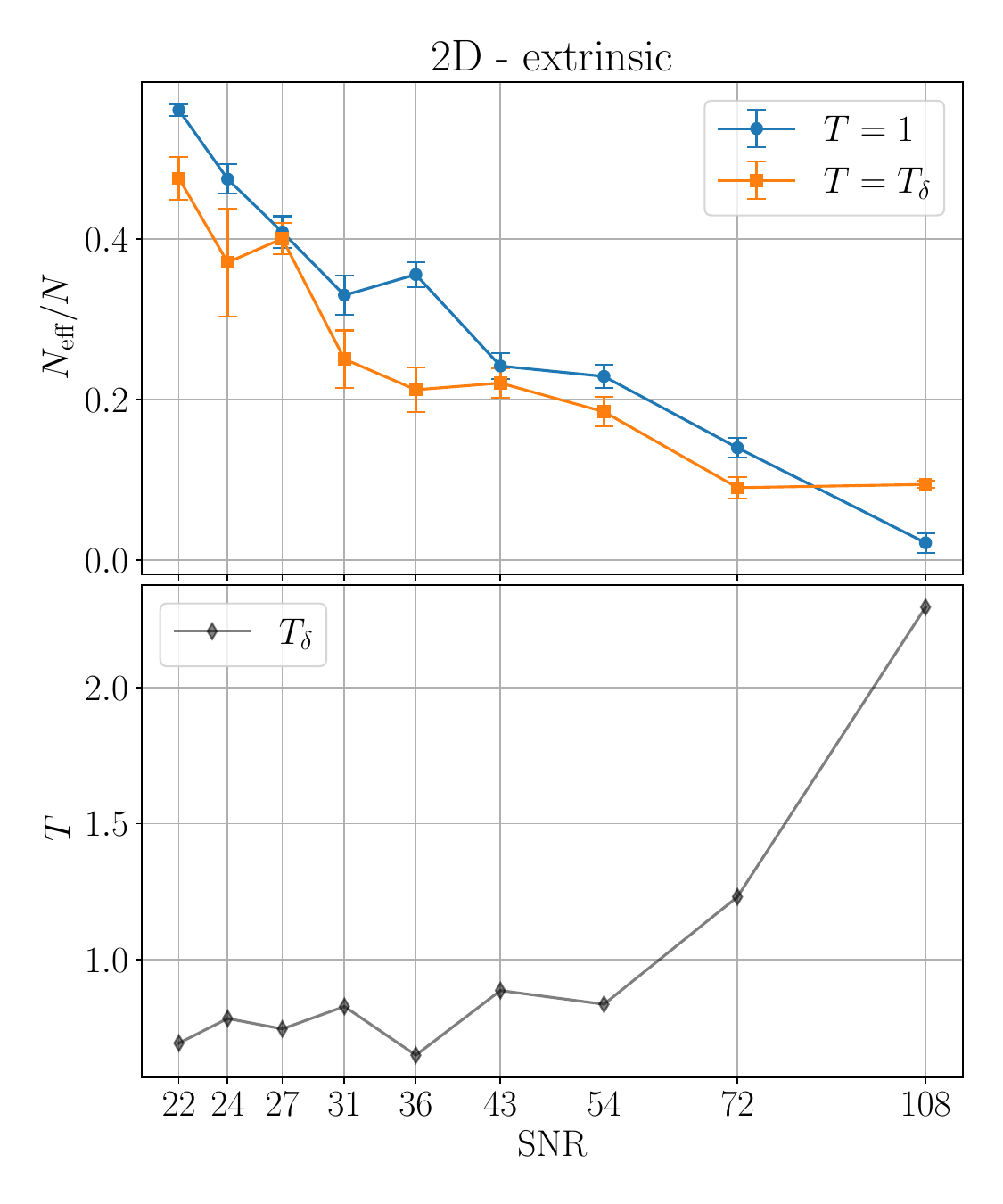}
    \caption{Result of tempering for the 2D extrinsic problem, where the inferred parameters are the luminosity distance and inclination angle.}
    \label{fig:extrinsic-2D}
\end{figure}

\begin{figure*}[t]
    \centering
    \includegraphics[width = 0.95\columnwidth]{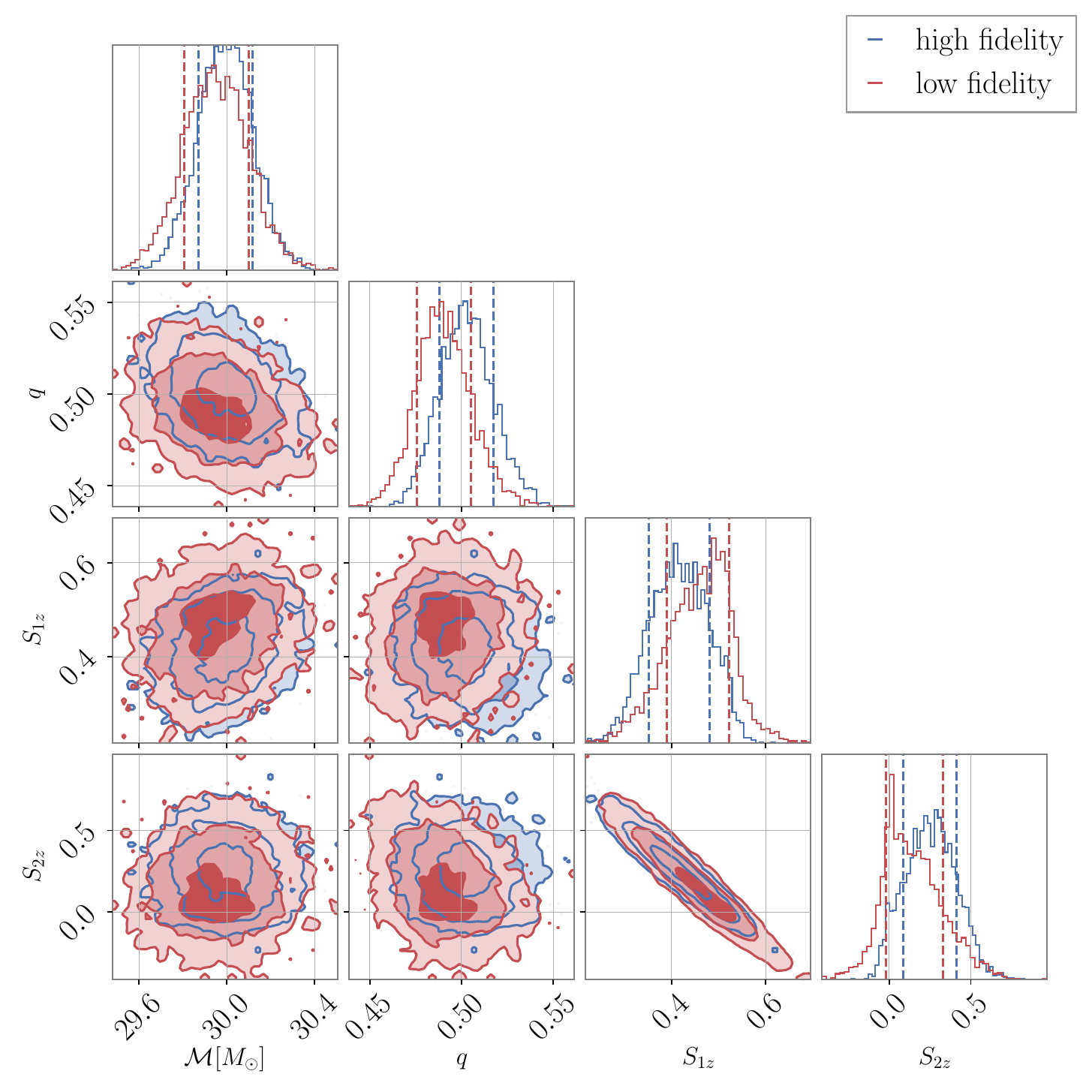}
    \includegraphics[width = 0.95\columnwidth]{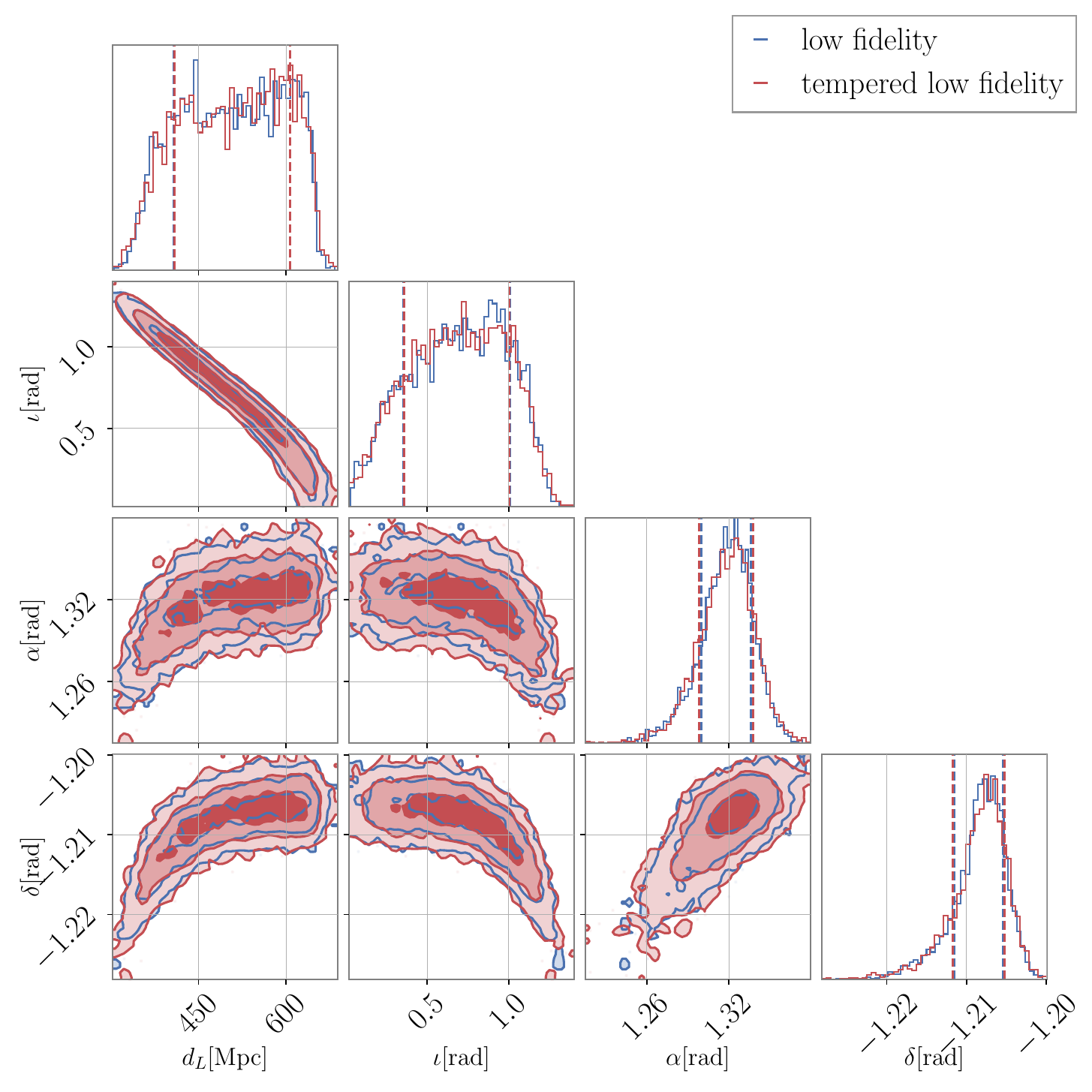}
    \caption{4D corner plots for $d_L=400$ (SNR 54). 
    For the intrinsic parameters we compare recovery of the high-fidelity injection with high- and low-fidelity models. For the extrinsic parameters we compare the low-fidelity recovery with the tempered low-fidelity recovery.
    }
    \label{fig:corner-4D}
\end{figure*}

\begin{figure}[tb!]
    \centering
    \includegraphics[scale=0.45]{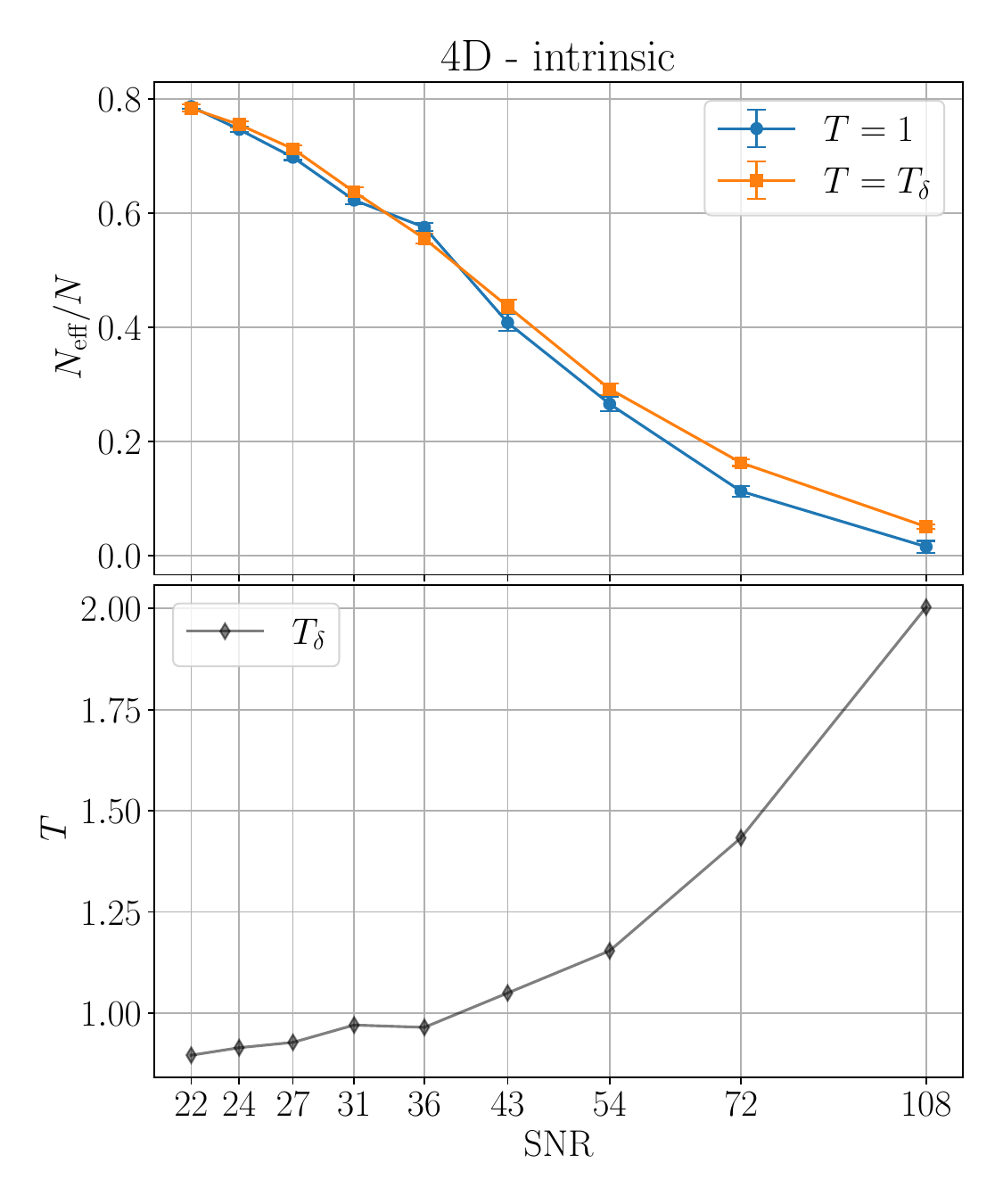}
    \caption{Result of tempering for the 4D intrinsic problem, where the inferred parameters are the mass ratio, chirp mass, and spin magnitudes.}
    \label{fig:intrinsic-4D}
\end{figure}

\begin{figure}[tb!]
    \centering
    \includegraphics[scale=0.45]{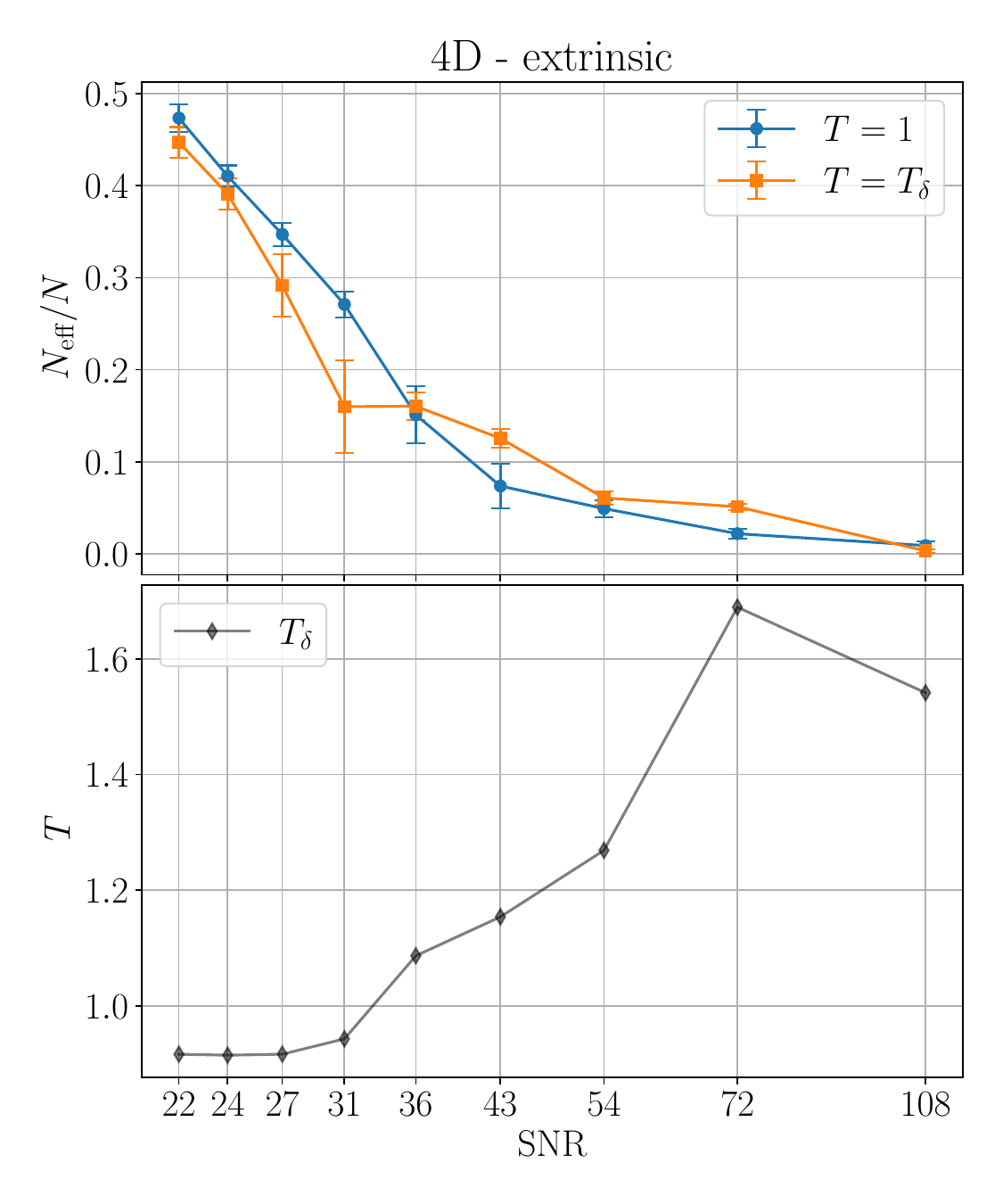}
    \caption{Result of tempering for the 4D extrinsic problem, where the inferred parameters are the luminosity distance, inclination angle, and sky position.}
    \label{fig:extrinsic-4D}
\end{figure}

We begin by studying lower dimensional GW inference problems in order to understand the effects of tempering.
For these 2D and 4D parameter estimation experiments, we fix all but a few of the parameters at their injection values. 
Figures \ref{fig:intrinsic-2D}-\ref{fig:extrinsic-4D} show a comparison between the importance sampling efficiency for an untempered low-fidelity posterior and the efficiency for a posterior tempered at our approximate temperature $T_\delta$. 
This comparison is plotted as a function of SNR, and the corresponding values of $T_\delta$ are also provided.

Figure~\ref{fig:intrinsic-2D} shows the impact of tempering for a 2-dimensional inference problem in which only the chirp mass $\mathcal{M}$ and the mass ratio $q$ are sampled over. 
The posteriors for this simple problem are in the Gaussian regime for the range of SNRs we explore, ${\rm SNR} > 20$.
As such the results are well-modeled by our Gaussian expectations.
In fact the efficiency $N_{\rm eff}/N$ of importance sampling the low-fidelity posteriors with the high-fidelity model is fit well by a Gaussian as a function of SNR.
This is the expected behavior when $q$ and $p$ are both Gaussian, 
and a fit to Eq.~\eqref{eq:GaussianEff} reveals $(\sigma_q^2 - \sigma_p^2/2)^{1/2}/\mu \approx 38/{\rm SNR}$ in this case.
We also see from the top panel of Fig.~\ref{fig:intrinsic-2D} that
there is a range of SNRs for which tempering at $T_\delta$ produces marked improvement in the efficiency.

Figure~\ref{fig:extrinsic-2D} shows the analogous result for a case in which we sample over the extrinsic parameters luminosity distance $d_L$ and the inclination $\iota$. 
In this case the expectations from our Gaussian model do not hold at these SNR values.
The approximate temperatures recovered from our method are below $1$, and the resulting tempered distributions mostly hurt the efficiency until the highest SNR case.

We turn next to analogous experiments for 4D.
Figure~\ref{fig:corner-4D} shows corner plots for the recovery on an SNR 54 signal for two cases.
In the first (left panel) we sample over the four intrinsic parameters of the aligned-spin models, the chirp mass $\mathcal{M}$, the mass ratio $q$, and the dimensionless aligned-spin components $S_{1z}$ and $S_{2z}$.
Here the corner plots illustrate the difference between the high- and low-fidelity posteriors.
In the second (right panel) we sample over four extrinsic parameters, specifically the luminosity distance $d_L$, the inclination $\iota$, and the sky position given by right ascension $\alpha$ and declination $\delta$.
Here the corner plots instead illustrate the effect of tempering, which is modest since $T = 1.27$
In the case of the intrinsic parameters, we see that the low-fidelity posteriors are mostly biased with respect to the high-fidelity case, with similar widths and shapes in the marginals. 
The posteriors appear to be fairly Gaussian, indicating that we expect improvement in $N_{\rm eff}/N$ with tempering.
Meanwhile, the posteriors for the intrinsic parameters are more complicated, with correlations that vary across parameter space.
In this case our intuition from the Gaussian examples may not apply directly.

Figure~\ref{fig:intrinsic-4D} illustrates the effect of tempering the 4D inference over the instrisic parameters.
As in the 2D case, the efficiency is fit well by a Gaussian, with $(\sigma_q^2 - \sigma_p^2/2)^{1/2}/\mu \approx 34/{\rm SNR}$.
The 4D intrinsic case shows systematic improvement in the efficiency, but it is more modest than the improvement seen in 2D.
Fig.~\ref{fig:extrinsic-4D} illustrates the effect when sampling over the intrinsic parameters.
This case does not show consistent improvement with tempering, with some improvement seen at higher SNRs.
Similar to the 2D case, the initial temperature estimates $T_\delta$ are below 1, but generally climb with SNR.

These low-dimensional experiments are useful for guiding our expectations and intuition for higher-dimensional problems of practical interest.
We see that for well-behaved cases, such as in the intrinsic parameter inferences, our approximation for the optimal temperature leads to improvements in efficiencies, and the behavior of both the efficiency and the impact of tempering follows our expectations from the Gaussian models.
However, our intuition from the simplest models does not appear to apply to the extrinsic parameters, and these show a case where tempering can hurt the efficiency of MFIS.

\subsection{Full aligned-spin CBC parameter estimation}
\label{sec:11DInjections}

\begin{figure}[tb!]
    \centering
    \includegraphics[scale=0.5]{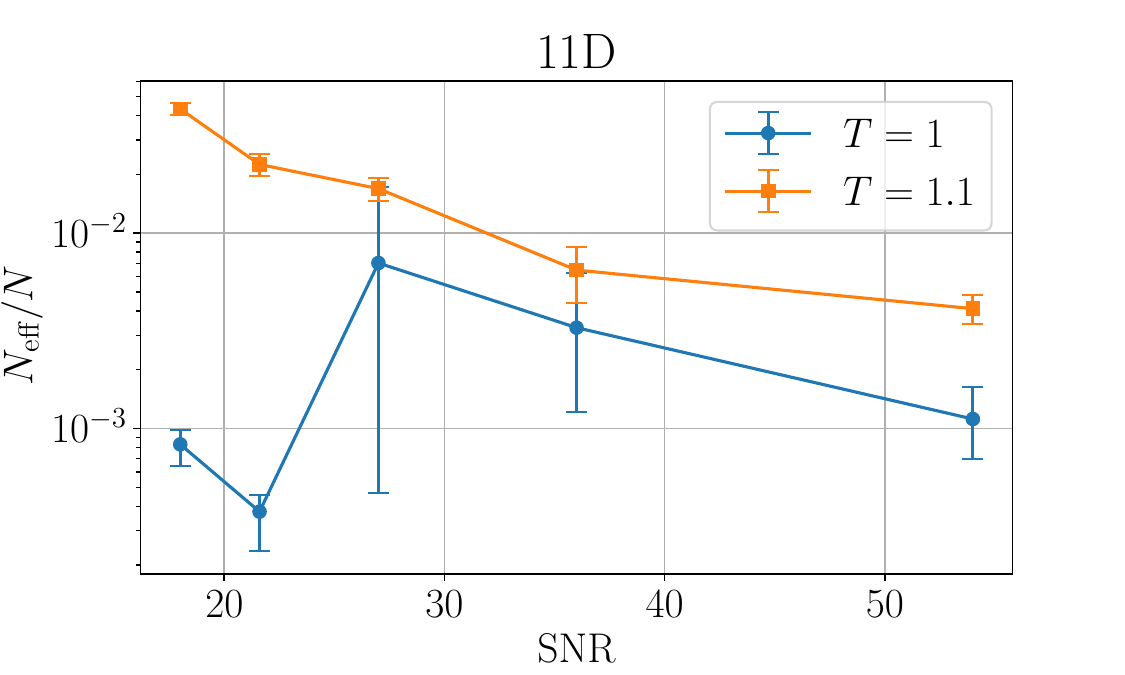}
    \caption{Result of tempering for the 11-dimensional aligned-spin injection. For this experiment, we tempered all the runs at $T=1.1$ instead of using the approximation $T_\delta$.}
    \label{fig:11D-injection}
\end{figure}

Having seen mixed results for tempering in our lower-dimensional tests, we turn to a more realistic case of the 11D recovery of an injected GW signal from an aligned-spin binary black hole coalescence. 
As with our 2D and 4D tests, our injected signal has no added noise (we assume a zero-noise realization of the random detector noise).
We find a very poor efficiency for our lower-fidelity recover of the high-fidelity injection.
This is seen in Fig.~\ref{fig:11D-injection} across a range of SNR values, where the $T=1$ recovery commonly has efficiencies of $\sim 1\%$.
The efficiencies are roughly flat across the SNRs tested.
Our low efficiencies are consistent with the poor efficiency of the zero-noise injection and recovery presented in~\cite{Payne:2019wmy} using similar analysis choices.\footnote{See Table 1 of~\cite{Payne:2019wmy}, and note that the differences in their waveform models are expected to be even larger than ours, likely accounting for another factor of $\sim10$ decrease in the efficiency.}

Initially, we were concerned that these low efficiencies would mean that our estimates for $T_\delta$ would be unreliable, given the small $N_{\rm eff}$ for the samples we drew from $q$.
Therefore, inspired by the observation that in many cases any amount of tempering improves the efficiency of importance sampling in our Gaussian examples, for this experiment we opted for a different prescription and simply tempered each case by the same temperature $T = 1.1$, drawing a similar number $N \approx 2\times10^4$ samples.
The resulting efficiencies are seen in Fig.~\ref{fig:11D-injection}.
While these efficiencies are still poor overall, we see a several times improvement at moderate SNRs with no particular loss of efficiency at higher SNRs as compared to the $T = 1$ case.

The success of uniform tempering in 11D points towards another potential way to benefit from tempered MFIS.
Rather our proposed two-step process, first performing standard inference with $T = 1$ to get samples from $q$ and then estimating $T_\delta$ with these, prior experience or theoretical analysis can provide a proposed temperature for a single step of tempered MFIS.

\subsection{Parameter Estimation for GWTC-3 Events}

In order to test our method in a fully realistic situation, we apply it to two
events from GWTC-3.
Our goal is
to probe the regime in which the two posteriors are very similar as well as the regime in which they substantially differ. 
For the former scenario, we choose GW191222\_033537 (hereafter GW191222), 
a fairly typical binary black hole signal which favors equal masses, small effective spin parameter $\chi_{\rm eff}$, and no strong signs of orbital precession.
This is an example of an event for which the higher mode content of the gravitational wave is expected to be small, meaning our low and fidelity models produce similar results. This means that the overall efficiency of multifidelity importance sampling is higher, since $q$ is very similar to $p$, but also that the margin of improvement is smaller since there is little extra information coming from the high fidelity model. 

\begin{figure*}[t]
    \centering
    \includegraphics[width = 0.95\columnwidth]{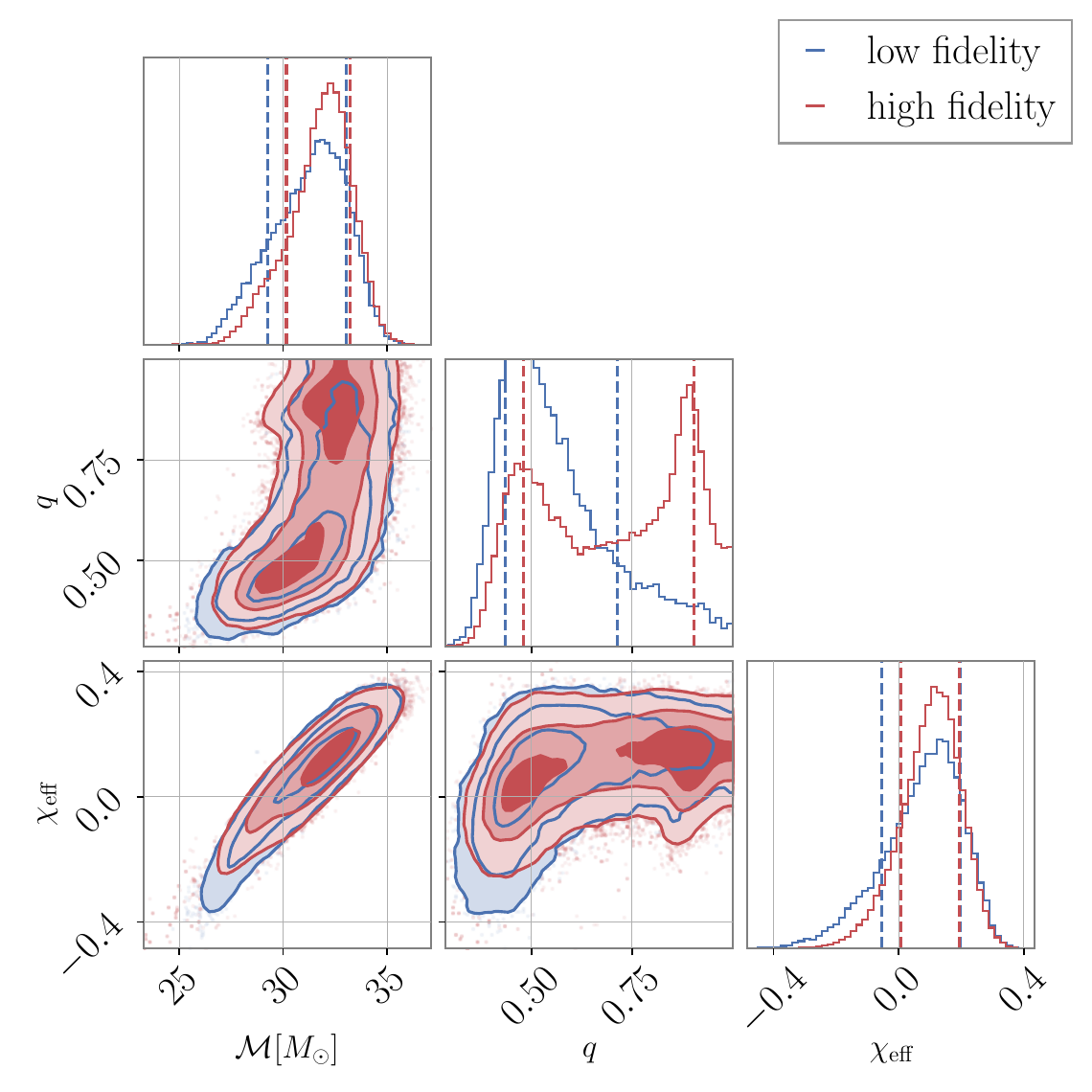}
    \includegraphics[width = 0.95\columnwidth]{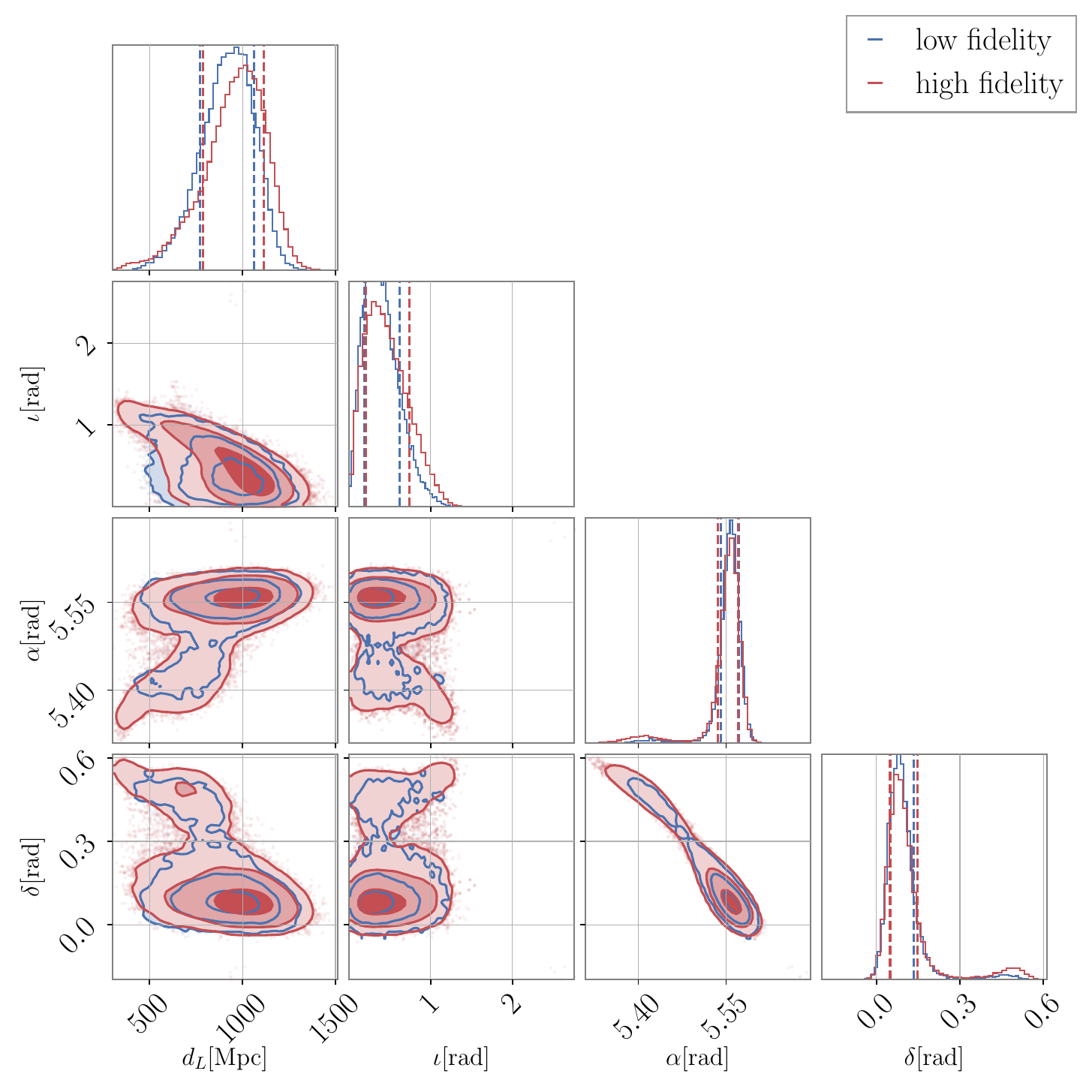}
    \caption{Corner plots for GW200129, illustrating recover with the high-fidelity model, which includes precession and higher mulitipolar emission, and the low-fidelity model which includes only precessiona and the dominant quadrupolar emission.}
    \label{fig:corner-GW200129}
\end{figure*}

GW200129\_065458 (hereafter GW200129) was selected for the opposite reason; it is a high SNR event where standard analysis with IMPhenomXPHM infers large component spins, clear orbital precession~\cite{KAGRA:2021vkt, Hannam:2021pit} and some posterior weight towards unequal masses $m_2/m_1\sim 1/2$ where higher modes make greater contributions to the signal.
Parameter estimation for this event is systematically different across signal models, e.g.~\cite{KAGRA:2021vkt,Hannam:2021pit,Varma:2022pld,Islam:2023zzj}, and it is also complicated by non-Gaussian noise in the raw data, which must be mitigated~\cite{KAGRA:2021vkt,Payne:2022spz}.

For these reasons we expect the choice of signal model to have greater impact for GW200129,
and therefore the distributions $p$ and $q$ inferred with and without higher modes to be more different.
We see evidence for this in the corner plots of Fig.~\ref{fig:corner-GW200129}, which shows marginalized 2D and 1D posteriors for GW200129 for selected intrinsic and extrinsic parameters. 
We show both the samples presented in GWTC-3 using the IMRPhenomXPHM model with a complete set of modes (high-fidelity), and our own low-fidelity recovery of this event.
The primary difference in the instrinsic parameters appears to be the absence in the low-fidelity recovery of a second mode at  larger $q$ values, which impacts both the effective spin $\chi_{\rm eff}$ and chirp mass $\mathcal M$ inferences.
While it is natural to attribute this to the difference in the models (and the presence or absence of multiple posterior modes is model-dependent for GW200129, see e.g.~\cite{KAGRA:2021vkt,Hannam:2021pit,Islam:2023zzj}), we cannot rule out that our four independent low-fidelity runs failed to recover this distinct region of probability.
The intrinsic parameters also show differences in recovery, notably less coverage of the tails of the high-fidelity posteriors by the low-fidelity in several regions.
The clear difference between this posteriors should result in lower overall efficiency, but greater potential for improvement with tempering.

\begin{figure}[tb!]
    \centering
    \includegraphics[scale=0.5]{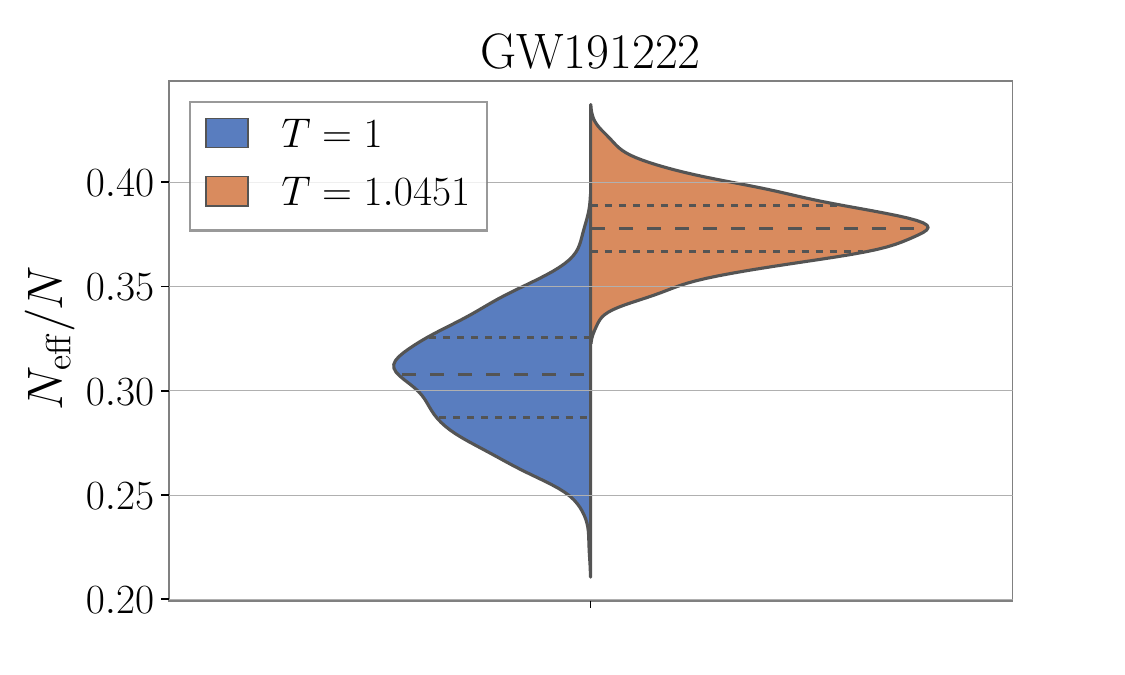}\\
    \includegraphics[scale=0.5]{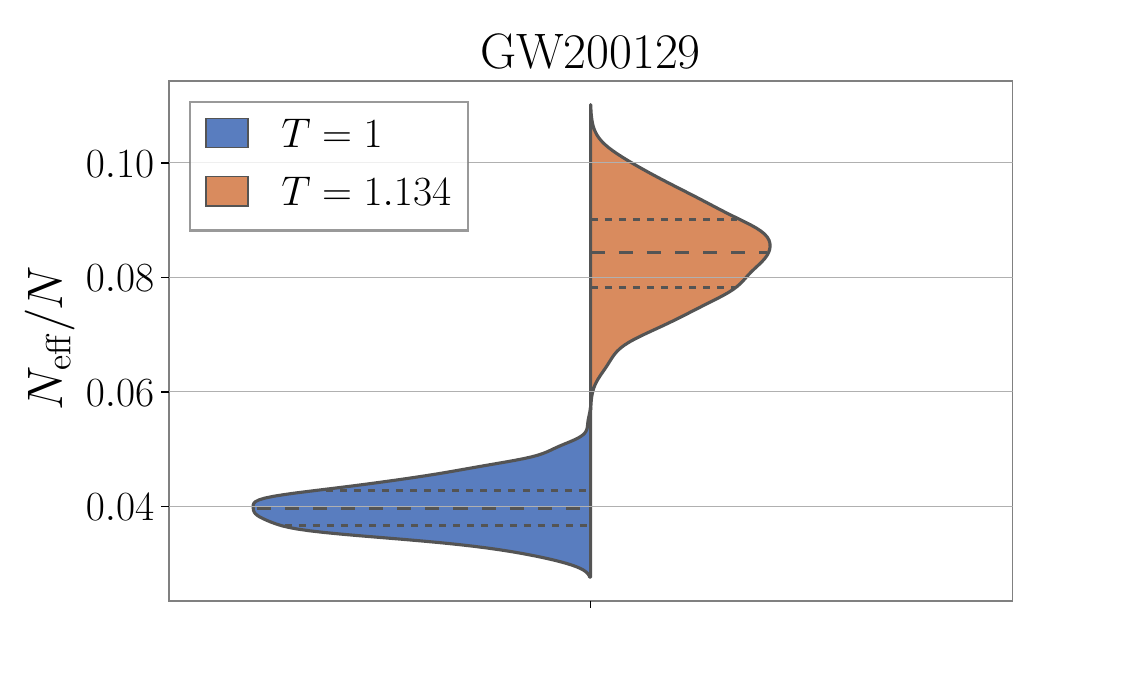}
    \caption{Result of tempering for the events GW191222 and GW200129. Note the improvement in efficiency, subject to uncertainty estimated through bootstrap resampling.}
    \label{fig:events}
\end{figure}

Figure \ref{fig:events} shows the observed improvement in efficiency for both events, in this case using violin plots to visualize the uncertainty on the efficiency as estimated from bootstrap resampling.
Notably importance sampling with real events provides generally better efficiencies than our 11D, zero-noise case even though the dimensionality is higher here.
This was observed also in the analysis of~\cite{Payne:2019wmy}.
These higher efficiencies, together with analysis settings that draw a larger number of samples for the $T=1$ inference for GW200129, allow us to reliably estimate $T_\delta$ for these real events and temper using it.

As expected, GW191222 has a much higher efficiency than GW200129.
For GW191222 the estimated optimal temperature $T_\delta$ is close to unity, and tempering improves the efficiency of MFIS, with the median of the tempered efficiency estimate well into the tail of the uncertainty of the standard analysis.
For GW200129, tempering provides a clear improvement, more than doubling the efficiency of importance sampling.

\section{Discussion}
MFIS is a promising framework for GW inference because in this domain, there exists a rich hierarchy of waveform models, with a range of computational costs and including a variety of physical effects. 
The high computational cost of some of the best models means that exploiting samples from cheaper models via importance sampling has the potential to greatly reduce computational costs and accelerate inference.
In practice, direct application of importance sampling can lead to very low efficiencies, as seen in the existing literature, e.g.~\cite{Payne:2019wmy,Dax:2022pxd}.

Our goal in this work is to present a method to improve the efficiency of importance sampling generally, and explore its application to GW inference. 
We introduce the idea of tempered MFIS, where the biasing distribution is tempered in order to provide better coverage of the target distribution. 
Further, we derive a practical estimate for the temperature needed to improve the efficiency, at the cost of generating an initial set of samples from the untempered biasing distribution.

By carefully investigating this idea of tempering in controlled settings (such as our Gaussian experiments), we arrive at a principled methodology for improving importance sampling efficiency via the relatively cheap calculation of our approximately optimal temperature. 
In doing so, we extract several other general insights: low efficiency is often caused by samples drawn from poorly overlapped tails, importance sampling is highly sensitive to the shapes of the distributions, and the overall efficiency scales poorly with dimension.

Furthermore, the mixed success of applying our tempering method in the GW experiments suggests that this is not a one-size-fits-all solution, and that care must be taken in understanding and modifying these distributions. 
However, the cases in which the efficiency did improve in our experiments motivates the future pursuit of other ways to modify biasing distributions, the goal being a more robust and reliable method for battling low efficiency.
Two of our results are especially promising: first that our estimate for the optimal temperature improves the efficiency of importance sampling for real GW data, and secondly that an principled guess for a temperature uniformly improves the efficiency for simulated signals injected into zero noise across a range of signal strengths, in one case by more than an order of magnitude.

Finally, we use bootstrap resampling techniques to estimate uncertainties in quantities estimated from samples such as the efficiency and temperature.
The application of resampling methods to generate frequentist-based uncertainties on quantities like the effective number of samples is of independent interest in GW data analysis.
There are a variety of such resampling methods, and some such as jackknife resampling allow one to not only estimate these uncertainties but also correct for biases present in the estimators~\cite{efron1982jackknife}. 
Future work may pursue these methods to better capture and account for the uncertainties inherent in our sample-based efficiency and temperature estimates.

Our method may be useful in various applications of importance sampling to GWs. It would be especially interesting to understand if tempering can be applied to methods such as those used in~\cite{Dax:2022pxd}, where importance sampling is applied to correct samples drawn using a learned normalizing flow. 
The success of tempering may lead to simple ideas to improve the construction of these machine-learning based methods.

\acknowledgements

We thank Ethan Payne and Dingcheng Luo for useful discussions. 
We especially thank Colm Talbot for useful discussions and technical advice throughout this project.
We also thank Carl-Johan Haster for a careful reading of this manuscript and for helpful comments.
B.S.~ and O.G.~were supported by DOE grant DE-SC0019303 during a portion of this work, and B.S.~by a CNS Catalyst Grant at UT Austin during this work.
A.Z.~was supported by NSF Grants PHY-2207594 and PHY-2308833 while carrying out this work. 
P.C.~ was partially supported by NSF Grants  \#2245674, \#2245111, and \#2325631.
O.G.~\R{\dots}.
This paper has preprint numbers UT-WI-19-2024 and LIGO-P2400206.

This research has made use of data or software obtained from the Gravitational Wave Open Science Center~\cite{GWOSC}, a service of the LIGO Scientific Collaboration, the Virgo Collaboration, and KAGRA. This material is based upon work supported by NSF's LIGO Laboratory which is a major facility fully funded by the National Science Foundation, as well as the Science and Technology Facilities Council (STFC) of the United Kingdom, the Max-Planck-Society (MPS), and the State of Niedersachsen/Germany for support of the construction of Advanced LIGO and construction and operation of the GEO600 detector. Additional support for Advanced LIGO was provided by the Australian Research Council. Virgo is funded, through the European Gravitational Observatory (EGO), by the French Centre National de Recherche Scientifique (CNRS), the Italian Istituto Nazionale di Fisica Nucleare (INFN) and the Dutch Nikhef, with contributions by institutions from Belgium, Germany, Greece, Hungary, Ireland, Japan, Monaco, Poland, Portugal, Spain. KAGRA is supported by Ministry of Education, Culture, Sports, Science and Technology (MEXT), Japan Society for the Promotion of Science (JSPS) in Japan; National Research Foundation (NRF) and Ministry of Science and ICT (MSIT) in Korea; Academia Sinica (AS) and National Science and Technology Council (NSTC) in Taiwan.
The authors are grateful for computational resources provided by the LIGO Laboratory and supported by National Science Foundation Grants PHY-0757058 and PHY-0823459.

This work makes use of the \texttt{lalsuite}~\cite{lalsuite}, \texttt{gwpy}~\cite{duncan_macleod_2024_846265}, \texttt{bilby}~\cite{Ashton:2018jfp,Romero-Shaw:2019itr,greg_ashton_2019_2602178} \texttt{dynesty}~\cite{Speagle:2019ivv,sergey_koposov_2023_8408702}, and \texttt{pesummary}~\cite{Hoy:2020vys,charlie_hoy_2021_5237238} software packages.

\appendix

\section{Multi-dimensional Gaussian example}
\label{sec:Gaussian_appx}

In this appendix we collect results on the efficiency of importance sampling in the context of multi-dimensional Gaussians.
Without loss of generality, place the mean of $p$ at the origin, letting $p = \mathcal N(0,\Sigma_p)$ and $q = \mathcal N(\mu,\Sigma_q)$.
We could further rotate and rescale our coordinates to make $p$ a unit Gaussian, but for clarity we retain $\Sigma_p$ explicitly.
Provided that 
\begin{align}
    \Gamma & \defeq 2 \Sigma_p^{-1} - \Sigma_q^{-1} \,.
\end{align} is positive definite, these densities can be inserted into Eq.~\eqref{eq:ChiSquare} and the Gaussian integral resolved.
Let $\Sigma \defeq \Gamma^{-1}$, then
\begin{align}
    \label{eq:ChiSqMultivariate}
    \chi^2(p||q)+1 & = \frac{|\Sigma_q|}{|\Sigma_p|^{1/2}}
    \frac{\exp\left[\frac 12 \mu^\top(\Sigma_q^{-1} + \Sigma_q^{-1} \Sigma \Sigma_q^{-1}) \mu \right]}
    {|2\Sigma_q - \Sigma_p|^{1/2}} 
    \notag \\
    & = \frac{|\Sigma_q|}{|\Sigma_p|^{1/2}}
    \frac{\exp\left[\mu^\top (2\Sigma_q - \Sigma_p)^{-1} \mu \right]}
    {|2\Sigma_q - \Sigma_p|^{1/2}}  \,.
\end{align}
In the second line we have used a variant of the Woodbury identity~\cite{Woodbury:1950}, specifically
\begin{align}
    (A + B)^{-1} & = A^{-1} - A^{-1}B(\mathbb{1} + A^{-1}B)^{-1} A^{-1} \,.
\end{align}
Tempering then makes the replacement $\Sigma_q \to T \Sigma_q$ in Eq.~\eqref{eq:ChiSqMultivariate}.

In the case of isotropic $n$-dimensional Gaussians, the expressions simplify further.
With $\Sigma_p = \sigma_p^2 \mathbb{1}$, $\Sigma_q = \sigma_q^2 \mathbb{1}$ we also have $\Sigma = \sigma_p^2 \sigma_q^2/(2 \sigma_q^2 - \sigma_p^2) \mathbb{1}$ and so
\begin{align}
    \label{eq:chisq_gaussian_iso}
     \chi^2(p||q)+1 &= 
     \left(\frac{T \sigma_q^2}
     {\sigma_p\sqrt{2T\sigma_q^2 - \sigma_p^2}}
     \right)^n
      \,
      \exp\left[\frac{\mu^\top \mu}{2 T \sigma_q^2 - \sigma_p^2} \right]
      \,.
\end{align}
This result agrees with the $n=1$ result from Eq.~\eqref{eq:chisq_gaussian}.
Further, if the bias $\mu$ as $O(1)$ entries in all $n$ dimensions, then $\mu^\top \mu \sim O(n)$ and hence the result Eq.~\eqref{eq:chisq_gaussian_iso} is just Eq.~\eqref{eq:chisq_gaussian} raised to the power $n$.
If on the other hand there are some unbiased directions, the efficiency is penalized by an effective dimension that is less than $n$.

Continuing consideration of the isotropic Gaussian case, we can solve for the optimal temperature
\begin{align}
    T_* & = \frac{3n\sigma_p^2 + 2\mu^\top \mu + \sqrt{n^2\sigma_p^4 + 12 n \sigma_p^2\mu^\top \mu + 4(\mu^\top \mu)^2}}
    {4 n \sigma_q^2}\,.
\end{align}
Again under the assumption that $\mu^\top \mu \sim O(n)$, we see that the factors of $n$ cancel out of every term, resulting in essentially the same optimal temperature as for the $n=1$ case.

From this we conclude that as the dimension of the densities under consideration increases, we expect the efficiency of importance sampling to decrease rapidly, scaling as $\epsilon_{1D}^n$ for $n$ relevant dimensions.
Meanwhile, the best temperatures for tempering will remain similar to the estimates for $n=1$.

\section{Approximate optimal temperature derivation}
\label{sec:approx_deriv}
Let $p$ and $q$ be probability densities over a space $X$, such that the measure corresponding to $p$ is absolutely continuous with respect to the analogous measure for $q$.
The $\chi^2$-divergence between $p$ and $q$ is
\begin{align}
    \chi^2(p||q) 
    = \int\frac{p^2}{q}dx - 1 \,,
\end{align}
Recall that $p$ and $q$ are densities and therefore functions of $x\in X$. 
We write these functions without this explicit argument for visual clarity.

It is more convenient for our derivation to work with the inverse temperature $\beta = 1/T$ in this Appendix, rather than $T$.
The \textit{tempered density} $q_\beta$ is therefore defined as
\begin{align}
    q_\beta & \defeq \frac{q^\beta}{Z_\beta}\,,
    & Z_\beta & = \int q^\beta dx\,.
\end{align}
Our goal is to find $\beta_*$ such that
\begin{equation}
    \beta_* = \argmin_\beta \chi^2(p || q_\beta) \,,
\end{equation}
The value of $\beta$ that minimizes the $\chi^2$-divergence maximizes the efficiency of using $q_\beta$ as a biasing density for importance sampling. 

Since we cannot find $\beta_*$ in general, we derive an approximate expression $\beta_\delta$ by assuming that $p$ and $q$ are approximately the same and that $\beta_* \approx \beta_\delta$ is close to unity, in the following sense.
We define a bookkeeping parameter $\epsilon$ that tracks small quantities, and write
\begin{align}
    p & = q(1 + \epsilon r) \,, \\
    \beta & = 1 + \epsilon \, \delta \beta \,.
\end{align}
We track quantities at leading order in $\epsilon$ to solve for the $\delta \beta$ that minimizes $\chi^2$ at this order. 
This bookkeeping parameter falls out of the final solution, which is linear in the small $r$ deviation between $p$ and $q$.

We start by taking the derivative of $\chi^2(p||q_\beta)$ with respect to $\beta$.
We have
\begin{align}
    \chi^2(p||q_\beta) &= \int\frac{p^2}{q_\beta}dx - 1
    = Z_\beta \int\frac{p^2}{q^\beta}dx - 1 \,,
\end{align}

so that
\begin{align}
    \frac{d}{d\beta}& \chi^2(p||q_\beta) = Z_\beta\int\frac{d}{d\beta}\frac{p^2}{q^\beta}dx + \int\frac{p^2}{q^\beta}dx\frac{dZ_\beta}{d\beta}\\
    &= -\int q^\beta dx\int\frac{p^2}{q^\beta}\log{q}\,dx + \int\frac{p^2}{q^\beta}dx\int q^\beta\log{q}\,dx \,.\label{deriv}
\end{align}
Using the substitution $\beta = 1 + \epsilon\, \delta\beta$ and Taylor expanding $q^\beta$ and $p^2/q^\beta$ around $\epsilon \, \delta\beta=0$, we find
\begin{align}
    q^\beta &= q^{1 + \epsilon\delta\beta}
    = q(1 + \epsilon\delta\beta\log{q} 
    + \mathcal{O}(\epsilon^2))
\end{align}
and additionally, using $p = q(1 + \epsilon r)$,
\begin{align}
    \frac{p^2}{q^\beta} &= \frac{p^2}{q(1 + \epsilon\delta\beta\log{q} + \mathcal{O}(\epsilon^2)}\\
    &= \frac{p^2}{q}(1 - \epsilon\delta\beta\log{q} + \mathcal{O}(\epsilon^2))\\
    &= \frac{q^2(1 + \epsilon r)^2}{q}(1 - \epsilon\delta\beta\log{q} + \mathcal{O}(\epsilon^2))\\
    &= q(1 - \epsilon\delta\beta\log{q} + 2\epsilon r + \mathcal{O}(\epsilon^2)) \,.
\end{align}
Thus, keeping terms only to first order in $\epsilon$, we have
\begin{align}
    &q^\beta \approx q(1 + \epsilon\delta\beta\log{q}) \,,\\
    &\frac{p^2}{q^\beta} \approx q(1 - \epsilon\delta\beta\log{q} + 2\epsilon r) \,.
\end{align}
Plugging these approximations back into Eq.~\eqref{deriv}, we find that up to first order in $\epsilon$,
\begin{widetext} 
\begin{align}
\begin{split}
    \frac{d}{d\beta}&\chi^2(p||q_\beta) 
    \approx
    -\left(1 + \epsilon\delta\beta\int\log{q}\,dx\right)
    \int q(1 - \epsilon\delta\beta\log{q} + 2\epsilon r)\log{q}\,dx 
    \\ &\quad
    + \int q(1 - \epsilon\delta\beta\log{q} + 2\epsilon r)dx \int q(1 + \epsilon\delta\beta\log{q})\log{q}\,dx\\
    &= -(1 + \epsilon\delta\beta\E_q[\log{q}])(\E_q[\log{q}] - \epsilon\delta\beta\E_q[(\log{q})^2] + 2\epsilon\E_q[r\log{q}])
    + (1 - \epsilon\delta\beta\E_q[\log{q}])(\E_q[\log{q}] + \epsilon\delta\beta\E_q[(\log{q})^2])
    \\
    &\approx -(\E_q[\log{q}] - \epsilon\delta\beta\E_q[(\log{q})^2] + 2\epsilon\E_q[r\log{q}] + \epsilon\delta\beta\left(\E_q[\log{q}]\right)^2) 
    + \E_q[\log{q}] + \epsilon\delta\beta\E_q[(\log{q})^2] - \epsilon\delta\beta(\E_q[\log{q}])^2\\
    &= 2\epsilon\delta\beta\E_q[(\log{q})^2] - 2\epsilon\delta\beta(\E_q[\log{q}])^2 - 2\epsilon\E_q[r\log{q}]
     \,.
\end{split}
\end{align}
\end{widetext}
The result is
\begin{align}
   \frac{d}{d\beta}&\chi^2(p||q_\beta) \approx 
   2\epsilon (\delta\beta \, \text{Var}_q[\log{q}] - \E_q[r\log{q}]) \,.
\end{align}
Next, since our goal is to find the value of $\beta$ that minimizes $\chi^2(p||q_\beta)$, we set this expression equal to zero and solve for $\delta \beta$.
We find
\begin{equation}
    \beta_\delta = 1 + \frac{\E_q[r\log{q}]}{\text{Var}_q[\log{q}]} \,,
\end{equation}
where we denote the inverse temperature $\beta_\delta$ 
to distinguish this approximation from the true minimizer $\beta_*$.

Finally, we recall our goal is to compute an approximation $T_\delta$ for the optimal temperature $T_*$. 
At the order we have worked, there are two natural choices for representing $T_\delta$ in terms of $\delta \beta$. 
We can re-expand in small $\epsilon$, $T_\delta \approx 1 - \epsilon \, \delta \beta$, or use a resummed version $T_\delta \approx 1/(1 + \epsilon \, \delta\beta)$.
These agree to leading order in $\epsilon$ but can differ appreciably for moderate temperatures.
By applying both to our one-dimensional Gaussian example from Sec.~\ref{sec:Tempering}, we find that the former choice (re-expanding in small quantities) dramatically outperforms the latter resummed estimate, which suffers from divergences at moderate biases.
Remembering then that $r = (1/\epsilon)(p/q - 1)$, we obtain our result Eq.~\eqref{eq:approx}.

This approach to estimating $T_*$ boils down to seeking the minimum of $\chi^2(p||q)$ using a single Newton-Raphson step starting from $\beta = 1$.
If $\beta_*$ is sufficiently close to $\beta = 1$ the method is guaranteed to give a temperature $\beta_\delta$ closer to $\beta_*$, but as is standard for applications of Newton's method, if $\beta = 1$ is too far from $\beta_*$ this method can fail catastrophically. 
Note that while further iterations would improve the temperature estimate in the convergent case, each iterate requires sampling from $q_T$ and so may be prohibitively expensive.

\section{Approximate optimal temperature in the Gaussian case}
\label{sec:approx_gauss}
In the case that $p$ and $q$ are both Gaussian, we can find the optimal temperature $T_*$ explicitly, as given in Eq.~\eqref{eq:gauss_exact}.
We now aim to use our approximation as given in Eq.~\eqref{eq:gauss_approx} to obtain 
$T_\delta$
in terms of $\mu$, $\sigma_p$, and $\sigma_q$.

In this case,
\begin{align}
    & p(x) = \frac{1}{\sqrt{2\pi}\sigma_p}\exp\left(-\frac12\frac{x^2}{\sigma_p^2}\right) \,,\\
    &q(x) = \frac{1}{\sqrt{2\pi}\sigma_q}\exp\left(-\frac12\frac{(x-\mu)^2}{\sigma_q^2}\right) \,.
\end{align}
We start by noting that
\begin{align}
    &\E_q[(p/q-1)\log{q}] = \int p\log{q}\,dx - \int q\log{q}\,dx \,,\\
    &\text{Var}_q[\log{q}] = \int q(\log{q})^2\,dx - \left(\int q\log{q}\,dx\right)^2 \,.
\end{align}
\begin{widetext}
Next,
\begin{align*}
    \int q\log{q}\,dx &= \frac{1}{\sqrt{2\pi}\sigma_q}\int\left(-\frac12\frac{(x-\mu)^2}{\sigma_q^2} - \log{\sqrt{2\pi}\sigma_q}\right)\exp\left(-\frac12\frac{(x-\mu)^2}{\sigma_q^2}\right)dx\\
    &= -\frac{1}{2\sqrt{2\pi}\sigma_q^3}\int(x-\mu)^2\exp\left(-\frac12\frac{(x-\mu)^2}{\sigma_q^2}\right)dx - \frac{\log{\sqrt{2\pi}\sigma_q}}{\sqrt{2\pi}\sigma_q}\int\exp\left(-\frac12\frac{(x-\mu)^2}{\sigma_q^2}\right)dx\\
    &= -\frac{1}{2\sqrt{2\pi}\sigma_q^3}(\sqrt{2\pi}\sigma_q^3) - \frac{\log{\sqrt{2\pi}\sigma_q}}{\sqrt{2\pi}\sigma_q}(\sqrt{2\pi}\sigma_q)\\
    &= -\frac12 - \log{\sqrt{2\pi}\sigma_q} \,.
\end{align*}
Furthermore,
\begin{align*}
\begin{split}
    \int q(\log{q})^2dx &= \frac{1}{\sqrt{2\pi}\sigma_q}\int\left(\frac{(x-\mu)^4}{4\sigma_q^4} + \frac{(x-\mu)^2}{\sigma_q^2}\log{\sqrt{2\pi}\sigma_q} + (\log{\sqrt{2\pi}\sigma_q})^2\right)\exp\left(-\frac12\frac{(x-\mu)^2}{\sigma_q^2}\right)dx\\
    &= \frac{1}{4\sqrt{2\pi}\sigma_q^5}\int(x-\mu)^4\exp\left(-\frac12\frac{(x-\mu)^2}{\sigma_q^2}\right)dx + \frac{\log{\sqrt{2\pi}\sigma_q}}{\sqrt{2\pi}\sigma_q^3}\int(x-\mu)^2\exp\left(-\frac12\frac{(x-\mu)^2}{\sigma_q^2}\right)dx \\&\quad+ \frac{(\log{\sqrt{2\pi}\sigma_q})^2}{\sqrt{2\pi}\sigma_q}\exp\left(-\frac12\frac{(x-\mu)^2}{\sigma_q^2}\right)dx\\
    &= \frac{1}{4\sqrt{2\pi}\sigma_q^5}(3\sqrt{2\pi}\sigma_q^5) + \frac{\log{\sqrt{2\pi}\sigma_q}}{\sqrt{2\pi}\sigma_q^3}(\sqrt{2\pi}\sigma_q^3) + \frac{(\log{\sqrt{2\pi}\sigma_q})^2}{\sqrt{2\pi}\sigma_q}(\sqrt{2\pi}\sigma_q)\\
    &= \frac{3}{4} + \log{\sqrt{2\pi}\sigma_q} + (\log{\sqrt{2\pi}\sigma_q})^2 \,.
\end{split}
\end{align*}
Finally,
\begin{align*}
    \int p\log{q}\,dx &= \frac{1}{\sqrt{2\pi}\sigma_p}\int\left(-\frac12\frac{(x-\mu)^2}{\sigma_q^2} - \log{\sqrt{2\pi}\sigma_q}\right)\exp\left(-\frac12\frac{x^2}{\sigma_p^2}\right)dx\\
    &= -\frac{1}{2\sqrt{2\pi}\sigma_p\sigma_q^2}\int(x-\mu)^2\exp\left(-\frac12\frac{x^2}{\sigma_p^2}\right)dx - \frac{\log{\sqrt{2\pi}\sigma_q}}{\sqrt{2\pi}\sigma_p}\int\exp\left(-\frac12\frac{x^2}{\sigma_p^2}\right)dx\\
    &= -\frac{1}{2\sqrt{2\pi}\sigma_p\sigma_q^2}(\sqrt{2\pi}\sigma_p(\sigma_p^2+\mu^2)) - \frac{\log{\sqrt{2\pi}\sigma_q}}{\sqrt{2\pi}\sigma_p}(\sqrt{2\pi}\sigma_p)\\
    &= -\frac{\sigma_p^2 + \mu^2}{2\sigma_q^2} - \log{\sqrt{2\pi}\sigma_q} \,.
\end{align*}
Putting it all together, several cancellations yield the simplified results
\begin{align}
    \E_q[(p/q-1)\log{q}] &= -\frac{\sigma_p^2 + \mu^2}{2\sigma_q^2} - \log{\sqrt{2\pi}\sigma_q} + \frac12 + \log{\sqrt{2\pi}\sigma_q}
    = \frac12\left(1 - \frac{\sigma_p^2+\mu^2}{\sigma_q^2}\right) \,,\\
    \text{Var}_q[\log{q}] &= \frac{3}{4} + \log{\sqrt{2\pi}\sigma_q} + (\log{\sqrt{2\pi}\sigma_q})^2 - (\frac14 + \log{\sqrt{2\pi}\sigma_q} + (\log{\sqrt{2\pi}\sigma_q})^2)
    = \frac12 \,.
\end{align}
\end{widetext}
Thus,
\begin{equation}
    \delta \beta \approx 1 - \frac{\sigma_p^2 + \mu^2}{\sigma_q^2} \,.
\end{equation}
By re-expanding in the expected smallness of $\delta \beta$, we have finally
\begin{equation}
    T_\delta = \frac{\sigma_p^2 + \mu^2}{\sigma_q^2} \,,
\end{equation}
which is used in Sec.~\ref{sec:OptimalAppx}.

These steps can be generalized with some effort for generic multi-dimensional Gaussians, giving at the same level of approximation
\begin{align}
    T_\delta = \frac{1}{n} \left( {\rm Tr}[ \Sigma_q^{-1} \Sigma_p] + \mu^\top \Sigma_q^{-1} \mu \right)\,,
\end{align}
which demonstrates that this temperature estimate tends to remain $O(1)$ even as $n$ grows, as required for our approximation.

\bibliography{TMFIS.bbl}

\begin{thebibliography}{100}%
\makeatletter
\providecommand \@ifxundefined [1]{%
 \@ifx{#1\undefined}
}%
\providecommand \@ifnum [1]{%
 \ifnum #1\expandafter \@firstoftwo
 \else \expandafter \@secondoftwo
 \fi
}%
\providecommand \@ifx [1]{%
 \ifx #1\expandafter \@firstoftwo
 \else \expandafter \@secondoftwo
 \fi
}%
\providecommand \natexlab [1]{#1}%
\providecommand \enquote  [1]{``#1''}%
\providecommand \bibnamefont  [1]{#1}%
\providecommand \bibfnamefont [1]{#1}%
\providecommand \citenamefont [1]{#1}%
\providecommand \href@noop [0]{\@secondoftwo}%
\providecommand \href [0]{\begingroup \@sanitize@url \@href}%
\providecommand \@href[1]{\@@startlink{#1}\@@href}%
\providecommand \@@href[1]{\endgroup#1\@@endlink}%
\providecommand \@sanitize@url [0]{\catcode `\\12\catcode `\$12\catcode
  `\&12\catcode `\#12\catcode `\^12\catcode `\_12\catcode `\%12\relax}%
\providecommand \@@startlink[1]{}%
\providecommand \@@endlink[0]{}%
\providecommand \url  [0]{\begingroup\@sanitize@url \@url }%
\providecommand \@url [1]{\endgroup\@href {#1}{\urlprefix }}%
\providecommand \urlprefix  [0]{URL }%
\providecommand \Eprint [0]{\href }%
\providecommand \doibase [0]{http://dx.doi.org/}%
\providecommand \selectlanguage [0]{\@gobble}%
\providecommand \bibinfo  [0]{\@secondoftwo}%
\providecommand \bibfield  [0]{\@secondoftwo}%
\providecommand \translation [1]{[#1]}%
\providecommand \BibitemOpen [0]{}%
\providecommand \bibitemStop [0]{}%
\providecommand \bibitemNoStop [0]{.\EOS\space}%
\providecommand \EOS [0]{\spacefactor3000\relax}%
\providecommand \BibitemShut  [1]{\csname bibitem#1\endcsname}%
\let\auto@bib@innerbib\@empty
\bibitem [{\citenamefont {Abbott}\ \emph
  {et~al.}(2016{\natexlab{a}})\citenamefont {Abbott} \emph
  {et~al.}}]{LIGOScientific:2016aoc}%
  \BibitemOpen
  \bibfield  {author} {\bibinfo {author} {\bibfnamefont {B.~P.}\ \bibnamefont
  {Abbott}} \emph {et~al.} (\bibinfo {collaboration} {LIGO Scientific,
  Virgo}),\ }\bibfield  {title} {\enquote {\bibinfo {title} {{Observation of
  Gravitational Waves from a Binary Black Hole Merger}},}\ }\href {\doibase
  10.1103/PhysRevLett.116.061102} {\bibfield  {journal} {\bibinfo  {journal}
  {Phys. Rev. Lett.}\ }\textbf {\bibinfo {volume} {116}},\ \bibinfo {pages}
  {061102} (\bibinfo {year} {2016}{\natexlab{a}})},\ \Eprint
  {http://arxiv.org/abs/1602.03837} {arXiv:1602.03837 [gr-qc]} \BibitemShut
  {NoStop}%
\bibitem [{\citenamefont {Abbott}\ \emph
  {et~al.}(2016{\natexlab{b}})\citenamefont {Abbott} \emph
  {et~al.}}]{LIGOScientific:2016sjg}%
  \BibitemOpen
  \bibfield  {author} {\bibinfo {author} {\bibfnamefont {B.~P.}\ \bibnamefont
  {Abbott}} \emph {et~al.} (\bibinfo {collaboration} {LIGO Scientific,
  Virgo}),\ }\bibfield  {title} {\enquote {\bibinfo {title} {{GW151226:
  Observation of Gravitational Waves from a 22-Solar-Mass Binary Black Hole
  Coalescence}},}\ }\href {\doibase 10.1103/PhysRevLett.116.241103} {\bibfield
  {journal} {\bibinfo  {journal} {Phys. Rev. Lett.}\ }\textbf {\bibinfo
  {volume} {116}},\ \bibinfo {pages} {241103} (\bibinfo {year}
  {2016}{\natexlab{b}})},\ \Eprint {http://arxiv.org/abs/1606.04855}
  {arXiv:1606.04855 [gr-qc]} \BibitemShut {NoStop}%
\bibitem [{\citenamefont {Abbott}\ \emph
  {et~al.}(2016{\natexlab{c}})\citenamefont {Abbott} \emph
  {et~al.}}]{LIGOScientific:2016dsl}%
  \BibitemOpen
  \bibfield  {author} {\bibinfo {author} {\bibfnamefont {B.~P.}\ \bibnamefont
  {Abbott}} \emph {et~al.} (\bibinfo {collaboration} {LIGO Scientific,
  Virgo}),\ }\bibfield  {title} {\enquote {\bibinfo {title} {{Binary Black Hole
  Mergers in the first Advanced LIGO Observing Run}},}\ }\href {\doibase
  10.1103/PhysRevX.6.041015} {\bibfield  {journal} {\bibinfo  {journal} {Phys.
  Rev. X}\ }\textbf {\bibinfo {volume} {6}},\ \bibinfo {pages} {041015}
  (\bibinfo {year} {2016}{\natexlab{c}})},\ \bibinfo {note} {[Erratum:
  Phys.Rev.X 8, 039903 (2018)]},\ \Eprint {http://arxiv.org/abs/1606.04856}
  {arXiv:1606.04856 [gr-qc]} \BibitemShut {NoStop}%
\bibitem [{\citenamefont {Abbott}\ \emph
  {et~al.}(2017{\natexlab{a}})\citenamefont {Abbott} \emph
  {et~al.}}]{LIGOScientific:2017bnn}%
  \BibitemOpen
  \bibfield  {author} {\bibinfo {author} {\bibfnamefont {Benjamin~P.}\
  \bibnamefont {Abbott}} \emph {et~al.} (\bibinfo {collaboration} {LIGO
  Scientific, VIRGO}),\ }\bibfield  {title} {\enquote {\bibinfo {title}
  {{GW170104: Observation of a 50-Solar-Mass Binary Black Hole Coalescence at
  Redshift 0.2}},}\ }\href {\doibase 10.1103/PhysRevLett.118.221101} {\bibfield
   {journal} {\bibinfo  {journal} {Phys. Rev. Lett.}\ }\textbf {\bibinfo
  {volume} {118}},\ \bibinfo {pages} {221101} (\bibinfo {year}
  {2017}{\natexlab{a}})},\ \bibinfo {note} {[Erratum: Phys.Rev.Lett. 121,
  129901 (2018)]},\ \Eprint {http://arxiv.org/abs/1706.01812} {arXiv:1706.01812
  [gr-qc]} \BibitemShut {NoStop}%
\bibitem [{\citenamefont {Abbott}\ \emph
  {et~al.}(2017{\natexlab{b}})\citenamefont {Abbott} \emph
  {et~al.}}]{LIGOScientific:2017ycc}%
  \BibitemOpen
  \bibfield  {author} {\bibinfo {author} {\bibfnamefont {B.~P.}\ \bibnamefont
  {Abbott}} \emph {et~al.} (\bibinfo {collaboration} {LIGO Scientific,
  Virgo}),\ }\bibfield  {title} {\enquote {\bibinfo {title} {{GW170814: A
  Three-Detector Observation of Gravitational Waves from a Binary Black Hole
  Coalescence}},}\ }\href {\doibase 10.1103/PhysRevLett.119.141101} {\bibfield
  {journal} {\bibinfo  {journal} {Phys. Rev. Lett.}\ }\textbf {\bibinfo
  {volume} {119}},\ \bibinfo {pages} {141101} (\bibinfo {year}
  {2017}{\natexlab{b}})},\ \Eprint {http://arxiv.org/abs/1709.09660}
  {arXiv:1709.09660 [gr-qc]} \BibitemShut {NoStop}%
\bibitem [{\citenamefont {Abbott}\ \emph
  {et~al.}(2017{\natexlab{c}})\citenamefont {Abbott} \emph
  {et~al.}}]{LIGOScientific:2017vwq}%
  \BibitemOpen
  \bibfield  {author} {\bibinfo {author} {\bibfnamefont {B.~P.}\ \bibnamefont
  {Abbott}} \emph {et~al.} (\bibinfo {collaboration} {LIGO Scientific,
  Virgo}),\ }\bibfield  {title} {\enquote {\bibinfo {title} {{GW170817:
  Observation of Gravitational Waves from a Binary Neutron Star Inspiral}},}\
  }\href {\doibase 10.1103/PhysRevLett.119.161101} {\bibfield  {journal}
  {\bibinfo  {journal} {Phys. Rev. Lett.}\ }\textbf {\bibinfo {volume} {119}},\
  \bibinfo {pages} {161101} (\bibinfo {year} {2017}{\natexlab{c}})},\ \Eprint
  {http://arxiv.org/abs/1710.05832} {arXiv:1710.05832 [gr-qc]} \BibitemShut
  {NoStop}%
\bibitem [{\citenamefont {Abbott}\ \emph
  {et~al.}(2017{\natexlab{d}})\citenamefont {Abbott} \emph
  {et~al.}}]{LIGOScientific:2017vox}%
  \BibitemOpen
  \bibfield  {author} {\bibinfo {author} {\bibfnamefont {B.~. P.~.}\
  \bibnamefont {Abbott}} \emph {et~al.} (\bibinfo {collaboration} {LIGO
  Scientific, Virgo}),\ }\bibfield  {title} {\enquote {\bibinfo {title}
  {{GW170608: Observation of a 19-solar-mass Binary Black Hole Coalescence}},}\
  }\href {\doibase 10.3847/2041-8213/aa9f0c} {\bibfield  {journal} {\bibinfo
  {journal} {Astrophys. J. Lett.}\ }\textbf {\bibinfo {volume} {851}},\
  \bibinfo {pages} {L35} (\bibinfo {year} {2017}{\natexlab{d}})},\ \Eprint
  {http://arxiv.org/abs/1711.05578} {arXiv:1711.05578 [astro-ph.HE]}
  \BibitemShut {NoStop}%
\bibitem [{\citenamefont {Abbott}\ \emph {et~al.}(2019)\citenamefont {Abbott}
  \emph {et~al.}}]{LIGOScientific:2018mvr}%
  \BibitemOpen
  \bibfield  {author} {\bibinfo {author} {\bibfnamefont {B.~P.}\ \bibnamefont
  {Abbott}} \emph {et~al.} (\bibinfo {collaboration} {LIGO Scientific,
  Virgo}),\ }\bibfield  {title} {\enquote {\bibinfo {title} {{GWTC-1: A
  Gravitational-Wave Transient Catalog of Compact Binary Mergers Observed by
  LIGO and Virgo during the First and Second Observing Runs}},}\ }\href
  {\doibase 10.1103/PhysRevX.9.031040} {\bibfield  {journal} {\bibinfo
  {journal} {Phys. Rev. X}\ }\textbf {\bibinfo {volume} {9}},\ \bibinfo {pages}
  {031040} (\bibinfo {year} {2019})},\ \Eprint
  {http://arxiv.org/abs/1811.12907} {arXiv:1811.12907 [astro-ph.HE]}
  \BibitemShut {NoStop}%
\bibitem [{\citenamefont {Abbott}\ \emph
  {et~al.}(2020{\natexlab{a}})\citenamefont {Abbott} \emph
  {et~al.}}]{LIGOScientific:2020aai}%
  \BibitemOpen
  \bibfield  {author} {\bibinfo {author} {\bibfnamefont {B.~P.}\ \bibnamefont
  {Abbott}} \emph {et~al.} (\bibinfo {collaboration} {LIGO Scientific,
  Virgo}),\ }\bibfield  {title} {\enquote {\bibinfo {title} {{GW190425:
  Observation of a Compact Binary Coalescence with Total Mass $\sim 3.4
  M_{\odot}$}},}\ }\href {\doibase 10.3847/2041-8213/ab75f5} {\bibfield
  {journal} {\bibinfo  {journal} {Astrophys. J. Lett.}\ }\textbf {\bibinfo
  {volume} {892}},\ \bibinfo {pages} {L3} (\bibinfo {year}
  {2020}{\natexlab{a}})},\ \Eprint {http://arxiv.org/abs/2001.01761}
  {arXiv:2001.01761 [astro-ph.HE]} \BibitemShut {NoStop}%
\bibitem [{\citenamefont {Abbott}\ \emph
  {et~al.}(2020{\natexlab{b}})\citenamefont {Abbott} \emph
  {et~al.}}]{LIGOScientific:2020stg}%
  \BibitemOpen
  \bibfield  {author} {\bibinfo {author} {\bibfnamefont {R.}~\bibnamefont
  {Abbott}} \emph {et~al.} (\bibinfo {collaboration} {LIGO Scientific,
  Virgo}),\ }\bibfield  {title} {\enquote {\bibinfo {title} {{GW190412:
  Observation of a Binary-Black-Hole Coalescence with Asymmetric Masses}},}\
  }\href {\doibase 10.1103/PhysRevD.102.043015} {\bibfield  {journal} {\bibinfo
   {journal} {Phys. Rev. D}\ }\textbf {\bibinfo {volume} {102}},\ \bibinfo
  {pages} {043015} (\bibinfo {year} {2020}{\natexlab{b}})},\ \Eprint
  {http://arxiv.org/abs/2004.08342} {arXiv:2004.08342 [astro-ph.HE]}
  \BibitemShut {NoStop}%
\bibitem [{\citenamefont {Abbott}\ \emph
  {et~al.}(2020{\natexlab{c}})\citenamefont {Abbott} \emph
  {et~al.}}]{LIGOScientific:2020zkf}%
  \BibitemOpen
  \bibfield  {author} {\bibinfo {author} {\bibfnamefont {R.}~\bibnamefont
  {Abbott}} \emph {et~al.} (\bibinfo {collaboration} {LIGO Scientific,
  Virgo}),\ }\bibfield  {title} {\enquote {\bibinfo {title} {{GW190814:
  Gravitational Waves from the Coalescence of a 23 Solar Mass Black Hole with a
  2.6 Solar Mass Compact Object}},}\ }\href {\doibase 10.3847/2041-8213/ab960f}
  {\bibfield  {journal} {\bibinfo  {journal} {Astrophys. J. Lett.}\ }\textbf
  {\bibinfo {volume} {896}},\ \bibinfo {pages} {L44} (\bibinfo {year}
  {2020}{\natexlab{c}})},\ \Eprint {http://arxiv.org/abs/2006.12611}
  {arXiv:2006.12611 [astro-ph.HE]} \BibitemShut {NoStop}%
\bibitem [{\citenamefont {Abbott}\ \emph
  {et~al.}(2020{\natexlab{d}})\citenamefont {Abbott} \emph
  {et~al.}}]{LIGOScientific:2020iuh}%
  \BibitemOpen
  \bibfield  {author} {\bibinfo {author} {\bibfnamefont {R.}~\bibnamefont
  {Abbott}} \emph {et~al.} (\bibinfo {collaboration} {LIGO Scientific,
  Virgo}),\ }\bibfield  {title} {\enquote {\bibinfo {title} {{GW190521: A
  Binary Black Hole Merger with a Total Mass of $150 M_{\odot}$}},}\ }\href
  {\doibase 10.1103/PhysRevLett.125.101102} {\bibfield  {journal} {\bibinfo
  {journal} {Phys. Rev. Lett.}\ }\textbf {\bibinfo {volume} {125}},\ \bibinfo
  {pages} {101102} (\bibinfo {year} {2020}{\natexlab{d}})},\ \Eprint
  {http://arxiv.org/abs/2009.01075} {arXiv:2009.01075 [gr-qc]} \BibitemShut
  {NoStop}%
\bibitem [{\citenamefont {Abbott}\ \emph
  {et~al.}(2021{\natexlab{a}})\citenamefont {Abbott} \emph
  {et~al.}}]{LIGOScientific:2020ibl}%
  \BibitemOpen
  \bibfield  {author} {\bibinfo {author} {\bibfnamefont {R.}~\bibnamefont
  {Abbott}} \emph {et~al.} (\bibinfo {collaboration} {LIGO Scientific,
  Virgo}),\ }\bibfield  {title} {\enquote {\bibinfo {title} {{GWTC-2: Compact
  Binary Coalescences Observed by LIGO and Virgo During the First Half of the
  Third Observing Run}},}\ }\href {\doibase 10.1103/PhysRevX.11.021053}
  {\bibfield  {journal} {\bibinfo  {journal} {Phys. Rev. X}\ }\textbf {\bibinfo
  {volume} {11}},\ \bibinfo {pages} {021053} (\bibinfo {year}
  {2021}{\natexlab{a}})},\ \Eprint {http://arxiv.org/abs/2010.14527}
  {arXiv:2010.14527 [gr-qc]} \BibitemShut {NoStop}%
\bibitem [{\citenamefont {Abbott}\ \emph
  {et~al.}(2021{\natexlab{b}})\citenamefont {Abbott} \emph
  {et~al.}}]{LIGOScientific:2021qlt}%
  \BibitemOpen
  \bibfield  {author} {\bibinfo {author} {\bibfnamefont {R.}~\bibnamefont
  {Abbott}} \emph {et~al.} (\bibinfo {collaboration} {LIGO Scientific, KAGRA,
  VIRGO}),\ }\bibfield  {title} {\enquote {\bibinfo {title} {{Observation of
  Gravitational Waves from Two Neutron Star\textendash{}Black Hole
  Coalescences}},}\ }\href {\doibase 10.3847/2041-8213/ac082e} {\bibfield
  {journal} {\bibinfo  {journal} {Astrophys. J. Lett.}\ }\textbf {\bibinfo
  {volume} {915}},\ \bibinfo {pages} {L5} (\bibinfo {year}
  {2021}{\natexlab{b}})},\ \Eprint {http://arxiv.org/abs/2106.15163}
  {arXiv:2106.15163 [astro-ph.HE]} \BibitemShut {NoStop}%
\bibitem [{\citenamefont {Abbott}\ \emph {et~al.}(2024)\citenamefont {Abbott}
  \emph {et~al.}}]{LIGOScientific:2021usb}%
  \BibitemOpen
  \bibfield  {author} {\bibinfo {author} {\bibfnamefont {R.}~\bibnamefont
  {Abbott}} \emph {et~al.} (\bibinfo {collaboration} {LIGO Scientific,
  VIRGO}),\ }\bibfield  {title} {\enquote {\bibinfo {title} {{GWTC-2.1: Deep
  extended catalog of compact binary coalescences observed by LIGO and Virgo
  during the first half of the third observing run}},}\ }\href {\doibase
  10.1103/PhysRevD.109.022001} {\bibfield  {journal} {\bibinfo  {journal}
  {Phys. Rev. D}\ }\textbf {\bibinfo {volume} {109}},\ \bibinfo {pages}
  {022001} (\bibinfo {year} {2024})},\ \Eprint
  {http://arxiv.org/abs/2108.01045} {arXiv:2108.01045 [gr-qc]} \BibitemShut
  {NoStop}%
\bibitem [{\citenamefont {Abbott}\ \emph
  {et~al.}(2023{\natexlab{a}})\citenamefont {Abbott} \emph
  {et~al.}}]{KAGRA:2021vkt}%
  \BibitemOpen
  \bibfield  {author} {\bibinfo {author} {\bibfnamefont {R.}~\bibnamefont
  {Abbott}} \emph {et~al.} (\bibinfo {collaboration} {LIGO Scientific, Virgo,
  KAGRA}),\ }\bibfield  {title} {\enquote {\bibinfo {title} {{GWTC-3: Compact
  Binary Coalescences Observed by LIGO and Virgo during the Second Part of the
  Third Observing Run}},}\ }\href {\doibase 10.1103/PhysRevX.13.041039}
  {\bibfield  {journal} {\bibinfo  {journal} {Phys. Rev. X}\ }\textbf {\bibinfo
  {volume} {13}},\ \bibinfo {pages} {041039} (\bibinfo {year}
  {2023}{\natexlab{a}})},\ \Eprint {http://arxiv.org/abs/2111.03606}
  {arXiv:2111.03606 [gr-qc]} \BibitemShut {NoStop}%
\bibitem [{LIG(2024)}]{LIGOScientific:2024elc}%
  \BibitemOpen
  \bibfield  {title} {\enquote {\bibinfo {title} {{Observation of Gravitational
  Waves from the Coalescence of a $2.5-4.5~M_\odot$ Compact Object and a
  Neutron Star}},}\ }\href@noop {} {\  (\bibinfo {year} {2024})},\ \Eprint
  {http://arxiv.org/abs/2404.04248} {arXiv:2404.04248 [astro-ph.HE]}
  \BibitemShut {NoStop}%
\bibitem [{\citenamefont {Nitz}\ \emph {et~al.}(2019)\citenamefont {Nitz},
  \citenamefont {Capano}, \citenamefont {Nielsen}, \citenamefont {Reyes},
  \citenamefont {White}, \citenamefont {Brown},\ and\ \citenamefont
  {Krishnan}}]{Nitz:2018imz}%
  \BibitemOpen
  \bibfield  {author} {\bibinfo {author} {\bibfnamefont {Alexander~H.}\
  \bibnamefont {Nitz}}, \bibinfo {author} {\bibfnamefont {Collin}\ \bibnamefont
  {Capano}}, \bibinfo {author} {\bibfnamefont {Alex~B.}\ \bibnamefont
  {Nielsen}}, \bibinfo {author} {\bibfnamefont {Steven}\ \bibnamefont {Reyes}},
  \bibinfo {author} {\bibfnamefont {Rebecca}\ \bibnamefont {White}}, \bibinfo
  {author} {\bibfnamefont {Duncan~A.}\ \bibnamefont {Brown}}, \ and\ \bibinfo
  {author} {\bibfnamefont {Badri}\ \bibnamefont {Krishnan}},\ }\bibfield
  {title} {\enquote {\bibinfo {title} {{1-OGC: The first open
  gravitational-wave catalog of binary mergers from analysis of public Advanced
  LIGO data}},}\ }\href {\doibase 10.3847/1538-4357/ab0108} {\bibfield
  {journal} {\bibinfo  {journal} {Astrophys. J.}\ }\textbf {\bibinfo {volume}
  {872}},\ \bibinfo {pages} {195} (\bibinfo {year} {2019})},\ \Eprint
  {http://arxiv.org/abs/1811.01921} {arXiv:1811.01921 [gr-qc]} \BibitemShut
  {NoStop}%
\bibitem [{\citenamefont {Nitz}\ \emph {et~al.}(2020)\citenamefont {Nitz},
  \citenamefont {Dent}, \citenamefont {Davies}, \citenamefont {Kumar},
  \citenamefont {Capano}, \citenamefont {Harry}, \citenamefont {Mozzon},
  \citenamefont {Nuttall}, \citenamefont {Lundgren},\ and\ \citenamefont
  {T\'apai}}]{Nitz:2020oeq}%
  \BibitemOpen
  \bibfield  {author} {\bibinfo {author} {\bibfnamefont {Alexander~H.}\
  \bibnamefont {Nitz}}, \bibinfo {author} {\bibfnamefont {Thomas}\ \bibnamefont
  {Dent}}, \bibinfo {author} {\bibfnamefont {Gareth~S.}\ \bibnamefont
  {Davies}}, \bibinfo {author} {\bibfnamefont {Sumit}\ \bibnamefont {Kumar}},
  \bibinfo {author} {\bibfnamefont {Collin~D.}\ \bibnamefont {Capano}},
  \bibinfo {author} {\bibfnamefont {Ian}\ \bibnamefont {Harry}}, \bibinfo
  {author} {\bibfnamefont {Simone}\ \bibnamefont {Mozzon}}, \bibinfo {author}
  {\bibfnamefont {Laura}\ \bibnamefont {Nuttall}}, \bibinfo {author}
  {\bibfnamefont {Andrew}\ \bibnamefont {Lundgren}}, \ and\ \bibinfo {author}
  {\bibfnamefont {M\'arton}\ \bibnamefont {T\'apai}},\ }\bibfield  {title}
  {\enquote {\bibinfo {title} {{2-OGC: Open Gravitational-wave Catalog of
  binary mergers from analysis of public Advanced LIGO and Virgo data}},}\
  }\href {\doibase 10.3847/1538-4357/ab733f} {\bibfield  {journal} {\bibinfo
  {journal} {Astrophys. J.}\ }\textbf {\bibinfo {volume} {891}},\ \bibinfo
  {pages} {123} (\bibinfo {year} {2020})},\ \Eprint
  {http://arxiv.org/abs/1910.05331} {arXiv:1910.05331 [astro-ph.HE]}
  \BibitemShut {NoStop}%
\bibitem [{\citenamefont {Nitz}\ \emph {et~al.}(2021)\citenamefont {Nitz},
  \citenamefont {Capano}, \citenamefont {Kumar}, \citenamefont {Wang},
  \citenamefont {Kastha}, \citenamefont {Sch\"afer}, \citenamefont
  {Dhurkunde},\ and\ \citenamefont {Cabero}}]{Nitz:2021uxj}%
  \BibitemOpen
  \bibfield  {author} {\bibinfo {author} {\bibfnamefont {Alexander~H.}\
  \bibnamefont {Nitz}}, \bibinfo {author} {\bibfnamefont {Collin~D.}\
  \bibnamefont {Capano}}, \bibinfo {author} {\bibfnamefont {Sumit}\
  \bibnamefont {Kumar}}, \bibinfo {author} {\bibfnamefont {Yi-Fan}\
  \bibnamefont {Wang}}, \bibinfo {author} {\bibfnamefont {Shilpa}\ \bibnamefont
  {Kastha}}, \bibinfo {author} {\bibfnamefont {Marlin}\ \bibnamefont
  {Sch\"afer}}, \bibinfo {author} {\bibfnamefont {Rahul}\ \bibnamefont
  {Dhurkunde}}, \ and\ \bibinfo {author} {\bibfnamefont {Miriam}\ \bibnamefont
  {Cabero}},\ }\bibfield  {title} {\enquote {\bibinfo {title} {{3-OGC: Catalog
  of Gravitational Waves from Compact-binary Mergers}},}\ }\href {\doibase
  10.3847/1538-4357/ac1c03} {\bibfield  {journal} {\bibinfo  {journal}
  {Astrophys. J.}\ }\textbf {\bibinfo {volume} {922}},\ \bibinfo {pages} {76}
  (\bibinfo {year} {2021})},\ \Eprint {http://arxiv.org/abs/2105.09151}
  {arXiv:2105.09151 [astro-ph.HE]} \BibitemShut {NoStop}%
\bibitem [{\citenamefont {Nitz}\ \emph {et~al.}(2023)\citenamefont {Nitz},
  \citenamefont {Kumar}, \citenamefont {Wang}, \citenamefont {Kastha},
  \citenamefont {Wu}, \citenamefont {Sch\"afer}, \citenamefont {Dhurkunde},\
  and\ \citenamefont {Capano}}]{Nitz:2021zwj}%
  \BibitemOpen
  \bibfield  {author} {\bibinfo {author} {\bibfnamefont {Alexander~H.}\
  \bibnamefont {Nitz}}, \bibinfo {author} {\bibfnamefont {Sumit}\ \bibnamefont
  {Kumar}}, \bibinfo {author} {\bibfnamefont {Yi-Fan}\ \bibnamefont {Wang}},
  \bibinfo {author} {\bibfnamefont {Shilpa}\ \bibnamefont {Kastha}}, \bibinfo
  {author} {\bibfnamefont {Shichao}\ \bibnamefont {Wu}}, \bibinfo {author}
  {\bibfnamefont {Marlin}\ \bibnamefont {Sch\"afer}}, \bibinfo {author}
  {\bibfnamefont {Rahul}\ \bibnamefont {Dhurkunde}}, \ and\ \bibinfo {author}
  {\bibfnamefont {Collin~D.}\ \bibnamefont {Capano}},\ }\bibfield  {title}
  {\enquote {\bibinfo {title} {{4-OGC: Catalog of Gravitational Waves from
  Compact Binary Mergers}},}\ }\href {\doibase 10.3847/1538-4357/aca591}
  {\bibfield  {journal} {\bibinfo  {journal} {Astrophys. J.}\ }\textbf
  {\bibinfo {volume} {946}},\ \bibinfo {pages} {59} (\bibinfo {year} {2023})},\
  \Eprint {http://arxiv.org/abs/2112.06878} {arXiv:2112.06878 [astro-ph.HE]}
  \BibitemShut {NoStop}%
\bibitem [{\citenamefont {Zackay}\ \emph {et~al.}(2019)\citenamefont {Zackay},
  \citenamefont {Venumadhav}, \citenamefont {Dai}, \citenamefont {Roulet},\
  and\ \citenamefont {Zaldarriaga}}]{Zackay:2019tzo}%
  \BibitemOpen
  \bibfield  {author} {\bibinfo {author} {\bibfnamefont {Barak}\ \bibnamefont
  {Zackay}}, \bibinfo {author} {\bibfnamefont {Tejaswi}\ \bibnamefont
  {Venumadhav}}, \bibinfo {author} {\bibfnamefont {Liang}\ \bibnamefont {Dai}},
  \bibinfo {author} {\bibfnamefont {Javier}\ \bibnamefont {Roulet}}, \ and\
  \bibinfo {author} {\bibfnamefont {Matias}\ \bibnamefont {Zaldarriaga}},\
  }\bibfield  {title} {\enquote {\bibinfo {title} {{Highly spinning and aligned
  binary black hole merger in the Advanced LIGO first observing run}},}\ }\href
  {\doibase 10.1103/PhysRevD.100.023007} {\bibfield  {journal} {\bibinfo
  {journal} {Phys. Rev. D}\ }\textbf {\bibinfo {volume} {100}},\ \bibinfo
  {pages} {023007} (\bibinfo {year} {2019})},\ \Eprint
  {http://arxiv.org/abs/1902.10331} {arXiv:1902.10331 [astro-ph.HE]}
  \BibitemShut {NoStop}%
\bibitem [{\citenamefont {Venumadhav}\ \emph {et~al.}(2019)\citenamefont
  {Venumadhav}, \citenamefont {Zackay}, \citenamefont {Roulet}, \citenamefont
  {Dai},\ and\ \citenamefont {Zaldarriaga}}]{Venumadhav:2019tad}%
  \BibitemOpen
  \bibfield  {author} {\bibinfo {author} {\bibfnamefont {Tejaswi}\ \bibnamefont
  {Venumadhav}}, \bibinfo {author} {\bibfnamefont {Barak}\ \bibnamefont
  {Zackay}}, \bibinfo {author} {\bibfnamefont {Javier}\ \bibnamefont {Roulet}},
  \bibinfo {author} {\bibfnamefont {Liang}\ \bibnamefont {Dai}}, \ and\
  \bibinfo {author} {\bibfnamefont {Matias}\ \bibnamefont {Zaldarriaga}},\
  }\bibfield  {title} {\enquote {\bibinfo {title} {{New search pipeline for
  compact binary mergers: Results for binary black holes in the first observing
  run of Advanced LIGO}},}\ }\href {\doibase 10.1103/PhysRevD.100.023011}
  {\bibfield  {journal} {\bibinfo  {journal} {Phys. Rev. D}\ }\textbf {\bibinfo
  {volume} {100}},\ \bibinfo {pages} {023011} (\bibinfo {year} {2019})},\
  \Eprint {http://arxiv.org/abs/1902.10341} {arXiv:1902.10341 [astro-ph.IM]}
  \BibitemShut {NoStop}%
\bibitem [{\citenamefont {Venumadhav}\ \emph {et~al.}(2020)\citenamefont
  {Venumadhav}, \citenamefont {Zackay}, \citenamefont {Roulet}, \citenamefont
  {Dai},\ and\ \citenamefont {Zaldarriaga}}]{Venumadhav:2019lyq}%
  \BibitemOpen
  \bibfield  {author} {\bibinfo {author} {\bibfnamefont {Tejaswi}\ \bibnamefont
  {Venumadhav}}, \bibinfo {author} {\bibfnamefont {Barak}\ \bibnamefont
  {Zackay}}, \bibinfo {author} {\bibfnamefont {Javier}\ \bibnamefont {Roulet}},
  \bibinfo {author} {\bibfnamefont {Liang}\ \bibnamefont {Dai}}, \ and\
  \bibinfo {author} {\bibfnamefont {Matias}\ \bibnamefont {Zaldarriaga}},\
  }\bibfield  {title} {\enquote {\bibinfo {title} {{New binary black hole
  mergers in the second observing run of Advanced LIGO and Advanced Virgo}},}\
  }\href {\doibase 10.1103/PhysRevD.101.083030} {\bibfield  {journal} {\bibinfo
   {journal} {Phys. Rev. D}\ }\textbf {\bibinfo {volume} {101}},\ \bibinfo
  {pages} {083030} (\bibinfo {year} {2020})},\ \Eprint
  {http://arxiv.org/abs/1904.07214} {arXiv:1904.07214 [astro-ph.HE]}
  \BibitemShut {NoStop}%
\bibitem [{\citenamefont {Zackay}\ \emph {et~al.}(2021)\citenamefont {Zackay},
  \citenamefont {Dai}, \citenamefont {Venumadhav}, \citenamefont {Roulet},\
  and\ \citenamefont {Zaldarriaga}}]{Zackay:2019btq}%
  \BibitemOpen
  \bibfield  {author} {\bibinfo {author} {\bibfnamefont {Barak}\ \bibnamefont
  {Zackay}}, \bibinfo {author} {\bibfnamefont {Liang}\ \bibnamefont {Dai}},
  \bibinfo {author} {\bibfnamefont {Tejaswi}\ \bibnamefont {Venumadhav}},
  \bibinfo {author} {\bibfnamefont {Javier}\ \bibnamefont {Roulet}}, \ and\
  \bibinfo {author} {\bibfnamefont {Matias}\ \bibnamefont {Zaldarriaga}},\
  }\bibfield  {title} {\enquote {\bibinfo {title} {{Detecting gravitational
  waves with disparate detector responses: Two new binary black hole
  mergers}},}\ }\href {\doibase 10.1103/PhysRevD.104.063030} {\bibfield
  {journal} {\bibinfo  {journal} {Phys. Rev. D}\ }\textbf {\bibinfo {volume}
  {104}},\ \bibinfo {pages} {063030} (\bibinfo {year} {2021})},\ \Eprint
  {http://arxiv.org/abs/1910.09528} {arXiv:1910.09528 [astro-ph.HE]}
  \BibitemShut {NoStop}%
\bibitem [{\citenamefont {Olsen}\ \emph {et~al.}(2022)\citenamefont {Olsen},
  \citenamefont {Venumadhav}, \citenamefont {Mushkin}, \citenamefont {Roulet},
  \citenamefont {Zackay},\ and\ \citenamefont {Zaldarriaga}}]{Olsen:2022pin}%
  \BibitemOpen
  \bibfield  {author} {\bibinfo {author} {\bibfnamefont {Seth}\ \bibnamefont
  {Olsen}}, \bibinfo {author} {\bibfnamefont {Tejaswi}\ \bibnamefont
  {Venumadhav}}, \bibinfo {author} {\bibfnamefont {Jonathan}\ \bibnamefont
  {Mushkin}}, \bibinfo {author} {\bibfnamefont {Javier}\ \bibnamefont
  {Roulet}}, \bibinfo {author} {\bibfnamefont {Barak}\ \bibnamefont {Zackay}},
  \ and\ \bibinfo {author} {\bibfnamefont {Matias}\ \bibnamefont
  {Zaldarriaga}},\ }\bibfield  {title} {\enquote {\bibinfo {title} {{New binary
  black hole mergers in the LIGO-Virgo O3a data}},}\ }\href {\doibase
  10.1103/PhysRevD.106.043009} {\bibfield  {journal} {\bibinfo  {journal}
  {Phys. Rev. D}\ }\textbf {\bibinfo {volume} {106}},\ \bibinfo {pages}
  {043009} (\bibinfo {year} {2022})},\ \Eprint
  {http://arxiv.org/abs/2201.02252} {arXiv:2201.02252 [astro-ph.HE]}
  \BibitemShut {NoStop}%
\bibitem [{\citenamefont {Mehta}\ \emph {et~al.}(2023)\citenamefont {Mehta},
  \citenamefont {Olsen}, \citenamefont {Wadekar}, \citenamefont {Roulet},
  \citenamefont {Venumadhav}, \citenamefont {Mushkin}, \citenamefont {Zackay},\
  and\ \citenamefont {Zaldarriaga}}]{Mehta:2023zlk}%
  \BibitemOpen
  \bibfield  {author} {\bibinfo {author} {\bibfnamefont {Ajit~Kumar}\
  \bibnamefont {Mehta}}, \bibinfo {author} {\bibfnamefont {Seth}\ \bibnamefont
  {Olsen}}, \bibinfo {author} {\bibfnamefont {Digvijay}\ \bibnamefont
  {Wadekar}}, \bibinfo {author} {\bibfnamefont {Javier}\ \bibnamefont
  {Roulet}}, \bibinfo {author} {\bibfnamefont {Tejaswi}\ \bibnamefont
  {Venumadhav}}, \bibinfo {author} {\bibfnamefont {Jonathan}\ \bibnamefont
  {Mushkin}}, \bibinfo {author} {\bibfnamefont {Barak}\ \bibnamefont {Zackay}},
  \ and\ \bibinfo {author} {\bibfnamefont {Matias}\ \bibnamefont
  {Zaldarriaga}},\ }\bibfield  {title} {\enquote {\bibinfo {title} {{New binary
  black hole mergers in the LIGO-Virgo O3b data}},}\ }\href@noop {} {\
  (\bibinfo {year} {2023})},\ \Eprint {http://arxiv.org/abs/2311.06061}
  {arXiv:2311.06061 [gr-qc]} \BibitemShut {NoStop}%
\bibitem [{\citenamefont {Wadekar}\ \emph {et~al.}(2023)\citenamefont
  {Wadekar}, \citenamefont {Roulet}, \citenamefont {Venumadhav}, \citenamefont
  {Mehta}, \citenamefont {Zackay}, \citenamefont {Mushkin}, \citenamefont
  {Olsen},\ and\ \citenamefont {Zaldarriaga}}]{Wadekar:2023gea}%
  \BibitemOpen
  \bibfield  {author} {\bibinfo {author} {\bibfnamefont {Digvijay}\
  \bibnamefont {Wadekar}}, \bibinfo {author} {\bibfnamefont {Javier}\
  \bibnamefont {Roulet}}, \bibinfo {author} {\bibfnamefont {Tejaswi}\
  \bibnamefont {Venumadhav}}, \bibinfo {author} {\bibfnamefont {Ajit~Kumar}\
  \bibnamefont {Mehta}}, \bibinfo {author} {\bibfnamefont {Barak}\ \bibnamefont
  {Zackay}}, \bibinfo {author} {\bibfnamefont {Jonathan}\ \bibnamefont
  {Mushkin}}, \bibinfo {author} {\bibfnamefont {Seth}\ \bibnamefont {Olsen}}, \
  and\ \bibinfo {author} {\bibfnamefont {Matias}\ \bibnamefont {Zaldarriaga}},\
  }\bibfield  {title} {\enquote {\bibinfo {title} {{New black hole mergers in
  the LIGO-Virgo O3 data from a gravitational wave search including
  higher-order harmonics}},}\ }\href@noop {} {\  (\bibinfo {year} {2023})},\
  \Eprint {http://arxiv.org/abs/2312.06631} {arXiv:2312.06631 [gr-qc]}
  \BibitemShut {NoStop}%
\bibitem [{\citenamefont {Aasi}\ \emph {et~al.}(2015)\citenamefont {Aasi} \emph
  {et~al.}}]{LIGOScientific:2014pky}%
  \BibitemOpen
  \bibfield  {author} {\bibinfo {author} {\bibfnamefont {J.}~\bibnamefont
  {Aasi}} \emph {et~al.} (\bibinfo {collaboration} {LIGO Scientific}),\
  }\bibfield  {title} {\enquote {\bibinfo {title} {{Advanced LIGO}},}\ }\href
  {\doibase 10.1088/0264-9381/32/7/074001} {\bibfield  {journal} {\bibinfo
  {journal} {Class. Quant. Grav.}\ }\textbf {\bibinfo {volume} {32}},\ \bibinfo
  {pages} {074001} (\bibinfo {year} {2015})},\ \Eprint
  {http://arxiv.org/abs/1411.4547} {arXiv:1411.4547 [gr-qc]} \BibitemShut
  {NoStop}%
\bibitem [{\citenamefont {Acernese}\ \emph {et~al.}(2015)\citenamefont
  {Acernese} \emph {et~al.}}]{VIRGO:2014yos}%
  \BibitemOpen
  \bibfield  {author} {\bibinfo {author} {\bibfnamefont {F.}~\bibnamefont
  {Acernese}} \emph {et~al.} (\bibinfo {collaboration} {VIRGO}),\ }\bibfield
  {title} {\enquote {\bibinfo {title} {{Advanced Virgo: a second-generation
  interferometric gravitational wave detector}},}\ }\href {\doibase
  10.1088/0264-9381/32/2/024001} {\bibfield  {journal} {\bibinfo  {journal}
  {Class. Quant. Grav.}\ }\textbf {\bibinfo {volume} {32}},\ \bibinfo {pages}
  {024001} (\bibinfo {year} {2015})},\ \Eprint {http://arxiv.org/abs/1408.3978}
  {arXiv:1408.3978 [gr-qc]} \BibitemShut {NoStop}%
\bibitem [{\citenamefont {Akutsu}\ \emph {et~al.}(2021)\citenamefont {Akutsu}
  \emph {et~al.}}]{KAGRA:2020tym}%
  \BibitemOpen
  \bibfield  {author} {\bibinfo {author} {\bibfnamefont {T.}~\bibnamefont
  {Akutsu}} \emph {et~al.} (\bibinfo {collaboration} {KAGRA}),\ }\bibfield
  {title} {\enquote {\bibinfo {title} {{Overview of KAGRA: Detector design and
  construction history}},}\ }\href {\doibase 10.1093/ptep/ptaa125} {\bibfield
  {journal} {\bibinfo  {journal} {PTEP}\ }\textbf {\bibinfo {volume} {2021}},\
  \bibinfo {pages} {05A101} (\bibinfo {year} {2021})},\ \Eprint
  {http://arxiv.org/abs/2005.05574} {arXiv:2005.05574 [physics.ins-det]}
  \BibitemShut {NoStop}%
\bibitem [{\citenamefont {Veitch}\ \emph {et~al.}(2015)\citenamefont {Veitch}
  \emph {et~al.}}]{Veitch:2014wba}%
  \BibitemOpen
  \bibfield  {author} {\bibinfo {author} {\bibfnamefont {J.}~\bibnamefont
  {Veitch}} \emph {et~al.},\ }\bibfield  {title} {\enquote {\bibinfo {title}
  {{Parameter estimation for compact binaries with ground-based
  gravitational-wave observations using the LALInference software library}},}\
  }\href {\doibase 10.1103/PhysRevD.91.042003} {\bibfield  {journal} {\bibinfo
  {journal} {Phys. Rev. D}\ }\textbf {\bibinfo {volume} {91}},\ \bibinfo
  {pages} {042003} (\bibinfo {year} {2015})},\ \Eprint
  {http://arxiv.org/abs/1409.7215} {arXiv:1409.7215 [gr-qc]} \BibitemShut
  {NoStop}%
\bibitem [{\citenamefont {Thrane}\ and\ \citenamefont
  {Talbot}(2019)}]{Thrane:2018qnx}%
  \BibitemOpen
  \bibfield  {author} {\bibinfo {author} {\bibfnamefont {Eric}\ \bibnamefont
  {Thrane}}\ and\ \bibinfo {author} {\bibfnamefont {Colm}\ \bibnamefont
  {Talbot}},\ }\bibfield  {title} {\enquote {\bibinfo {title} {{An introduction
  to Bayesian inference in gravitational-wave astronomy: parameter estimation,
  model selection, and hierarchical models}},}\ }\href {\doibase
  10.1017/pasa.2019.2} {\bibfield  {journal} {\bibinfo  {journal} {Publ.
  Astron. Soc. Austral.}\ }\textbf {\bibinfo {volume} {36}},\ \bibinfo {pages}
  {e010} (\bibinfo {year} {2019})},\ \bibinfo {note} {[Erratum:
  Publ.Astron.Soc.Austral. 37, e036 (2020)]},\ \Eprint
  {http://arxiv.org/abs/1809.02293} {arXiv:1809.02293 [astro-ph.IM]}
  \BibitemShut {NoStop}%
\bibitem [{\citenamefont {Ashton}\ \emph
  {et~al.}(2019{\natexlab{a}})\citenamefont {Ashton} \emph
  {et~al.}}]{Ashton:2018jfp}%
  \BibitemOpen
  \bibfield  {author} {\bibinfo {author} {\bibfnamefont {Gregory}\ \bibnamefont
  {Ashton}} \emph {et~al.},\ }\bibfield  {title} {\enquote {\bibinfo {title}
  {{BILBY: A user-friendly Bayesian inference library for gravitational-wave
  astronomy}},}\ }\href {\doibase 10.3847/1538-4365/ab06fc} {\bibfield
  {journal} {\bibinfo  {journal} {Astrophys. J. Suppl.}\ }\textbf {\bibinfo
  {volume} {241}},\ \bibinfo {pages} {27} (\bibinfo {year}
  {2019}{\natexlab{a}})},\ \Eprint {http://arxiv.org/abs/1811.02042}
  {arXiv:1811.02042 [astro-ph.IM]} \BibitemShut {NoStop}%
\bibitem [{\citenamefont {Romero-Shaw}\ \emph {et~al.}(2020)\citenamefont
  {Romero-Shaw} \emph {et~al.}}]{Romero-Shaw:2020owr}%
  \BibitemOpen
  \bibfield  {author} {\bibinfo {author} {\bibfnamefont {I.~M.}\ \bibnamefont
  {Romero-Shaw}} \emph {et~al.},\ }\bibfield  {title} {\enquote {\bibinfo
  {title} {{Bayesian inference for compact binary coalescences with BILBY:
  validation and application to the first LIGO\textendash{}Virgo
  gravitational-wave transient catalogue}},}\ }\href {\doibase
  10.1093/mnras/staa2850} {\bibfield  {journal} {\bibinfo  {journal} {Mon. Not.
  Roy. Astron. Soc.}\ }\textbf {\bibinfo {volume} {499}},\ \bibinfo {pages}
  {3295--3319} (\bibinfo {year} {2020})},\ \Eprint
  {http://arxiv.org/abs/2006.00714} {arXiv:2006.00714 [astro-ph.IM]}
  \BibitemShut {NoStop}%
\bibitem [{\citenamefont {Field}\ \emph {et~al.}(2014)\citenamefont {Field},
  \citenamefont {Galley}, \citenamefont {Hesthaven}, \citenamefont {Kaye},\
  and\ \citenamefont {Tiglio}}]{Field:2013cfa}%
  \BibitemOpen
  \bibfield  {author} {\bibinfo {author} {\bibfnamefont {Scott~E.}\
  \bibnamefont {Field}}, \bibinfo {author} {\bibfnamefont {Chad~R.}\
  \bibnamefont {Galley}}, \bibinfo {author} {\bibfnamefont {Jan~S.}\
  \bibnamefont {Hesthaven}}, \bibinfo {author} {\bibfnamefont {Jason}\
  \bibnamefont {Kaye}}, \ and\ \bibinfo {author} {\bibfnamefont {Manuel}\
  \bibnamefont {Tiglio}},\ }\bibfield  {title} {\enquote {\bibinfo {title}
  {{Fast prediction and evaluation of gravitational waveforms using surrogate
  models}},}\ }\href {\doibase 10.1103/PhysRevX.4.031006} {\bibfield  {journal}
  {\bibinfo  {journal} {Phys. Rev. X}\ }\textbf {\bibinfo {volume} {4}},\
  \bibinfo {pages} {031006} (\bibinfo {year} {2014})},\ \Eprint
  {http://arxiv.org/abs/1308.3565} {arXiv:1308.3565 [gr-qc]} \BibitemShut
  {NoStop}%
\bibitem [{\citenamefont {Healy}\ \emph {et~al.}(2020)\citenamefont {Healy},
  \citenamefont {Lousto}, \citenamefont {Lange},\ and\ \citenamefont
  {O'Shaughnessy}}]{Healy:2020jjs}%
  \BibitemOpen
  \bibfield  {author} {\bibinfo {author} {\bibfnamefont {James}\ \bibnamefont
  {Healy}}, \bibinfo {author} {\bibfnamefont {Carlos~O.}\ \bibnamefont
  {Lousto}}, \bibinfo {author} {\bibfnamefont {Jacob}\ \bibnamefont {Lange}}, \
  and\ \bibinfo {author} {\bibfnamefont {Richard}\ \bibnamefont
  {O'Shaughnessy}},\ }\bibfield  {title} {\enquote {\bibinfo {title}
  {{Application of the third RIT binary black hole simulations catalog to
  parameter estimation of gravitational waves signals from the LIGO-Virgo O1/O2
  observational runs}},}\ }\href {\doibase 10.1103/PhysRevD.102.124053}
  {\bibfield  {journal} {\bibinfo  {journal} {Phys. Rev. D}\ }\textbf {\bibinfo
  {volume} {102}},\ \bibinfo {pages} {124053} (\bibinfo {year} {2020})},\
  \Eprint {http://arxiv.org/abs/2010.00108} {arXiv:2010.00108 [gr-qc]}
  \BibitemShut {NoStop}%
\bibitem [{\citenamefont {Ajith}\ \emph {et~al.}(2011)\citenamefont {Ajith}
  \emph {et~al.}}]{Ajith:2009bn}%
  \BibitemOpen
  \bibfield  {author} {\bibinfo {author} {\bibfnamefont {P.}~\bibnamefont
  {Ajith}} \emph {et~al.},\ }\bibfield  {title} {\enquote {\bibinfo {title}
  {{Inspiral-merger-ringdown waveforms for black-hole binaries with
  non-precessing spins}},}\ }\href {\doibase 10.1103/PhysRevLett.106.241101}
  {\bibfield  {journal} {\bibinfo  {journal} {Phys. Rev. Lett.}\ }\textbf
  {\bibinfo {volume} {106}},\ \bibinfo {pages} {241101} (\bibinfo {year}
  {2011})},\ \Eprint {http://arxiv.org/abs/0909.2867} {arXiv:0909.2867 [gr-qc]}
  \BibitemShut {NoStop}%
\bibitem [{\citenamefont {Hannam}\ \emph {et~al.}(2014)\citenamefont {Hannam},
  \citenamefont {Schmidt}, \citenamefont {Boh\'e}, \citenamefont {Haegel},
  \citenamefont {Husa}, \citenamefont {Ohme}, \citenamefont {Pratten},\ and\
  \citenamefont {P\"urrer}}]{Hannam:2013oca}%
  \BibitemOpen
  \bibfield  {author} {\bibinfo {author} {\bibfnamefont {Mark}\ \bibnamefont
  {Hannam}}, \bibinfo {author} {\bibfnamefont {Patricia}\ \bibnamefont
  {Schmidt}}, \bibinfo {author} {\bibfnamefont {Alejandro}\ \bibnamefont
  {Boh\'e}}, \bibinfo {author} {\bibfnamefont {Le\"\i{}la}\ \bibnamefont
  {Haegel}}, \bibinfo {author} {\bibfnamefont {Sascha}\ \bibnamefont {Husa}},
  \bibinfo {author} {\bibfnamefont {Frank}\ \bibnamefont {Ohme}}, \bibinfo
  {author} {\bibfnamefont {Geraint}\ \bibnamefont {Pratten}}, \ and\ \bibinfo
  {author} {\bibfnamefont {Michael}\ \bibnamefont {P\"urrer}},\ }\bibfield
  {title} {\enquote {\bibinfo {title} {{Simple Model of Complete Precessing
  Black-Hole-Binary Gravitational Waveforms}},}\ }\href {\doibase
  10.1103/PhysRevLett.113.151101} {\bibfield  {journal} {\bibinfo  {journal}
  {Phys. Rev. Lett.}\ }\textbf {\bibinfo {volume} {113}},\ \bibinfo {pages}
  {151101} (\bibinfo {year} {2014})},\ \Eprint {http://arxiv.org/abs/1308.3271}
  {arXiv:1308.3271 [gr-qc]} \BibitemShut {NoStop}%
\bibitem [{\citenamefont {Pratten}\ \emph {et~al.}(2020)\citenamefont
  {Pratten}, \citenamefont {Husa}, \citenamefont {Garcia-Quiros}, \citenamefont
  {Colleoni}, \citenamefont {Ramos-Buades}, \citenamefont {Estelles},\ and\
  \citenamefont {Jaume}}]{Pratten:2020fqn}%
  \BibitemOpen
  \bibfield  {author} {\bibinfo {author} {\bibfnamefont {Geraint}\ \bibnamefont
  {Pratten}}, \bibinfo {author} {\bibfnamefont {Sascha}\ \bibnamefont {Husa}},
  \bibinfo {author} {\bibfnamefont {Cecilio}\ \bibnamefont {Garcia-Quiros}},
  \bibinfo {author} {\bibfnamefont {Marta}\ \bibnamefont {Colleoni}}, \bibinfo
  {author} {\bibfnamefont {Antoni}\ \bibnamefont {Ramos-Buades}}, \bibinfo
  {author} {\bibfnamefont {Hector}\ \bibnamefont {Estelles}}, \ and\ \bibinfo
  {author} {\bibfnamefont {Rafel}\ \bibnamefont {Jaume}},\ }\bibfield  {title}
  {\enquote {\bibinfo {title} {{Setting the cornerstone for a family of models
  for gravitational waves from compact binaries: The dominant harmonic for
  nonprecessing quasicircular black holes}},}\ }\href {\doibase
  10.1103/PhysRevD.102.064001} {\bibfield  {journal} {\bibinfo  {journal}
  {Phys. Rev. D}\ }\textbf {\bibinfo {volume} {102}},\ \bibinfo {pages}
  {064001} (\bibinfo {year} {2020})},\ \Eprint
  {http://arxiv.org/abs/2001.11412} {arXiv:2001.11412 [gr-qc]} \BibitemShut
  {NoStop}%
\bibitem [{\citenamefont {Garc\'\i{}a-Quir\'os}\ \emph
  {et~al.}(2020)\citenamefont {Garc\'\i{}a-Quir\'os}, \citenamefont {Colleoni},
  \citenamefont {Husa}, \citenamefont {Estell\'es}, \citenamefont {Pratten},
  \citenamefont {Ramos-Buades}, \citenamefont {Mateu-Lucena},\ and\
  \citenamefont {Jaume}}]{Garcia-Quiros:2020qpx}%
  \BibitemOpen
  \bibfield  {author} {\bibinfo {author} {\bibfnamefont {Cecilio}\ \bibnamefont
  {Garc\'\i{}a-Quir\'os}}, \bibinfo {author} {\bibfnamefont {Marta}\
  \bibnamefont {Colleoni}}, \bibinfo {author} {\bibfnamefont {Sascha}\
  \bibnamefont {Husa}}, \bibinfo {author} {\bibfnamefont {H\'ector}\
  \bibnamefont {Estell\'es}}, \bibinfo {author} {\bibfnamefont {Geraint}\
  \bibnamefont {Pratten}}, \bibinfo {author} {\bibfnamefont {Antoni}\
  \bibnamefont {Ramos-Buades}}, \bibinfo {author} {\bibfnamefont {Maite}\
  \bibnamefont {Mateu-Lucena}}, \ and\ \bibinfo {author} {\bibfnamefont
  {Rafel}\ \bibnamefont {Jaume}},\ }\bibfield  {title} {\enquote {\bibinfo
  {title} {{Multimode frequency-domain model for the gravitational wave signal
  from nonprecessing black-hole binaries}},}\ }\href {\doibase
  10.1103/PhysRevD.102.064002} {\bibfield  {journal} {\bibinfo  {journal}
  {Phys. Rev. D}\ }\textbf {\bibinfo {volume} {102}},\ \bibinfo {pages}
  {064002} (\bibinfo {year} {2020})},\ \Eprint
  {http://arxiv.org/abs/2001.10914} {arXiv:2001.10914 [gr-qc]} \BibitemShut
  {NoStop}%
\bibitem [{\citenamefont {Pratten}\ \emph {et~al.}(2021)\citenamefont {Pratten}
  \emph {et~al.}}]{Pratten:2020ceb}%
  \BibitemOpen
  \bibfield  {author} {\bibinfo {author} {\bibfnamefont {Geraint}\ \bibnamefont
  {Pratten}} \emph {et~al.},\ }\bibfield  {title} {\enquote {\bibinfo {title}
  {{Computationally efficient models for the dominant and subdominant harmonic
  modes of precessing binary black holes}},}\ }\href {\doibase
  10.1103/PhysRevD.103.104056} {\bibfield  {journal} {\bibinfo  {journal}
  {Phys. Rev. D}\ }\textbf {\bibinfo {volume} {103}},\ \bibinfo {pages}
  {104056} (\bibinfo {year} {2021})},\ \Eprint
  {http://arxiv.org/abs/2004.06503} {arXiv:2004.06503 [gr-qc]} \BibitemShut
  {NoStop}%
\bibitem [{\citenamefont {Estell\'es}\ \emph {et~al.}(2022)\citenamefont
  {Estell\'es}, \citenamefont {Colleoni}, \citenamefont {Garc\'\i{}a-Quir\'os},
  \citenamefont {Husa}, \citenamefont {Keitel}, \citenamefont {Mateu-Lucena},
  \citenamefont {Planas},\ and\ \citenamefont
  {Ramos-Buades}}]{Estelles:2021gvs}%
  \BibitemOpen
  \bibfield  {author} {\bibinfo {author} {\bibfnamefont {H\'ector}\
  \bibnamefont {Estell\'es}}, \bibinfo {author} {\bibfnamefont {Marta}\
  \bibnamefont {Colleoni}}, \bibinfo {author} {\bibfnamefont {Cecilio}\
  \bibnamefont {Garc\'\i{}a-Quir\'os}}, \bibinfo {author} {\bibfnamefont
  {Sascha}\ \bibnamefont {Husa}}, \bibinfo {author} {\bibfnamefont {David}\
  \bibnamefont {Keitel}}, \bibinfo {author} {\bibfnamefont {Maite}\
  \bibnamefont {Mateu-Lucena}}, \bibinfo {author} {\bibfnamefont {Maria
  de~Lluc}\ \bibnamefont {Planas}}, \ and\ \bibinfo {author} {\bibfnamefont
  {Antoni}\ \bibnamefont {Ramos-Buades}},\ }\bibfield  {title} {\enquote
  {\bibinfo {title} {{New twists in compact binary waveform modeling: A fast
  time-domain model for precession}},}\ }\href {\doibase
  10.1103/PhysRevD.105.084040} {\bibfield  {journal} {\bibinfo  {journal}
  {Phys. Rev. D}\ }\textbf {\bibinfo {volume} {105}},\ \bibinfo {pages}
  {084040} (\bibinfo {year} {2022})},\ \Eprint
  {http://arxiv.org/abs/2105.05872} {arXiv:2105.05872 [gr-qc]} \BibitemShut
  {NoStop}%
\bibitem [{\citenamefont {Yu}\ \emph {et~al.}(2023)\citenamefont {Yu},
  \citenamefont {Roulet}, \citenamefont {Venumadhav}, \citenamefont {Zackay},\
  and\ \citenamefont {Zaldarriaga}}]{Yu:2023lml}%
  \BibitemOpen
  \bibfield  {author} {\bibinfo {author} {\bibfnamefont {Hang}\ \bibnamefont
  {Yu}}, \bibinfo {author} {\bibfnamefont {Javier}\ \bibnamefont {Roulet}},
  \bibinfo {author} {\bibfnamefont {Tejaswi}\ \bibnamefont {Venumadhav}},
  \bibinfo {author} {\bibfnamefont {Barak}\ \bibnamefont {Zackay}}, \ and\
  \bibinfo {author} {\bibfnamefont {Matias}\ \bibnamefont {Zaldarriaga}},\
  }\bibfield  {title} {\enquote {\bibinfo {title} {{Accurate and efficient
  waveform model for precessing binary black holes}},}\ }\href {\doibase
  10.1103/PhysRevD.108.064059} {\bibfield  {journal} {\bibinfo  {journal}
  {Phys. Rev. D}\ }\textbf {\bibinfo {volume} {108}},\ \bibinfo {pages}
  {064059} (\bibinfo {year} {2023})},\ \Eprint
  {http://arxiv.org/abs/2306.08774} {arXiv:2306.08774 [gr-qc]} \BibitemShut
  {NoStop}%
\bibitem [{\citenamefont {Thompson}\ \emph {et~al.}(2024)\citenamefont
  {Thompson}, \citenamefont {Hamilton}, \citenamefont {London}, \citenamefont
  {Ghosh}, \citenamefont {Kolitsidou}, \citenamefont {Hoy},\ and\ \citenamefont
  {Hannam}}]{Thompson:2023ase}%
  \BibitemOpen
  \bibfield  {author} {\bibinfo {author} {\bibfnamefont {Jonathan~E.}\
  \bibnamefont {Thompson}}, \bibinfo {author} {\bibfnamefont {Eleanor}\
  \bibnamefont {Hamilton}}, \bibinfo {author} {\bibfnamefont {Lionel}\
  \bibnamefont {London}}, \bibinfo {author} {\bibfnamefont {Shrobana}\
  \bibnamefont {Ghosh}}, \bibinfo {author} {\bibfnamefont {Panagiota}\
  \bibnamefont {Kolitsidou}}, \bibinfo {author} {\bibfnamefont {Charlie}\
  \bibnamefont {Hoy}}, \ and\ \bibinfo {author} {\bibfnamefont {Mark}\
  \bibnamefont {Hannam}},\ }\bibfield  {title} {\enquote {\bibinfo {title}
  {{PhenomXO4a: a phenomenological gravitational-wave model for precessing
  black-hole binaries with higher multipoles and asymmetries}},}\ }\href
  {\doibase 10.1103/PhysRevD.109.063012} {\bibfield  {journal} {\bibinfo
  {journal} {Phys. Rev. D}\ }\textbf {\bibinfo {volume} {109}},\ \bibinfo
  {pages} {063012} (\bibinfo {year} {2024})},\ \Eprint
  {http://arxiv.org/abs/2312.10025} {arXiv:2312.10025 [gr-qc]} \BibitemShut
  {NoStop}%
\bibitem [{\citenamefont {Buonanno}\ and\ \citenamefont
  {Damour}(1999)}]{Buonanno:1998gg}%
  \BibitemOpen
  \bibfield  {author} {\bibinfo {author} {\bibfnamefont {A.}~\bibnamefont
  {Buonanno}}\ and\ \bibinfo {author} {\bibfnamefont {T.}~\bibnamefont
  {Damour}},\ }\bibfield  {title} {\enquote {\bibinfo {title} {{Effective
  one-body approach to general relativistic two-body dynamics}},}\ }\href
  {\doibase 10.1103/PhysRevD.59.084006} {\bibfield  {journal} {\bibinfo
  {journal} {Phys. Rev. D}\ }\textbf {\bibinfo {volume} {59}},\ \bibinfo
  {pages} {084006} (\bibinfo {year} {1999})},\ \Eprint
  {http://arxiv.org/abs/gr-qc/9811091} {arXiv:gr-qc/9811091} \BibitemShut
  {NoStop}%
\bibitem [{\citenamefont {Buonanno}\ and\ \citenamefont
  {Damour}(2000)}]{Buonanno:2000ef}%
  \BibitemOpen
  \bibfield  {author} {\bibinfo {author} {\bibfnamefont {Alessandra}\
  \bibnamefont {Buonanno}}\ and\ \bibinfo {author} {\bibfnamefont {Thibault}\
  \bibnamefont {Damour}},\ }\bibfield  {title} {\enquote {\bibinfo {title}
  {{Transition from inspiral to plunge in binary black hole coalescences}},}\
  }\href {\doibase 10.1103/PhysRevD.62.064015} {\bibfield  {journal} {\bibinfo
  {journal} {Phys. Rev. D}\ }\textbf {\bibinfo {volume} {62}},\ \bibinfo
  {pages} {064015} (\bibinfo {year} {2000})},\ \Eprint
  {http://arxiv.org/abs/gr-qc/0001013} {arXiv:gr-qc/0001013} \BibitemShut
  {NoStop}%
\bibitem [{\citenamefont {Damour}(2001)}]{Damour:2001tu}%
  \BibitemOpen
  \bibfield  {author} {\bibinfo {author} {\bibfnamefont {Thibault}\
  \bibnamefont {Damour}},\ }\bibfield  {title} {\enquote {\bibinfo {title}
  {{Coalescence of two spinning black holes: an effective one-body
  approach}},}\ }\href {\doibase 10.1103/PhysRevD.64.124013} {\bibfield
  {journal} {\bibinfo  {journal} {Phys. Rev. D}\ }\textbf {\bibinfo {volume}
  {64}},\ \bibinfo {pages} {124013} (\bibinfo {year} {2001})},\ \Eprint
  {http://arxiv.org/abs/gr-qc/0103018} {arXiv:gr-qc/0103018} \BibitemShut
  {NoStop}%
\bibitem [{\citenamefont {Taracchini}\ \emph {et~al.}(2014)\citenamefont
  {Taracchini} \emph {et~al.}}]{Taracchini:2013rva}%
  \BibitemOpen
  \bibfield  {author} {\bibinfo {author} {\bibfnamefont {Andrea}\ \bibnamefont
  {Taracchini}} \emph {et~al.},\ }\bibfield  {title} {\enquote {\bibinfo
  {title} {{Effective-one-body model for black-hole binaries with generic mass
  ratios and spins}},}\ }\href {\doibase 10.1103/PhysRevD.89.061502} {\bibfield
   {journal} {\bibinfo  {journal} {Phys. Rev. D}\ }\textbf {\bibinfo {volume}
  {89}},\ \bibinfo {pages} {061502} (\bibinfo {year} {2014})},\ \Eprint
  {http://arxiv.org/abs/1311.2544} {arXiv:1311.2544 [gr-qc]} \BibitemShut
  {NoStop}%
\bibitem [{\citenamefont {Nagar}\ \emph {et~al.}(2021)\citenamefont {Nagar},
  \citenamefont {Bonino},\ and\ \citenamefont {Rettegno}}]{Nagar:2021gss}%
  \BibitemOpen
  \bibfield  {author} {\bibinfo {author} {\bibfnamefont {Alessandro}\
  \bibnamefont {Nagar}}, \bibinfo {author} {\bibfnamefont {Alice}\ \bibnamefont
  {Bonino}}, \ and\ \bibinfo {author} {\bibfnamefont {Piero}\ \bibnamefont
  {Rettegno}},\ }\bibfield  {title} {\enquote {\bibinfo {title} {{Effective
  one-body multipolar waveform model for spin-aligned, quasicircular,
  eccentric, hyperbolic black hole binaries}},}\ }\href {\doibase
  10.1103/PhysRevD.103.104021} {\bibfield  {journal} {\bibinfo  {journal}
  {Phys. Rev. D}\ }\textbf {\bibinfo {volume} {103}},\ \bibinfo {pages}
  {104021} (\bibinfo {year} {2021})},\ \Eprint
  {http://arxiv.org/abs/2101.08624} {arXiv:2101.08624 [gr-qc]} \BibitemShut
  {NoStop}%
\bibitem [{\citenamefont {Nagar}\ \emph {et~al.}(2023)\citenamefont {Nagar},
  \citenamefont {Rettegno}, \citenamefont {Gamba}, \citenamefont {Albanesi},
  \citenamefont {Albertini},\ and\ \citenamefont {Bernuzzi}}]{Nagar:2023zxh}%
  \BibitemOpen
  \bibfield  {author} {\bibinfo {author} {\bibfnamefont {Alessandro}\
  \bibnamefont {Nagar}}, \bibinfo {author} {\bibfnamefont {Piero}\ \bibnamefont
  {Rettegno}}, \bibinfo {author} {\bibfnamefont {Rossella}\ \bibnamefont
  {Gamba}}, \bibinfo {author} {\bibfnamefont {Simone}\ \bibnamefont
  {Albanesi}}, \bibinfo {author} {\bibfnamefont {Angelica}\ \bibnamefont
  {Albertini}}, \ and\ \bibinfo {author} {\bibfnamefont {Sebastiano}\
  \bibnamefont {Bernuzzi}},\ }\bibfield  {title} {\enquote {\bibinfo {title}
  {{Analytic systematics in next generation of effective-one-body gravitational
  waveform models for future observations}},}\ }\href {\doibase
  10.1103/PhysRevD.108.124018} {\bibfield  {journal} {\bibinfo  {journal}
  {Phys. Rev. D}\ }\textbf {\bibinfo {volume} {108}},\ \bibinfo {pages}
  {124018} (\bibinfo {year} {2023})},\ \Eprint
  {http://arxiv.org/abs/2304.09662} {arXiv:2304.09662 [gr-qc]} \BibitemShut
  {NoStop}%
\bibitem [{\citenamefont {Pompili}\ \emph {et~al.}(2023)\citenamefont {Pompili}
  \emph {et~al.}}]{Pompili:2023tna}%
  \BibitemOpen
  \bibfield  {author} {\bibinfo {author} {\bibfnamefont {Lorenzo}\ \bibnamefont
  {Pompili}} \emph {et~al.},\ }\bibfield  {title} {\enquote {\bibinfo {title}
  {{Laying the foundation of the effective-one-body waveform models SEOBNRv5:
  Improved accuracy and efficiency for spinning nonprecessing binary black
  holes}},}\ }\href {\doibase 10.1103/PhysRevD.108.124035} {\bibfield
  {journal} {\bibinfo  {journal} {Phys. Rev. D}\ }\textbf {\bibinfo {volume}
  {108}},\ \bibinfo {pages} {124035} (\bibinfo {year} {2023})},\ \Eprint
  {http://arxiv.org/abs/2303.18039} {arXiv:2303.18039 [gr-qc]} \BibitemShut
  {NoStop}%
\bibitem [{\citenamefont {Ramos-Buades}\ \emph {et~al.}(2023)\citenamefont
  {Ramos-Buades}, \citenamefont {Buonanno}, \citenamefont {Estell\'es},
  \citenamefont {Khalil}, \citenamefont {Mihaylov}, \citenamefont {Ossokine},
  \citenamefont {Pompili},\ and\ \citenamefont
  {Shiferaw}}]{Ramos-Buades:2023ehm}%
  \BibitemOpen
  \bibfield  {author} {\bibinfo {author} {\bibfnamefont {Antoni}\ \bibnamefont
  {Ramos-Buades}}, \bibinfo {author} {\bibfnamefont {Alessandra}\ \bibnamefont
  {Buonanno}}, \bibinfo {author} {\bibfnamefont {H\'ector}\ \bibnamefont
  {Estell\'es}}, \bibinfo {author} {\bibfnamefont {Mohammed}\ \bibnamefont
  {Khalil}}, \bibinfo {author} {\bibfnamefont {Deyan~P.}\ \bibnamefont
  {Mihaylov}}, \bibinfo {author} {\bibfnamefont {Serguei}\ \bibnamefont
  {Ossokine}}, \bibinfo {author} {\bibfnamefont {Lorenzo}\ \bibnamefont
  {Pompili}}, \ and\ \bibinfo {author} {\bibfnamefont {Mahlet}\ \bibnamefont
  {Shiferaw}},\ }\bibfield  {title} {\enquote {\bibinfo {title} {{Next
  generation of accurate and efficient multipolar precessing-spin
  effective-one-body waveforms for binary black holes}},}\ }\href {\doibase
  10.1103/PhysRevD.108.124037} {\bibfield  {journal} {\bibinfo  {journal}
  {Phys. Rev. D}\ }\textbf {\bibinfo {volume} {108}},\ \bibinfo {pages}
  {124037} (\bibinfo {year} {2023})},\ \Eprint
  {http://arxiv.org/abs/2303.18046} {arXiv:2303.18046 [gr-qc]} \BibitemShut
  {NoStop}%
\bibitem [{\citenamefont {Blackman}\ \emph {et~al.}(2015)\citenamefont
  {Blackman}, \citenamefont {Field}, \citenamefont {Galley}, \citenamefont
  {Szil\'agyi}, \citenamefont {Scheel}, \citenamefont {Tiglio},\ and\
  \citenamefont {Hemberger}}]{Blackman:2015pia}%
  \BibitemOpen
  \bibfield  {author} {\bibinfo {author} {\bibfnamefont {Jonathan}\
  \bibnamefont {Blackman}}, \bibinfo {author} {\bibfnamefont {Scott~E.}\
  \bibnamefont {Field}}, \bibinfo {author} {\bibfnamefont {Chad~R.}\
  \bibnamefont {Galley}}, \bibinfo {author} {\bibfnamefont {B\'ela}\
  \bibnamefont {Szil\'agyi}}, \bibinfo {author} {\bibfnamefont {Mark~A.}\
  \bibnamefont {Scheel}}, \bibinfo {author} {\bibfnamefont {Manuel}\
  \bibnamefont {Tiglio}}, \ and\ \bibinfo {author} {\bibfnamefont {Daniel~A.}\
  \bibnamefont {Hemberger}},\ }\bibfield  {title} {\enquote {\bibinfo {title}
  {{Fast and Accurate Prediction of Numerical Relativity Waveforms from Binary
  Black Hole Coalescences Using Surrogate Models}},}\ }\href {\doibase
  10.1103/PhysRevLett.115.121102} {\bibfield  {journal} {\bibinfo  {journal}
  {Phys. Rev. Lett.}\ }\textbf {\bibinfo {volume} {115}},\ \bibinfo {pages}
  {121102} (\bibinfo {year} {2015})},\ \Eprint
  {http://arxiv.org/abs/1502.07758} {arXiv:1502.07758 [gr-qc]} \BibitemShut
  {NoStop}%
\bibitem [{\citenamefont {Blackman}\ \emph {et~al.}(2017)\citenamefont
  {Blackman}, \citenamefont {Field}, \citenamefont {Scheel}, \citenamefont
  {Galley}, \citenamefont {Hemberger}, \citenamefont {Schmidt},\ and\
  \citenamefont {Smith}}]{Blackman:2017dfb}%
  \BibitemOpen
  \bibfield  {author} {\bibinfo {author} {\bibfnamefont {Jonathan}\
  \bibnamefont {Blackman}}, \bibinfo {author} {\bibfnamefont {Scott~E.}\
  \bibnamefont {Field}}, \bibinfo {author} {\bibfnamefont {Mark~A.}\
  \bibnamefont {Scheel}}, \bibinfo {author} {\bibfnamefont {Chad~R.}\
  \bibnamefont {Galley}}, \bibinfo {author} {\bibfnamefont {Daniel~A.}\
  \bibnamefont {Hemberger}}, \bibinfo {author} {\bibfnamefont {Patricia}\
  \bibnamefont {Schmidt}}, \ and\ \bibinfo {author} {\bibfnamefont {Rory}\
  \bibnamefont {Smith}},\ }\bibfield  {title} {\enquote {\bibinfo {title} {{A
  Surrogate Model of Gravitational Waveforms from Numerical Relativity
  Simulations of Precessing Binary Black Hole Mergers}},}\ }\href {\doibase
  10.1103/PhysRevD.95.104023} {\bibfield  {journal} {\bibinfo  {journal} {Phys.
  Rev. D}\ }\textbf {\bibinfo {volume} {95}},\ \bibinfo {pages} {104023}
  (\bibinfo {year} {2017})},\ \Eprint {http://arxiv.org/abs/1701.00550}
  {arXiv:1701.00550 [gr-qc]} \BibitemShut {NoStop}%
\bibitem [{\citenamefont {Varma}\ \emph
  {et~al.}(2019{\natexlab{a}})\citenamefont {Varma}, \citenamefont {Field},
  \citenamefont {Scheel}, \citenamefont {Blackman}, \citenamefont {Kidder},\
  and\ \citenamefont {Pfeiffer}}]{Varma:2018mmi}%
  \BibitemOpen
  \bibfield  {author} {\bibinfo {author} {\bibfnamefont {Vijay}\ \bibnamefont
  {Varma}}, \bibinfo {author} {\bibfnamefont {Scott~E.}\ \bibnamefont {Field}},
  \bibinfo {author} {\bibfnamefont {Mark~A.}\ \bibnamefont {Scheel}}, \bibinfo
  {author} {\bibfnamefont {Jonathan}\ \bibnamefont {Blackman}}, \bibinfo
  {author} {\bibfnamefont {Lawrence~E.}\ \bibnamefont {Kidder}}, \ and\
  \bibinfo {author} {\bibfnamefont {Harald~P.}\ \bibnamefont {Pfeiffer}},\
  }\bibfield  {title} {\enquote {\bibinfo {title} {{Surrogate model of
  hybridized numerical relativity binary black hole waveforms}},}\ }\href
  {\doibase 10.1103/PhysRevD.99.064045} {\bibfield  {journal} {\bibinfo
  {journal} {Phys. Rev. D}\ }\textbf {\bibinfo {volume} {99}},\ \bibinfo
  {pages} {064045} (\bibinfo {year} {2019}{\natexlab{a}})},\ \Eprint
  {http://arxiv.org/abs/1812.07865} {arXiv:1812.07865 [gr-qc]} \BibitemShut
  {NoStop}%
\bibitem [{\citenamefont {Varma}\ \emph
  {et~al.}(2019{\natexlab{b}})\citenamefont {Varma}, \citenamefont {Field},
  \citenamefont {Scheel}, \citenamefont {Blackman}, \citenamefont {Gerosa},
  \citenamefont {Stein}, \citenamefont {Kidder},\ and\ \citenamefont
  {Pfeiffer}}]{Varma:2019csw}%
  \BibitemOpen
  \bibfield  {author} {\bibinfo {author} {\bibfnamefont {Vijay}\ \bibnamefont
  {Varma}}, \bibinfo {author} {\bibfnamefont {Scott~E.}\ \bibnamefont {Field}},
  \bibinfo {author} {\bibfnamefont {Mark~A.}\ \bibnamefont {Scheel}}, \bibinfo
  {author} {\bibfnamefont {Jonathan}\ \bibnamefont {Blackman}}, \bibinfo
  {author} {\bibfnamefont {Davide}\ \bibnamefont {Gerosa}}, \bibinfo {author}
  {\bibfnamefont {Leo~C.}\ \bibnamefont {Stein}}, \bibinfo {author}
  {\bibfnamefont {Lawrence~E.}\ \bibnamefont {Kidder}}, \ and\ \bibinfo
  {author} {\bibfnamefont {Harald~P.}\ \bibnamefont {Pfeiffer}},\ }\bibfield
  {title} {\enquote {\bibinfo {title} {{Surrogate models for precessing binary
  black hole simulations with unequal masses}},}\ }\href {\doibase
  10.1103/PhysRevResearch.1.033015} {\bibfield  {journal} {\bibinfo  {journal}
  {Phys. Rev. Research.}\ }\textbf {\bibinfo {volume} {1}},\ \bibinfo {pages}
  {033015} (\bibinfo {year} {2019}{\natexlab{b}})},\ \Eprint
  {http://arxiv.org/abs/1905.09300} {arXiv:1905.09300 [gr-qc]} \BibitemShut
  {NoStop}%
\bibitem [{\citenamefont {Pathak}\ \emph {et~al.}(2024)\citenamefont {Pathak},
  \citenamefont {Reza},\ and\ \citenamefont {Sengupta}}]{Pathak:2024zgo}%
  \BibitemOpen
  \bibfield  {author} {\bibinfo {author} {\bibfnamefont {Lalit}\ \bibnamefont
  {Pathak}}, \bibinfo {author} {\bibfnamefont {Amit}\ \bibnamefont {Reza}}, \
  and\ \bibinfo {author} {\bibfnamefont {Anand~S.}\ \bibnamefont {Sengupta}},\
  }\bibfield  {title} {\enquote {\bibinfo {title} {{Fast and faithful
  interpolation of numerical relativity surrogate waveforms using meshfree
  approximation}},}\ }\href@noop {} {\  (\bibinfo {year} {2024})},\ \Eprint
  {http://arxiv.org/abs/2403.19162} {arXiv:2403.19162 [gr-qc]} \BibitemShut
  {NoStop}%
\bibitem [{\citenamefont {Afshordi}\ \emph {et~al.}(2023)\citenamefont
  {Afshordi} \emph {et~al.}}]{LISAConsortiumWaveformWorkingGroup:2023arg}%
  \BibitemOpen
  \bibfield  {author} {\bibinfo {author} {\bibfnamefont {Niayesh}\ \bibnamefont
  {Afshordi}} \emph {et~al.} (\bibinfo {collaboration} {LISA Consortium
  Waveform Working Group}),\ }\bibfield  {title} {\enquote {\bibinfo {title}
  {{Waveform Modelling for the Laser Interferometer Space Antenna}},}\
  }\href@noop {} {\  (\bibinfo {year} {2023})},\ \Eprint
  {http://arxiv.org/abs/2311.01300} {arXiv:2311.01300 [gr-qc]} \BibitemShut
  {NoStop}%
\bibitem [{\citenamefont {Peherstorfer}\ \emph {et~al.}(2018)\citenamefont
  {Peherstorfer}, \citenamefont {Willcox},\ and\ \citenamefont
  {Gunzburger}}]{peherstorfer_survey_2018}%
  \BibitemOpen
  \bibfield  {author} {\bibinfo {author} {\bibfnamefont {Benjamin}\
  \bibnamefont {Peherstorfer}}, \bibinfo {author} {\bibfnamefont {Karen}\
  \bibnamefont {Willcox}}, \ and\ \bibinfo {author} {\bibfnamefont {Max}\
  \bibnamefont {Gunzburger}},\ }\bibfield  {title} {\enquote {\bibinfo {title}
  {Survey of {Multifidelity} {Methods} in {Uncertainty} {Propagation},
  {Inference}, and {Optimization}},}\ }\href {\doibase 10.1137/16M1082469}
  {\bibfield  {journal} {\bibinfo  {journal} {SIAM Review}\ }\textbf {\bibinfo
  {volume} {60}},\ \bibinfo {pages} {550--591} (\bibinfo {year} {2018})},\
  \bibinfo {note} {publisher: Society for Industrial and Applied
  Mathematics}\BibitemShut {NoStop}%
\bibitem [{\citenamefont {Calder\'on~Bustillo}\ \emph
  {et~al.}(2017)\citenamefont {Calder\'on~Bustillo}, \citenamefont {Laguna},\
  and\ \citenamefont {Shoemaker}}]{CalderonBustillo:2016rlt}%
  \BibitemOpen
  \bibfield  {author} {\bibinfo {author} {\bibfnamefont {Juan}\ \bibnamefont
  {Calder\'on~Bustillo}}, \bibinfo {author} {\bibfnamefont {Pablo}\
  \bibnamefont {Laguna}}, \ and\ \bibinfo {author} {\bibfnamefont {Deirdre}\
  \bibnamefont {Shoemaker}},\ }\bibfield  {title} {\enquote {\bibinfo {title}
  {{Detectability of gravitational waves from binary black holes: Impact of
  precession and higher modes}},}\ }\href {\doibase 10.1103/PhysRevD.95.104038}
  {\bibfield  {journal} {\bibinfo  {journal} {Phys. Rev. D}\ }\textbf {\bibinfo
  {volume} {95}},\ \bibinfo {pages} {104038} (\bibinfo {year} {2017})},\
  \Eprint {http://arxiv.org/abs/1612.02340} {arXiv:1612.02340 [gr-qc]}
  \BibitemShut {NoStop}%
\bibitem [{\citenamefont {Lange}\ \emph {et~al.}(2018)\citenamefont {Lange},
  \citenamefont {O'Shaughnessy},\ and\ \citenamefont {Rizzo}}]{Lange:2018pyp}%
  \BibitemOpen
  \bibfield  {author} {\bibinfo {author} {\bibfnamefont {Jacob}\ \bibnamefont
  {Lange}}, \bibinfo {author} {\bibfnamefont {Richard}\ \bibnamefont
  {O'Shaughnessy}}, \ and\ \bibinfo {author} {\bibfnamefont {Monica}\
  \bibnamefont {Rizzo}},\ }\bibfield  {title} {\enquote {\bibinfo {title}
  {{Rapid and accurate parameter inference for coalescing, precessing compact
  binaries}},}\ }\href@noop {} {\  (\bibinfo {year} {2018})},\ \Eprint
  {http://arxiv.org/abs/1805.10457} {arXiv:1805.10457 [gr-qc]} \BibitemShut
  {NoStop}%
\bibitem [{\citenamefont {Kumar}\ \emph {et~al.}(2019)\citenamefont {Kumar},
  \citenamefont {Blackman}, \citenamefont {Field}, \citenamefont {Scheel},
  \citenamefont {Galley}, \citenamefont {Boyle}, \citenamefont {Kidder},
  \citenamefont {Pfeiffer}, \citenamefont {Szilagyi},\ and\ \citenamefont
  {Teukolsky}}]{Kumar:2018hml}%
  \BibitemOpen
  \bibfield  {author} {\bibinfo {author} {\bibfnamefont {Prayush}\ \bibnamefont
  {Kumar}}, \bibinfo {author} {\bibfnamefont {Jonathan}\ \bibnamefont
  {Blackman}}, \bibinfo {author} {\bibfnamefont {Scott~E.}\ \bibnamefont
  {Field}}, \bibinfo {author} {\bibfnamefont {Mark}\ \bibnamefont {Scheel}},
  \bibinfo {author} {\bibfnamefont {Chad~R.}\ \bibnamefont {Galley}}, \bibinfo
  {author} {\bibfnamefont {Michael}\ \bibnamefont {Boyle}}, \bibinfo {author}
  {\bibfnamefont {Lawrence~E.}\ \bibnamefont {Kidder}}, \bibinfo {author}
  {\bibfnamefont {Harald~P.}\ \bibnamefont {Pfeiffer}}, \bibinfo {author}
  {\bibfnamefont {Bela}\ \bibnamefont {Szilagyi}}, \ and\ \bibinfo {author}
  {\bibfnamefont {Saul~A.}\ \bibnamefont {Teukolsky}},\ }\bibfield  {title}
  {\enquote {\bibinfo {title} {{Constraining the parameters of GW150914 and
  GW170104 with numerical relativity surrogates}},}\ }\href {\doibase
  10.1103/PhysRevD.99.124005} {\bibfield  {journal} {\bibinfo  {journal} {Phys.
  Rev. D}\ }\textbf {\bibinfo {volume} {99}},\ \bibinfo {pages} {124005}
  (\bibinfo {year} {2019})},\ \Eprint {http://arxiv.org/abs/1808.08004}
  {arXiv:1808.08004 [gr-qc]} \BibitemShut {NoStop}%
\bibitem [{\citenamefont {Huang}\ \emph {et~al.}(2021)\citenamefont {Huang},
  \citenamefont {Haster}, \citenamefont {Vitale}, \citenamefont {Varma},
  \citenamefont {Foucart},\ and\ \citenamefont {Biscoveanu}}]{Huang:2020pba}%
  \BibitemOpen
  \bibfield  {author} {\bibinfo {author} {\bibfnamefont {Yiwen}\ \bibnamefont
  {Huang}}, \bibinfo {author} {\bibfnamefont {Carl-Johan}\ \bibnamefont
  {Haster}}, \bibinfo {author} {\bibfnamefont {Salvatore}\ \bibnamefont
  {Vitale}}, \bibinfo {author} {\bibfnamefont {Vijay}\ \bibnamefont {Varma}},
  \bibinfo {author} {\bibfnamefont {Francois}\ \bibnamefont {Foucart}}, \ and\
  \bibinfo {author} {\bibfnamefont {Sylvia}\ \bibnamefont {Biscoveanu}},\
  }\bibfield  {title} {\enquote {\bibinfo {title} {{Statistical and systematic
  uncertainties in extracting the source properties of neutron star - black
  hole binaries with gravitational waves}},}\ }\href {\doibase
  10.1103/PhysRevD.103.083001} {\bibfield  {journal} {\bibinfo  {journal}
  {Phys. Rev. D}\ }\textbf {\bibinfo {volume} {103}},\ \bibinfo {pages}
  {083001} (\bibinfo {year} {2021})},\ \Eprint
  {http://arxiv.org/abs/2005.11850} {arXiv:2005.11850 [gr-qc]} \BibitemShut
  {NoStop}%
\bibitem [{\citenamefont {Peherstorfer}\ \emph {et~al.}(2016)\citenamefont
  {Peherstorfer}, \citenamefont {Cui}, \citenamefont {Marzouk},\ and\
  \citenamefont {Willcox}}]{peherstorfer2016multifidelity}%
  \BibitemOpen
  \bibfield  {author} {\bibinfo {author} {\bibfnamefont {Benjamin}\
  \bibnamefont {Peherstorfer}}, \bibinfo {author} {\bibfnamefont {Tiangang}\
  \bibnamefont {Cui}}, \bibinfo {author} {\bibfnamefont {Youssef}\ \bibnamefont
  {Marzouk}}, \ and\ \bibinfo {author} {\bibfnamefont {Karen}\ \bibnamefont
  {Willcox}},\ }\bibfield  {title} {\enquote {\bibinfo {title} {Multifidelity
  importance sampling},}\ }\href@noop {} {\bibfield  {journal} {\bibinfo
  {journal} {Computer Methods in Applied Mechanics and Engineering}\ }\textbf
  {\bibinfo {volume} {300}},\ \bibinfo {pages} {490--509} (\bibinfo {year}
  {2016})}\BibitemShut {NoStop}%
\bibitem [{\citenamefont {Payne}\ \emph {et~al.}(2019)\citenamefont {Payne},
  \citenamefont {Talbot},\ and\ \citenamefont {Thrane}}]{Payne:2019wmy}%
  \BibitemOpen
  \bibfield  {author} {\bibinfo {author} {\bibfnamefont {Ethan}\ \bibnamefont
  {Payne}}, \bibinfo {author} {\bibfnamefont {Colm}\ \bibnamefont {Talbot}}, \
  and\ \bibinfo {author} {\bibfnamefont {Eric}\ \bibnamefont {Thrane}},\
  }\bibfield  {title} {\enquote {\bibinfo {title} {{Higher order
  gravitational-wave modes with likelihood reweighting}},}\ }\href {\doibase
  10.1103/PhysRevD.100.123017} {\bibfield  {journal} {\bibinfo  {journal}
  {Phys. Rev. D}\ }\textbf {\bibinfo {volume} {100}},\ \bibinfo {pages}
  {123017} (\bibinfo {year} {2019})},\ \Eprint
  {http://arxiv.org/abs/1905.05477} {arXiv:1905.05477 [astro-ph.IM]}
  \BibitemShut {NoStop}%
\bibitem [{\citenamefont {Payne}\ \emph {et~al.}(2020)\citenamefont {Payne},
  \citenamefont {Talbot}, \citenamefont {Lasky}, \citenamefont {Thrane},\ and\
  \citenamefont {Kissel}}]{Payne:2020myg}%
  \BibitemOpen
  \bibfield  {author} {\bibinfo {author} {\bibfnamefont {Ethan}\ \bibnamefont
  {Payne}}, \bibinfo {author} {\bibfnamefont {Colm}\ \bibnamefont {Talbot}},
  \bibinfo {author} {\bibfnamefont {Paul~D.}\ \bibnamefont {Lasky}}, \bibinfo
  {author} {\bibfnamefont {Eric}\ \bibnamefont {Thrane}}, \ and\ \bibinfo
  {author} {\bibfnamefont {Jeffrey~S.}\ \bibnamefont {Kissel}},\ }\bibfield
  {title} {\enquote {\bibinfo {title} {{Gravitational-wave astronomy with a
  physical calibration model}},}\ }\href {\doibase 10.1103/PhysRevD.102.122004}
  {\bibfield  {journal} {\bibinfo  {journal} {Phys. Rev. D}\ }\textbf {\bibinfo
  {volume} {102}},\ \bibinfo {pages} {122004} (\bibinfo {year} {2020})},\
  \Eprint {http://arxiv.org/abs/2009.10193} {arXiv:2009.10193 [astro-ph.IM]}
  \BibitemShut {NoStop}%
\bibitem [{\citenamefont {Dax}\ \emph {et~al.}(2023)\citenamefont {Dax},
  \citenamefont {Green}, \citenamefont {Gair}, \citenamefont {P\"urrer},
  \citenamefont {Wildberger}, \citenamefont {Macke}, \citenamefont {Buonanno},\
  and\ \citenamefont {Sch\"olkopf}}]{Dax:2022pxd}%
  \BibitemOpen
  \bibfield  {author} {\bibinfo {author} {\bibfnamefont {Maximilian}\
  \bibnamefont {Dax}}, \bibinfo {author} {\bibfnamefont {Stephen~R.}\
  \bibnamefont {Green}}, \bibinfo {author} {\bibfnamefont {Jonathan}\
  \bibnamefont {Gair}}, \bibinfo {author} {\bibfnamefont {Michael}\
  \bibnamefont {P\"urrer}}, \bibinfo {author} {\bibfnamefont {Jonas}\
  \bibnamefont {Wildberger}}, \bibinfo {author} {\bibfnamefont {Jakob~H.}\
  \bibnamefont {Macke}}, \bibinfo {author} {\bibfnamefont {Alessandra}\
  \bibnamefont {Buonanno}}, \ and\ \bibinfo {author} {\bibfnamefont {Bernhard}\
  \bibnamefont {Sch\"olkopf}},\ }\bibfield  {title} {\enquote {\bibinfo {title}
  {{Neural Importance Sampling for Rapid and Reliable Gravitational-Wave
  Inference}},}\ }\href {\doibase 10.1103/PhysRevLett.130.171403} {\bibfield
  {journal} {\bibinfo  {journal} {Phys. Rev. Lett.}\ }\textbf {\bibinfo
  {volume} {130}},\ \bibinfo {pages} {171403} (\bibinfo {year} {2023})},\
  \Eprint {http://arxiv.org/abs/2210.05686} {arXiv:2210.05686 [gr-qc]}
  \BibitemShut {NoStop}%
\bibitem [{\citenamefont {Agapiou}\ \emph {et~al.}(2017)\citenamefont
  {Agapiou}, \citenamefont {Papaspiliopoulos}, \citenamefont {Sanz-Alonso},\
  and\ \citenamefont {Stuart}}]{agapiou2017importance}%
  \BibitemOpen
  \bibfield  {author} {\bibinfo {author} {\bibfnamefont {Sergios}\ \bibnamefont
  {Agapiou}}, \bibinfo {author} {\bibfnamefont {Omiros}\ \bibnamefont
  {Papaspiliopoulos}}, \bibinfo {author} {\bibfnamefont {Daniel}\ \bibnamefont
  {Sanz-Alonso}}, \ and\ \bibinfo {author} {\bibfnamefont {Andrew~M}\
  \bibnamefont {Stuart}},\ }\bibfield  {title} {\enquote {\bibinfo {title}
  {Importance sampling: Intrinsic dimension and computational cost},}\
  }\href@noop {} {\bibfield  {journal} {\bibinfo  {journal} {Statistical
  Science}\ ,\ \bibinfo {pages} {405--431}} (\bibinfo {year}
  {2017})}\BibitemShut {NoStop}%
\bibitem [{\citenamefont {Alsup}\ and\ \citenamefont
  {Peherstorfer}(2021)}]{alsup_context-aware_2021}%
  \BibitemOpen
  \bibfield  {author} {\bibinfo {author} {\bibfnamefont {Terrence}\
  \bibnamefont {Alsup}}\ and\ \bibinfo {author} {\bibfnamefont {Benjamin}\
  \bibnamefont {Peherstorfer}},\ }\bibfield  {title} {\enquote {\bibinfo
  {title} {Context-aware surrogate modeling for balancing approximation and
  sampling costs in multi-fidelity importance sampling and {Bayesian} inverse
  problems},}\ }\href {http://arxiv.org/abs/2010.11708} {\bibfield  {journal}
  {\bibinfo  {journal} {arXiv:2010.11708 [cs, math, stat]}\ } (\bibinfo {year}
  {2021})},\ \bibinfo {note} {arXiv: 2010.11708}\BibitemShut {NoStop}%
\bibitem [{\citenamefont {Owen}(2013)}]{mcbook}%
  \BibitemOpen
  \bibfield  {author} {\bibinfo {author} {\bibfnamefont {Art~B.}\ \bibnamefont
  {Owen}},\ }\href@noop {} {\emph {\bibinfo {title} {Monte Carlo theory,
  methods and examples}}}\ (\bibinfo  {publisher}
  {\url{https://artowen.su.domains/mc/}},\ \bibinfo {year} {2013})\BibitemShut
  {NoStop}%
\bibitem [{\citenamefont {Swendsen}\ and\ \citenamefont
  {Wang}(1986)}]{Swendsen:1986vqb}%
  \BibitemOpen
  \bibfield  {author} {\bibinfo {author} {\bibfnamefont {Robert~H.}\
  \bibnamefont {Swendsen}}\ and\ \bibinfo {author} {\bibfnamefont {Jian-Sheng}\
  \bibnamefont {Wang}},\ }\bibfield  {title} {\enquote {\bibinfo {title}
  {{Replica Monte Carlo Simulation of Spin-Glasses}},}\ }\href {\doibase
  10.1103/PhysRevLett.57.2607} {\bibfield  {journal} {\bibinfo  {journal}
  {Phys. Rev. Lett.}\ }\textbf {\bibinfo {volume} {57}},\ \bibinfo {pages}
  {2607} (\bibinfo {year} {1986})}\BibitemShut {NoStop}%
\bibitem [{\citenamefont {Geyer}(1991)}]{Geyer:1991}%
  \BibitemOpen
  \bibfield  {author} {\bibinfo {author} {\bibfnamefont {Charles~J.}\
  \bibnamefont {Geyer}},\ }\bibfield  {title} {\enquote {\bibinfo {title}
  {Markov chain monte carlo maximum likelihood},}\ }in\ \href@noop {} {\emph
  {\bibinfo {booktitle} {Proc. 23rd Symp. Interface, Computing Science and
  Statistics}}},\ \bibinfo {editor} {edited by\ \bibinfo {editor}
  {\bibfnamefont {MK}~\bibnamefont {Elaine}}\ and\ \bibinfo {editor}
  {\bibfnamefont {MK}~\bibnamefont {Selma}}}\ (\bibinfo  {publisher} {Interface
  Foundation of North America, New York},\ \bibinfo {year} {1991})\BibitemShut
  {NoStop}%
\bibitem [{\citenamefont {Earl}\ and\ \citenamefont {Deem}(2005)}]{Earl:2005}%
  \BibitemOpen
  \bibfield  {author} {\bibinfo {author} {\bibfnamefont {David~J.}\
  \bibnamefont {Earl}}\ and\ \bibinfo {author} {\bibfnamefont {Michael~W.}\
  \bibnamefont {Deem}},\ }\bibfield  {title} {\enquote {\bibinfo {title}
  {Parallel tempering: Theory{,} applications{,} and new perspectives},}\
  }\href {\doibase 10.1039/B509983H} {\bibfield  {journal} {\bibinfo  {journal}
  {Phys. Chem. Chem. Phys.}\ }\textbf {\bibinfo {volume} {7}},\ \bibinfo
  {pages} {3910--3916} (\bibinfo {year} {2005})}\BibitemShut {NoStop}%
\bibitem [{\citenamefont {Vousden}\ \emph {et~al.}(2015)\citenamefont
  {Vousden}, \citenamefont {Farr},\ and\ \citenamefont
  {Mandel}}]{Vousden:2015}%
  \BibitemOpen
  \bibfield  {author} {\bibinfo {author} {\bibfnamefont {W.~D.}\ \bibnamefont
  {Vousden}}, \bibinfo {author} {\bibfnamefont {W.~M.}\ \bibnamefont {Farr}}, \
  and\ \bibinfo {author} {\bibfnamefont {I.}~\bibnamefont {Mandel}},\
  }\bibfield  {title} {\enquote {\bibinfo {title} {{Dynamic temperature
  selection for parallel tempering in Markov chain Monte Carlo simulations}},}\
  }\href {\doibase 10.1093/mnras/stv2422} {\bibfield  {journal} {\bibinfo
  {journal} {Monthly Notices of the Royal Astronomical Society}\ }\textbf
  {\bibinfo {volume} {455}},\ \bibinfo {pages} {1919--1937} (\bibinfo {year}
  {2015})},\ \Eprint
  {http://arxiv.org/abs/https://academic.oup.com/mnras/article-pdf/455/2/1919/18514064/stv2422.pdf}
  {https://academic.oup.com/mnras/article-pdf/455/2/1919/18514064/stv2422.pdf}
  \BibitemShut {NoStop}%
\bibitem [{\citenamefont {Abbott}\ \emph
  {et~al.}(2023{\natexlab{b}})\citenamefont {Abbott} \emph
  {et~al.}}]{KAGRA:2023pio}%
  \BibitemOpen
  \bibfield  {author} {\bibinfo {author} {\bibfnamefont {R.}~\bibnamefont
  {Abbott}} \emph {et~al.} (\bibinfo {collaboration} {KAGRA, VIRGO, LIGO
  Scientific}),\ }\bibfield  {title} {\enquote {\bibinfo {title} {{Open Data
  from the Third Observing Run of LIGO, Virgo, KAGRA, and GEO}},}\ }\href
  {\doibase 10.3847/1538-4365/acdc9f} {\bibfield  {journal} {\bibinfo
  {journal} {Astrophys. J. Suppl.}\ }\textbf {\bibinfo {volume} {267}},\
  \bibinfo {pages} {29} (\bibinfo {year} {2023}{\natexlab{b}})},\ \Eprint
  {http://arxiv.org/abs/2302.03676} {arXiv:2302.03676 [gr-qc]} \BibitemShut
  {NoStop}%
\bibitem [{\citenamefont {{{LIGO}, {Virgo,} and {KAGRA} Scientific
  Collaborations}}()}]{GWOSC}%
  \BibitemOpen
  \bibfield  {author} {\bibinfo {author} {\bibnamefont {{{LIGO}, {Virgo,} and
  {KAGRA} Scientific Collaborations}}},\ }\href@noop {} {\enquote {\bibinfo
  {title} {{Gravitational Wave Open Science Center}},}\ }\bibinfo
  {howpublished} {\url{https://www.gw-openscience.org/}}\BibitemShut {NoStop}%
\bibitem [{\citenamefont {{Higson}}\ \emph {et~al.}(2019)\citenamefont
  {{Higson}}, \citenamefont {{Handley}}, \citenamefont {{Hobson}},\ and\
  \citenamefont {{Lasenby}}}]{Higson2019}%
  \BibitemOpen
  \bibfield  {author} {\bibinfo {author} {\bibfnamefont {Edward}\ \bibnamefont
  {{Higson}}}, \bibinfo {author} {\bibfnamefont {Will}\ \bibnamefont
  {{Handley}}}, \bibinfo {author} {\bibfnamefont {Mike}\ \bibnamefont
  {{Hobson}}}, \ and\ \bibinfo {author} {\bibfnamefont {Anthony}\ \bibnamefont
  {{Lasenby}}},\ }\bibfield  {title} {\enquote {\bibinfo {title} {{Dynamic
  nested sampling: an improved algorithm for parameter estimation and evidence
  calculation}},}\ }\href {\doibase 10.1007/s11222-018-9844-0} {\bibfield
  {journal} {\bibinfo  {journal} {Statistics and Computing}\ }\textbf {\bibinfo
  {volume} {29}},\ \bibinfo {pages} {891--913} (\bibinfo {year} {2019})},\
  \Eprint {http://arxiv.org/abs/1704.03459} {arXiv:1704.03459 [stat.CO]}
  \BibitemShut {NoStop}%
\bibitem [{\citenamefont {Speagle}(2020)}]{Speagle:2019ivv}%
  \BibitemOpen
  \bibfield  {author} {\bibinfo {author} {\bibfnamefont {Joshua~S.}\
  \bibnamefont {Speagle}},\ }\bibfield  {title} {\enquote {\bibinfo {title}
  {{dynesty: a dynamic nested sampling package for estimating Bayesian
  posteriors and evidences}},}\ }\href {\doibase 10.1093/mnras/staa278}
  {\bibfield  {journal} {\bibinfo  {journal} {Mon. Not. Roy. Astron. Soc.}\
  }\textbf {\bibinfo {volume} {493}},\ \bibinfo {pages} {3132--3158} (\bibinfo
  {year} {2020})},\ \Eprint {http://arxiv.org/abs/1904.02180} {arXiv:1904.02180
  [astro-ph.IM]} \BibitemShut {NoStop}%
\bibitem [{\citenamefont {{Skilling}}(2004)}]{Skilling2004}%
  \BibitemOpen
  \bibfield  {author} {\bibinfo {author} {\bibfnamefont {John}\ \bibnamefont
  {{Skilling}}},\ }\bibfield  {title} {\enquote {\bibinfo {title} {{Nested
  Sampling}},}\ }in\ \href {\doibase 10.1063/1.1835238} {\emph {\bibinfo
  {booktitle} {Bayesian Inference and Maximum Entropy Methods in Science and
  Engineering: 24th International Workshop on Bayesian Inference and Maximum
  Entropy Methods in Science and Engineering}}},\ \bibinfo {series} {American
  Institute of Physics Conference Series}, Vol.\ \bibinfo {volume} {735},\
  \bibinfo {editor} {edited by\ \bibinfo {editor} {\bibfnamefont {Rainer}\
  \bibnamefont {{Fischer}}}, \bibinfo {editor} {\bibfnamefont {Roland}\
  \bibnamefont {{Preuss}}}, \ and\ \bibinfo {editor} {\bibfnamefont {Udo~Von}\
  \bibnamefont {{Toussaint}}}}\ (\bibinfo  {publisher} {AIP},\ \bibinfo {year}
  {2004})\ pp.\ \bibinfo {pages} {395--405}\BibitemShut {NoStop}%
\bibitem [{\citenamefont {Skilling}(2006)}]{Skilling2006}%
  \BibitemOpen
  \bibfield  {author} {\bibinfo {author} {\bibfnamefont {John}\ \bibnamefont
  {Skilling}},\ }\bibfield  {title} {\enquote {\bibinfo {title} {{Nested
  sampling for general Bayesian computation}},}\ }\href {\doibase
  10.1214/06-BA127} {\bibfield  {journal} {\bibinfo  {journal} {Bayesian
  Analysis}\ }\textbf {\bibinfo {volume} {1}},\ \bibinfo {pages} {833 -- 859}
  (\bibinfo {year} {2006})}\BibitemShut {NoStop}%
\bibitem [{\citenamefont {Hoy}\ and\ \citenamefont
  {Raymond}(2021)}]{Hoy:2020vys}%
  \BibitemOpen
  \bibfield  {author} {\bibinfo {author} {\bibfnamefont {Charlie}\ \bibnamefont
  {Hoy}}\ and\ \bibinfo {author} {\bibfnamefont {Vivien}\ \bibnamefont
  {Raymond}},\ }\bibfield  {title} {\enquote {\bibinfo {title} {{PESummary: the
  code agnostic Parameter Estimation Summary page builder}},}\ }\href {\doibase
  10.1016/j.softx.2021.100765} {\bibfield  {journal} {\bibinfo  {journal}
  {SoftwareX}\ }\textbf {\bibinfo {volume} {15}},\ \bibinfo {pages} {100765}
  (\bibinfo {year} {2021})},\ \Eprint {http://arxiv.org/abs/2006.06639}
  {arXiv:2006.06639 [astro-ph.IM]} \BibitemShut {NoStop}%
\bibitem [{\citenamefont {Ng}\ \emph {et~al.}(2018)\citenamefont {Ng},
  \citenamefont {Vitale}, \citenamefont {Zimmerman}, \citenamefont
  {Chatziioannou}, \citenamefont {Gerosa},\ and\ \citenamefont
  {Haster}}]{Ng:2018neg}%
  \BibitemOpen
  \bibfield  {author} {\bibinfo {author} {\bibfnamefont {Ken K.~Y.}\
  \bibnamefont {Ng}}, \bibinfo {author} {\bibfnamefont {Salvatore}\
  \bibnamefont {Vitale}}, \bibinfo {author} {\bibfnamefont {Aaron}\
  \bibnamefont {Zimmerman}}, \bibinfo {author} {\bibfnamefont {Katerina}\
  \bibnamefont {Chatziioannou}}, \bibinfo {author} {\bibfnamefont {Davide}\
  \bibnamefont {Gerosa}}, \ and\ \bibinfo {author} {\bibfnamefont {Carl-Johan}\
  \bibnamefont {Haster}},\ }\bibfield  {title} {\enquote {\bibinfo {title}
  {{Gravitational-wave astrophysics with effective-spin measurements:
  asymmetries and selection biases}},}\ }\href {\doibase
  10.1103/PhysRevD.98.083007} {\bibfield  {journal} {\bibinfo  {journal} {Phys.
  Rev. D}\ }\textbf {\bibinfo {volume} {98}},\ \bibinfo {pages} {083007}
  (\bibinfo {year} {2018})},\ \Eprint {http://arxiv.org/abs/1805.03046}
  {arXiv:1805.03046 [gr-qc]} \BibitemShut {NoStop}%
\bibitem [{ALI(2015)}]{ALIGODesignCurve}%
  \BibitemOpen
  \href {https://dcc.ligo.org/LIGO-T0900288/public} {\enquote {\bibinfo {title}
  {Advanced ligo anticipated sensitivity curves},}\ }\bibinfo {howpublished}
  {https://dcc.ligo.org/LIGO-T0900288/public} (\bibinfo {year}
  {2015})\BibitemShut {NoStop}%
\bibitem [{\citenamefont {Collaboration}\ \emph {et~al.}(2021)\citenamefont
  {Collaboration}, \citenamefont {Collaboration},\ and\ \citenamefont
  {Collaboration}}]{ligo_scientific_collaboration_and_virgo_2021_5546663}%
  \BibitemOpen
  \bibfield  {author} {\bibinfo {author} {\bibfnamefont {LIGO~Scientific}\
  \bibnamefont {Collaboration}}, \bibinfo {author} {\bibfnamefont {Virgo}\
  \bibnamefont {Collaboration}}, \ and\ \bibinfo {author} {\bibfnamefont
  {KAGRA}\ \bibnamefont {Collaboration}},\ }\href {\doibase
  10.5281/zenodo.5546663} {\enquote {\bibinfo {title} {{GWTC-3: Compact Binary
  Coalescences Observed by LIGO and Virgo During the Second Part of the Third
  Observing Run — Parameter estimation data release}},}\ } (\bibinfo {year}
  {2021})\BibitemShut {NoStop}%
\bibitem [{\citenamefont {Abbott}\ \emph
  {et~al.}(2016{\natexlab{d}})\citenamefont {Abbott} \emph
  {et~al.}}]{LIGOScientific:2016vlm}%
  \BibitemOpen
  \bibfield  {author} {\bibinfo {author} {\bibfnamefont {B.~P.}\ \bibnamefont
  {Abbott}} \emph {et~al.} (\bibinfo {collaboration} {LIGO Scientific,
  Virgo}),\ }\bibfield  {title} {\enquote {\bibinfo {title} {{Properties of the
  Binary Black Hole Merger GW150914}},}\ }\href {\doibase
  10.1103/PhysRevLett.116.241102} {\bibfield  {journal} {\bibinfo  {journal}
  {Phys. Rev. Lett.}\ }\textbf {\bibinfo {volume} {116}},\ \bibinfo {pages}
  {241102} (\bibinfo {year} {2016}{\natexlab{d}})},\ \Eprint
  {http://arxiv.org/abs/1602.03840} {arXiv:1602.03840 [gr-qc]} \BibitemShut
  {NoStop}%
\bibitem [{\citenamefont {Farr}\ \emph {et~al.}(2014)\citenamefont {Farr},
  \citenamefont {Farr},\ and\ \citenamefont {Littenberg}}]{FarrCalMarg2014}%
  \BibitemOpen
  \bibfield  {author} {\bibinfo {author} {\bibfnamefont {Will~M.}\ \bibnamefont
  {Farr}}, \bibinfo {author} {\bibfnamefont {Benjamin}\ \bibnamefont {Farr}}, \
  and\ \bibinfo {author} {\bibfnamefont {Tyson}\ \bibnamefont {Littenberg}},\
  }\href {https://dcc.ligo.org/LIGO-T1400682/public} {\enquote {\bibinfo
  {title} {Modelling calibration errors in cbc waveforms},}\ }\bibinfo
  {howpublished} {LIGO-T1400682} (\bibinfo {year} {2014})\BibitemShut {NoStop}%
\bibitem [{\citenamefont {Efron}(1982)}]{efron1982jackknife}%
  \BibitemOpen
  \bibfield  {author} {\bibinfo {author} {\bibfnamefont {Bradley}\ \bibnamefont
  {Efron}},\ }\href@noop {} {\emph {\bibinfo {title} {{The jackknife, the
  Bootstrap and Other Resampling Plans}}}}\ (\bibinfo  {publisher} {SIAM},\
  \bibinfo {year} {1982})\ \bibinfo {note} {{CBMS-NSF Regional Conference
  Series in Applied Mathematics}}\BibitemShut {NoStop}%
\bibitem [{\citenamefont {Hogg}\ \emph {et~al.}(2010)\citenamefont {Hogg},
  \citenamefont {Bovy},\ and\ \citenamefont {Lang}}]{Hogg:2010yz}%
  \BibitemOpen
  \bibfield  {author} {\bibinfo {author} {\bibfnamefont {David~W.}\
  \bibnamefont {Hogg}}, \bibinfo {author} {\bibfnamefont {Jo}~\bibnamefont
  {Bovy}}, \ and\ \bibinfo {author} {\bibfnamefont {Dustin}\ \bibnamefont
  {Lang}},\ }\bibfield  {title} {\enquote {\bibinfo {title} {{Data analysis
  recipes: Fitting a model to data}},}\ }\href@noop {} {\  (\bibinfo {year}
  {2010})},\ \Eprint {http://arxiv.org/abs/1008.4686} {arXiv:1008.4686
  [astro-ph.IM]} \BibitemShut {NoStop}%
\bibitem [{\citenamefont {Hannam}\ \emph {et~al.}(2022)\citenamefont {Hannam}
  \emph {et~al.}}]{Hannam:2021pit}%
  \BibitemOpen
  \bibfield  {author} {\bibinfo {author} {\bibfnamefont {Mark}\ \bibnamefont
  {Hannam}} \emph {et~al.},\ }\bibfield  {title} {\enquote {\bibinfo {title}
  {{General-relativistic precession in a black-hole binary}},}\ }\href
  {\doibase 10.1038/s41586-022-05212-z} {\bibfield  {journal} {\bibinfo
  {journal} {Nature}\ }\textbf {\bibinfo {volume} {610}},\ \bibinfo {pages}
  {652--655} (\bibinfo {year} {2022})},\ \Eprint
  {http://arxiv.org/abs/2112.11300} {arXiv:2112.11300 [gr-qc]} \BibitemShut
  {NoStop}%
\bibitem [{\citenamefont {Varma}\ \emph {et~al.}(2022)\citenamefont {Varma},
  \citenamefont {Biscoveanu}, \citenamefont {Islam}, \citenamefont {Shaik},
  \citenamefont {Haster}, \citenamefont {Isi}, \citenamefont {Farr},
  \citenamefont {Field},\ and\ \citenamefont {Vitale}}]{Varma:2022pld}%
  \BibitemOpen
  \bibfield  {author} {\bibinfo {author} {\bibfnamefont {Vijay}\ \bibnamefont
  {Varma}}, \bibinfo {author} {\bibfnamefont {Sylvia}\ \bibnamefont
  {Biscoveanu}}, \bibinfo {author} {\bibfnamefont {Tousif}\ \bibnamefont
  {Islam}}, \bibinfo {author} {\bibfnamefont {Feroz~H.}\ \bibnamefont {Shaik}},
  \bibinfo {author} {\bibfnamefont {Carl-Johan}\ \bibnamefont {Haster}},
  \bibinfo {author} {\bibfnamefont {Maximiliano}\ \bibnamefont {Isi}}, \bibinfo
  {author} {\bibfnamefont {Will~M.}\ \bibnamefont {Farr}}, \bibinfo {author}
  {\bibfnamefont {Scott~E.}\ \bibnamefont {Field}}, \ and\ \bibinfo {author}
  {\bibfnamefont {Salvatore}\ \bibnamefont {Vitale}},\ }\bibfield  {title}
  {\enquote {\bibinfo {title} {{Evidence of Large Recoil Velocity from a Black
  Hole Merger Signal}},}\ }\href {\doibase 10.1103/PhysRevLett.128.191102}
  {\bibfield  {journal} {\bibinfo  {journal} {Phys. Rev. Lett.}\ }\textbf
  {\bibinfo {volume} {128}},\ \bibinfo {pages} {191102} (\bibinfo {year}
  {2022})},\ \Eprint {http://arxiv.org/abs/2201.01302} {arXiv:2201.01302
  [astro-ph.HE]} \BibitemShut {NoStop}%
\bibitem [{\citenamefont {Islam}\ \emph {et~al.}(2023)\citenamefont {Islam},
  \citenamefont {Vajpeyi}, \citenamefont {Shaik}, \citenamefont {Haster},
  \citenamefont {Varma}, \citenamefont {Field}, \citenamefont {Lange},
  \citenamefont {O'Shaughnessy},\ and\ \citenamefont {Smith}}]{Islam:2023zzj}%
  \BibitemOpen
  \bibfield  {author} {\bibinfo {author} {\bibfnamefont {Tousif}\ \bibnamefont
  {Islam}}, \bibinfo {author} {\bibfnamefont {Avi}\ \bibnamefont {Vajpeyi}},
  \bibinfo {author} {\bibfnamefont {Feroz~H.}\ \bibnamefont {Shaik}}, \bibinfo
  {author} {\bibfnamefont {Carl-Johan}\ \bibnamefont {Haster}}, \bibinfo
  {author} {\bibfnamefont {Vijay}\ \bibnamefont {Varma}}, \bibinfo {author}
  {\bibfnamefont {Scott~E.}\ \bibnamefont {Field}}, \bibinfo {author}
  {\bibfnamefont {Jacob}\ \bibnamefont {Lange}}, \bibinfo {author}
  {\bibfnamefont {Richard}\ \bibnamefont {O'Shaughnessy}}, \ and\ \bibinfo
  {author} {\bibfnamefont {Rory}\ \bibnamefont {Smith}},\ }\bibfield  {title}
  {\enquote {\bibinfo {title} {{Analysis of GWTC-3 with fully precessing
  numerical relativity surrogate models}},}\ }\href@noop {} {\  (\bibinfo
  {year} {2023})},\ \Eprint {http://arxiv.org/abs/2309.14473} {arXiv:2309.14473
  [gr-qc]} \BibitemShut {NoStop}%
\bibitem [{\citenamefont {Payne}\ \emph {et~al.}(2022)\citenamefont {Payne},
  \citenamefont {Hourihane}, \citenamefont {Golomb}, \citenamefont {Udall},
  \citenamefont {Udall}, \citenamefont {Davis},\ and\ \citenamefont
  {Chatziioannou}}]{Payne:2022spz}%
  \BibitemOpen
  \bibfield  {author} {\bibinfo {author} {\bibfnamefont {Ethan}\ \bibnamefont
  {Payne}}, \bibinfo {author} {\bibfnamefont {Sophie}\ \bibnamefont
  {Hourihane}}, \bibinfo {author} {\bibfnamefont {Jacob}\ \bibnamefont
  {Golomb}}, \bibinfo {author} {\bibfnamefont {Rhiannon}\ \bibnamefont
  {Udall}}, \bibinfo {author} {\bibfnamefont {Richard}\ \bibnamefont {Udall}},
  \bibinfo {author} {\bibfnamefont {Derek}\ \bibnamefont {Davis}}, \ and\
  \bibinfo {author} {\bibfnamefont {Katerina}\ \bibnamefont {Chatziioannou}},\
  }\bibfield  {title} {\enquote {\bibinfo {title} {{Curious case of GW200129:
  Interplay between spin-precession inference and data-quality issues}},}\
  }\href {\doibase 10.1103/PhysRevD.106.104017} {\bibfield  {journal} {\bibinfo
   {journal} {Phys. Rev. D}\ }\textbf {\bibinfo {volume} {106}},\ \bibinfo
  {pages} {104017} (\bibinfo {year} {2022})},\ \Eprint
  {http://arxiv.org/abs/2206.11932} {arXiv:2206.11932 [gr-qc]} \BibitemShut
  {NoStop}%
\bibitem [{\citenamefont {{LIGO Scientific Collaboration}}\ \emph
  {et~al.}(2018)\citenamefont {{LIGO Scientific Collaboration}}, \citenamefont
  {{Virgo Collaboration}},\ and\ \citenamefont {{KAGRA
  Collaboration}}}]{lalsuite}%
  \BibitemOpen
  \bibfield  {author} {\bibinfo {author} {\bibnamefont {{LIGO Scientific
  Collaboration}}}, \bibinfo {author} {\bibnamefont {{Virgo Collaboration}}}, \
  and\ \bibinfo {author} {\bibnamefont {{KAGRA Collaboration}}},\ }\href
  {\doibase 10.7935/GT1W-FZ16} {\enquote {\bibinfo {title} {{LVK} {A}lgorithm
  {L}ibrary - {LALS}uite},}\ }\bibinfo {howpublished} {Free software (GPL)}
  (\bibinfo {year} {2018})\BibitemShut {NoStop}%
\bibitem [{\citenamefont {Macleod}\ \emph {et~al.}(2024)\citenamefont
  {Macleod}, \citenamefont {Coughlin}, \citenamefont {Southgate}, \citenamefont
  {Davis}, \citenamefont {Pitkin}, \citenamefont {rngeorge}, \citenamefont
  {paulaltin}, \citenamefont {Areeda}, \citenamefont {Godwin}, \citenamefont
  {Singer}, \citenamefont {Raymond}, \citenamefont {Quintero}, \citenamefont
  {aromerorodriguez}, \citenamefont {Massinger}, \citenamefont {Chanial},
  \citenamefont {Rozet}, \citenamefont {Goetz}, \citenamefont {Keitel},
  \citenamefont {Marx}, \citenamefont {Leinweber}, \citenamefont {Beroiz},\
  and\ \citenamefont {Badger}}]{duncan_macleod_2024_846265}%
  \BibitemOpen
  \bibfield  {author} {\bibinfo {author} {\bibfnamefont {Duncan}\ \bibnamefont
  {Macleod}}, \bibinfo {author} {\bibfnamefont {Scott}\ \bibnamefont
  {Coughlin}}, \bibinfo {author} {\bibfnamefont {Alex}\ \bibnamefont
  {Southgate}}, \bibinfo {author} {\bibfnamefont {Derek}\ \bibnamefont
  {Davis}}, \bibinfo {author} {\bibfnamefont {Matt}\ \bibnamefont {Pitkin}},
  \bibinfo {author} {\bibnamefont {rngeorge}}, \bibinfo {author} {\bibnamefont
  {paulaltin}}, \bibinfo {author} {\bibfnamefont {Joseph}\ \bibnamefont
  {Areeda}}, \bibinfo {author} {\bibfnamefont {Patrick}\ \bibnamefont
  {Godwin}}, \bibinfo {author} {\bibfnamefont {Leo}\ \bibnamefont {Singer}},
  \bibinfo {author} {\bibfnamefont {Vivien}\ \bibnamefont {Raymond}}, \bibinfo
  {author} {\bibfnamefont {Eric}\ \bibnamefont {Quintero}}, \bibinfo {author}
  {\bibnamefont {aromerorodriguez}}, \bibinfo {author} {\bibfnamefont {Thomas}\
  \bibnamefont {Massinger}}, \bibinfo {author} {\bibfnamefont {Pierre}\
  \bibnamefont {Chanial}}, \bibinfo {author} {\bibfnamefont {François}\
  \bibnamefont {Rozet}}, \bibinfo {author} {\bibfnamefont {Evan}\ \bibnamefont
  {Goetz}}, \bibinfo {author} {\bibfnamefont {David}\ \bibnamefont {Keitel}},
  \bibinfo {author} {\bibfnamefont {Ethan}\ \bibnamefont {Marx}}, \bibinfo
  {author} {\bibfnamefont {Katrin}\ \bibnamefont {Leinweber}}, \bibinfo
  {author} {\bibfnamefont {Martin}\ \bibnamefont {Beroiz}}, \ and\ \bibinfo
  {author} {\bibfnamefont {The~Gitter}\ \bibnamefont {Badger}},\ }\href
  {\doibase 10.5281/zenodo.846265} {\enquote {\bibinfo {title} {gwpy/gwpy: Gwpy
  3.0.8},}\ } (\bibinfo {year} {2024})\BibitemShut {NoStop}%
\bibitem [{\citenamefont {Romero-Shaw}\ \emph {et~al.}(2019)\citenamefont
  {Romero-Shaw}, \citenamefont {Lasky},\ and\ \citenamefont
  {Thrane}}]{Romero-Shaw:2019itr}%
  \BibitemOpen
  \bibfield  {author} {\bibinfo {author} {\bibfnamefont {Isobel~M.}\
  \bibnamefont {Romero-Shaw}}, \bibinfo {author} {\bibfnamefont {Paul~D.}\
  \bibnamefont {Lasky}}, \ and\ \bibinfo {author} {\bibfnamefont {Eric}\
  \bibnamefont {Thrane}},\ }\bibfield  {title} {\enquote {\bibinfo {title}
  {{Searching for Eccentricity: Signatures of Dynamical Formation in the First
  Gravitational-Wave Transient Catalogue of LIGO and Virgo}},}\ }\href
  {\doibase 10.1093/mnras/stz2996} {\bibfield  {journal} {\bibinfo  {journal}
  {Mon. Not. Roy. Astron. Soc.}\ }\textbf {\bibinfo {volume} {490}},\ \bibinfo
  {pages} {5210--5216} (\bibinfo {year} {2019})},\ \Eprint
  {http://arxiv.org/abs/1909.05466} {arXiv:1909.05466 [astro-ph.HE]}
  \BibitemShut {NoStop}%
\bibitem [{\citenamefont {Ashton}\ \emph
  {et~al.}(2019{\natexlab{b}})\citenamefont {Ashton}, \citenamefont {Hübner},
  \citenamefont {Lasky},\ and\ \citenamefont
  {Talbot}}]{greg_ashton_2019_2602178}%
  \BibitemOpen
  \bibfield  {author} {\bibinfo {author} {\bibfnamefont {Greg}\ \bibnamefont
  {Ashton}}, \bibinfo {author} {\bibfnamefont {Moritz}\ \bibnamefont
  {Hübner}}, \bibinfo {author} {\bibfnamefont {Paul}\ \bibnamefont {Lasky}}, \
  and\ \bibinfo {author} {\bibfnamefont {Colm}\ \bibnamefont {Talbot}},\ }\href
  {\doibase 10.5281/zenodo.2602178} {\enquote {\bibinfo {title} {Bilby: A
  user-friendly bayesian inference library},}\ } (\bibinfo {year}
  {2019}{\natexlab{b}})\BibitemShut {NoStop}%
\bibitem [{\citenamefont {Koposov}\ \emph {et~al.}(2023)\citenamefont
  {Koposov}, \citenamefont {Speagle}, \citenamefont {Barbary}, \citenamefont
  {Ashton}, \citenamefont {Bennett}, \citenamefont {Buchner}, \citenamefont
  {Scheffler}, \citenamefont {Cook}, \citenamefont {Talbot}, \citenamefont
  {Guillochon}, \citenamefont {Cubillos}, \citenamefont {Ramos}, \citenamefont
  {Johnson}, \citenamefont {Lang}, \citenamefont {Ilya}, \citenamefont
  {Dartiailh}, \citenamefont {Nitz}, \citenamefont {McCluskey},\ and\
  \citenamefont {Archibald}}]{sergey_koposov_2023_8408702}%
  \BibitemOpen
  \bibfield  {author} {\bibinfo {author} {\bibfnamefont {Sergey}\ \bibnamefont
  {Koposov}}, \bibinfo {author} {\bibfnamefont {Josh}\ \bibnamefont {Speagle}},
  \bibinfo {author} {\bibfnamefont {Kyle}\ \bibnamefont {Barbary}}, \bibinfo
  {author} {\bibfnamefont {Gregory}\ \bibnamefont {Ashton}}, \bibinfo {author}
  {\bibfnamefont {Ed}~\bibnamefont {Bennett}}, \bibinfo {author} {\bibfnamefont
  {Johannes}\ \bibnamefont {Buchner}}, \bibinfo {author} {\bibfnamefont {Carl}\
  \bibnamefont {Scheffler}}, \bibinfo {author} {\bibfnamefont {Ben}\
  \bibnamefont {Cook}}, \bibinfo {author} {\bibfnamefont {Colm}\ \bibnamefont
  {Talbot}}, \bibinfo {author} {\bibfnamefont {James}\ \bibnamefont
  {Guillochon}}, \bibinfo {author} {\bibfnamefont {Patricio}\ \bibnamefont
  {Cubillos}}, \bibinfo {author} {\bibfnamefont {Andrés~Asensio}\ \bibnamefont
  {Ramos}}, \bibinfo {author} {\bibfnamefont {Ben}\ \bibnamefont {Johnson}},
  \bibinfo {author} {\bibfnamefont {Dustin}\ \bibnamefont {Lang}}, \bibinfo
  {author} {\bibnamefont {Ilya}}, \bibinfo {author} {\bibfnamefont {Matthieu}\
  \bibnamefont {Dartiailh}}, \bibinfo {author} {\bibfnamefont {Alex}\
  \bibnamefont {Nitz}}, \bibinfo {author} {\bibfnamefont {Andrew}\ \bibnamefont
  {McCluskey}}, \ and\ \bibinfo {author} {\bibfnamefont {Anne}\ \bibnamefont
  {Archibald}},\ }\href {\doibase 10.5281/zenodo.8408702} {\enquote {\bibinfo
  {title} {joshspeagle/dynesty: v2.1.3},}\ } (\bibinfo {year}
  {2023})\BibitemShut {NoStop}%
\bibitem [{\citenamefont {Hoy}\ \emph {et~al.}(2021)\citenamefont {Hoy},
  \citenamefont {Raymond}, \citenamefont {Vijaykumar}, \citenamefont {Macleod},
  \citenamefont {Talbot}, \citenamefont {Pitkin}, \citenamefont {Fauchon},
  \citenamefont {Ashton}, \citenamefont {Sarin}, \citenamefont {Harry},
  \citenamefont {Veitch}, \citenamefont {Fischer},\ and\ \citenamefont
  {Khan}}]{charlie_hoy_2021_5237238}%
  \BibitemOpen
  \bibfield  {author} {\bibinfo {author} {\bibfnamefont {Charlie}\ \bibnamefont
  {Hoy}}, \bibinfo {author} {\bibfnamefont {Vivien}\ \bibnamefont {Raymond}},
  \bibinfo {author} {\bibfnamefont {Aditya}\ \bibnamefont {Vijaykumar}},
  \bibinfo {author} {\bibfnamefont {Duncan}\ \bibnamefont {Macleod}}, \bibinfo
  {author} {\bibfnamefont {Colm}\ \bibnamefont {Talbot}}, \bibinfo {author}
  {\bibfnamefont {Matt}\ \bibnamefont {Pitkin}}, \bibinfo {author}
  {\bibfnamefont {Edward}\ \bibnamefont {Fauchon}}, \bibinfo {author}
  {\bibfnamefont {Gregory}\ \bibnamefont {Ashton}}, \bibinfo {author}
  {\bibfnamefont {Nikhil}\ \bibnamefont {Sarin}}, \bibinfo {author}
  {\bibfnamefont {Ian}\ \bibnamefont {Harry}}, \bibinfo {author} {\bibfnamefont
  {John}\ \bibnamefont {Veitch}}, \bibinfo {author} {\bibfnamefont {Nils~Leif}\
  \bibnamefont {Fischer}}, \ and\ \bibinfo {author} {\bibfnamefont {Sebastian}\
  \bibnamefont {Khan}},\ }\href {\doibase 10.5281/zenodo.5237238} {\enquote
  {\bibinfo {title} {pesummary/pesummary: 0.13.0 release},}\ } (\bibinfo {year}
  {2021})\BibitemShut {NoStop}%
\bibitem [{\citenamefont {Woodbury}(1950)}]{Woodbury:1950}%
  \BibitemOpen
  \bibfield  {author} {\bibinfo {author} {\bibfnamefont {Max~A.}\ \bibnamefont
  {Woodbury}},\ }\href@noop {} {\emph {\bibinfo {title} {Inverting modified
  matrices}}}\ (\bibinfo  {publisher} {Princeton University, Princeton, NJ},\
  \bibinfo {year} {1950})\ p.~\bibinfo {pages} {4},\ \bibinfo {note}
  {statistical Research Group, Memo. Rep. no. 42,}\BibitemShut {NoStop}%
\end{thebibliography}%

\end{document}